\documentclass[12pt,showpacs,amsmath,amssymb,aps,floats,aps,floatfix,nofootinbib]{revtex4}%
\usepackage{epsfig}
\usepackage{dcolumn}
\usepackage{bm}
\usepackage{amsmath}
\usepackage{amsfonts}
\usepackage{amssymb}
\usepackage{graphicx}
\usepackage{latexsym}
\usepackage{color}
\def\nnd{\end{document}}

\def\be{\begin{equation}}
\def\ee{\end{equation}}

\newcommand{\bea}{\begin{eqnarray}}
\newcommand{\eea}{\end{eqnarray}}
\newcommand{\bwt}{\begin{widetext}}
\newcommand{\ewt}{\end{widetext}}

\def\u
\def\hZ{\widehat Z}

\def\eed{\end{document}}

\def\m_z{m_{\textrm {Z}}}

\renewcommand{\u}{\rm{u}}

\def\be{\beta}

\def\sl#1{#1\!\!\!\!/}
\def\rm#1{\textrm{#1}}

\makeatother
\begin{document}
\title{Light stop/sbottom pair production searches in the NMSSM}
\author{Xiao-Jun Bi$^1$}
\author{Qi-Shu Yan$^2$}
\author{Peng-Fei Yin$^1$}
\affiliation{$^1$Laboratory of Particle Astrophysics, Institute of High Energy Physics, Chinese Academy of Sciences,
Beijing 100049, China}
\affiliation{$^2$College of Physics Sciences, University of Chinese Academy of Sciences, Beijing 100049, China}

\begin{abstract}
In this work, we study the constraints on the scenario of light
stops and sbottoms in the next-to-minimal supersymmetric
standard model (NMSSM), especially by a 125 GeV Higgs boson
discovery and the LHC bounds on supersymmetry. The constraints from dark matter detections are also taken into account.
From the parameter scan, we find that the NMSSM can well accommodate a light Higgs boson
around 125 GeV and decay patterns.
We would like to stress that the LHC direct supersymmetry searches with b-tagging are very powerful and can set strong bounds on
many NMSSM parameter points with light stops and sbottoms. We find  $\tilde {t}\to b\tilde{\chi}^+_1$ is a very promising channel for light stop detection if the mass splitting between $\tilde{\chi}^+_1$ and $\tilde{\chi}^0_1$ is very small.
It is also pointed out that in order to close the parameter space of
light stops and sbottoms, new search strategies for signal channels such as $p p  \to
{\tilde t_1} {\tilde t_1} \to t {\bar t} h h \tilde{\chi}^0_1
\tilde{\chi}^0_1$ and $p p \to {\tilde b_1}
{\tilde b_1} \to t {\bar t} W^+ W^- \tilde{\chi}^0_1 \tilde{\chi}^0_1$ may be necessary.

\end{abstract}

\maketitle

\section{Introduction}

The searches from ATLAS
\cite{:2012gk}, CMS \cite{:2012gu}, and CDF \cite{Aaltonen:2012if}
as well as D0 \cite{:2012tf} established that there is a new
particle around $125 \sim 127$ GeV whose decay pattern is
consistent with the predicted Higgs boson of the standard model
(SM). If this new particle is a fundamental Higgs boson, it
is reasonable to ask whether it is the predicted
light CP-even Higgs boson in the supersymmetric models.
However, a Higgs boson of $125$ GeV seems a bit heavy for the
minimal supersymmetric standard model (MSSM). In the MSSM, the
tree-level mass of the lighter CP-even Higgs boson should always
be smaller than the mass of Z boson, while the loop effects of stop can
lift the Higgs boson mass up to 130 GeV or so. In order to avoid a
fine-tuning problem, natural supersymmetry (SUSY) requires that
the third generation squarks are light \cite{Kitano:2006gv,Papucci:2011wy}.

Compared with the MSSM, a Higgs boson at $125$ GeV  can be
realized more naturally next-to-minimal
supersymmetric standard model (NMSSM) without confronting the severe fine-tuning problem, since it can have a much larger
tree-level Higgs mass. This might be one of the reasons why the NMSSM is
appealing except that it solves the notorious $\mu$ problem
in the MSSM. In Ref. \cite{Gunion:2012zd}, the authors
studied the constrained NMSSM with all parameters defined at the
grand unification scale. It is found that the tension of the
constrained MSSM for a 125 GeV Higgs boson mass and a light SUSY
mass spectrum can be relaxed in the NMSSM with more general
parameters.

Another reason that may favor the NMSSM is the Higgs decay mode.
The experimental data of the Higgs boson decay modes have shown a
possible excess in the diphoton channel. This may be a hint of
new physics \cite{Carena:2011aa,Cheung:2011nv,Carmi:2012in}.
In Refs. \cite{Ellwanger:2012ke,Ellwanger:2011aa}, it is shown that
the NMSSM can accommodate the 125 GeV Higgs and enhance the partial width of $h \to \gamma \gamma$ by reducing the
partial width of $h \to b {\bar b}$ via the mixing effect. A comprehensive
study considering light third generation sparticles in the NMSSM also found that
the diphoton decay branching fraction can be enhanced \cite{King:2012is}.
Moreover, light charged sparticles such as stops, staus,
charginos, and charged Higgs may also significantly contribute to the Higgs diphoton
decay channel \cite{SchmidtHoberg:2012yy}. More
studies on diphoton enhancement in both the MSSM and NMSSM contexts can be
found in \cite{Djouadi:1996pb,Carena:2012gp,Carena:2011aa,Kang:2012sy,Blum:2012ii,Cao:2012fz,Benbrik:2012rm,Kang:2012bv,King:2012tr}.

The LHC is extensively searching for SUSY particles in various
channels. Up to now, the null result puts tension to many SUSY
models. In Ref. \cite{Fowlie:2011mb}, it is found that the LHC sparticle
search with 1 fb$^{-1}$ at $\sqrt{s}=7$ TeV exacerbated the
tension between the $\delta(g-2)_\mu^{\textrm{SUSY}}$ anomaly and
the branching fraction $BR(B \to X_s \gamma)$ in the constrained
minimal supersymmetric standard model (CMSSM). Of course, such
conclusion is based on the assumed grand unified theory (GUT) relations among the soft
breaking terms. When such relations are released as demonstrated
in \cite{Sekmen:2011cz}, the conclusion can be relaxed. For the
pMSSM with 19-dimensional parameter space, the current LHC search
cannot put very restrict constraints on supersymmetry in general.

The LHC direct SUSY searches have set strong constraints on the masses of
gluino and the first two generations of squarks in mSUGRA and some simplified models \cite{:2012rz,:2012mfa}.
Recently, there are many works on the detection of light third
generation squarks and gluinos below 1 TeV. It is pointed
out that if
the stop is light, it can be hidden from the
LHC searches and can avoid severe LHC constraints \cite{Papucci:2011wy,Bi:2011ha,Desai:2011th,Ajaib:2011hs}. Therefore the natural SUSY scenario is still alive.
More studies
\cite{Drees:2012dd,Cao:2012rz,Han:2012fw,Bai:2012gs,Plehn:2012pr,Kaplan:2012gd,Alves:2012ft,Brust:2012uf}
are trying to improve the sensitivity of light stop searches.
For example, in \cite{Han:2012fw,Bai:2012gs}, the authors considered the stop
dileptonic final state and explored a few kinematic observables to distinguish signal and background.
The authors of Ref. \cite{Plehn:2012pr} studied the top tagging technique for the stop
search with semileptonic and dileptonic modes at the LHC.
The hadronic top tagging technique has been examined in Ref.
\cite{Kaplan:2012gd}.
Interesting multiple lepton and jet
final states from the multiple top decays are investigated in
\cite{Bramante:2011xd}.
Moreover, light sbottom searches are investigated in
\cite{AdeelAjaib:2011ec,Alvarez:2012wf,Lee:2012sy}. In particular, the
sbottom-neutralino coannihilation scenario has been considered, and it was
found that the LHC has a good sensitivity to the parameter space by using
tagged a b-jet even for small mass splitting between the sbottom and
neutralino.

Since in the NMSSM the light stop and sbottom can be natural for a 125
GeV Higgs boson \cite{King:2012is,Ellwanger:2012ke},
we perform a systematic study on the constraints on the
light stop/sbottom scenario by using the LHC SUSY search results.
The constraints from B physics measurements and dark matter (DM)
detections as well as the Higgs boson mass on the NMSSM
parameter space are first considered (for more studies on the constraints on
the light stop/sbottom scenario, see \cite{Choudhury:2012tc,Espinosa:2012in,Wymant:2012zp}). Then we study the constraints
from various SUSY search channels at the LHC with
$\sqrt{s}=7$ TeV and $2\sim5$ fb$^{-1}$ of data, including the jets +
MET channel, associated monojet channel, and lepton + jets +
MET channel with and without tagged b jet(s). Since both stops and
sbottoms can decay into b jets, it is expected that b tagging
should play an important role to distinguish the signal and
background. Our results confirm this point and find that the direct SUSY searches are powerful to
exclude many parameter points with light stops and sbottoms up to 500 GeV.

The paper is organized as follows. In Sec. II, we briefly
describe the NMSSM and our parameter scanning strategy. We will concentrate on the contribution of the light
stop to the mass of the discovered 125 GeV Higgs boson
and the constraints from dark matter searches on the neutralino
sector. In Sec. III, we analyze the LHC bounds
from both ATLAS and CMS on the signatures of the
light stop pair and sbottom pair production. Section IV is the discussions and
conclusions.

\section{The NMSSM}

In the NMSSM a singlet superfield $S$ is introduced to solve the so-called ``$\mu$ problem". The superpotential of NMSSM related to this singlet superfield $S$ is given by \cite{Ellwanger:2009dp,Maniatis:2009re}:
\begin{equation}
W_{NMSSM}=\lambda \hat{S} \hat{H}_u \hat{H}_d + \frac{1}{3} \kappa \hat{S}^3 + ... ,
\end{equation}
where the dots denote the MSSM superpotential without the $\mu$ term. When the electroweak symmetry is broken, the effective $\mu$ term can be naturally generated via the vacuum expectation value of $S$ filed (labeled as $v_S$), and can be written as $\mu_{eff}= \lambda v_S$, which is expected to be of $\cal{O}$(100) GeV (of the same size as the rest of the soft breaking terms). The soft breaking terms in the Higgs sector \cite{Ellwanger:2009dp,Maniatis:2009re}
are extended as
\begin{equation}
V_{NMSSM}= \tilde{m}^2_{H_{u}} |H_u|^2 + \tilde{m}^2_{H_{d}} |H_d|^2 + \tilde{m}^2_S |S|^2 + (A_\lambda \lambda S H_u H_d + \frac{1}{3}
A_\kappa \kappa S^3) + H.c.
\end{equation}
Compared with the MSSM, the Higgs sector becomes richer and contains three CP-even Higgs bosons, {\it i.e.}, $H_1$, $H_2$, and $H_3$, and two CP-odd Higgs bosons, {\it i.e.}, $A_1$ and $A_2$. Five new parameters $\lambda$, $\kappa$, $A_\lambda$, $A_\kappa$, and $\mu_{eff}$ are added compared with the MSSM.

\subsection{ The Parameter space}
\label{scan}

We use NMSSMTools
\cite{Ellwanger:2004xm} to perform a scan over the parameter space
of the NMSSM. To obtain more generic conclusions, we consider the
parameters defined at the electroweak scale in our scan without
assuming the unification of the NMSSM parameters at the GUT scale.
We vary them in the ranges defined as follows:
\begin{eqnarray}
&&  10^{-4} < \kappa < 0.5, \;\;\; 1 <\tan \beta < 60, \;\;\; 50< \mu < 500 \textrm{GeV}, \;\;\; \nonumber \\
&& |A_\lambda|< 4 \textrm{TeV}, \;\;\; |A_\kappa| < 500 \textrm{GeV}, \;\;\; 10 \textrm{GeV} < M_1 < 1 \textrm{TeV}, 100 \textrm{GeV}< M_2 <1 \textrm{TeV}, \;\;\; \nonumber \\
&& 100 \textrm{GeV}< m_{Q_3} , m_{U_3} <2 \textrm{TeV}, \;\;\; |A_{U_3}|< 3 \textrm{TeV}, \;\;\; 100 \textrm{GeV}< m_{\tilde{l}}<1 \textrm{TeV}.
\end{eqnarray}
There are a few comments in order on the ranges of the NMSSM parameters.
\begin{itemize}
\item The large $\lambda$ is helpful to raise the SM-like Higgs mass at tree level and to ameliorate the fine-tuning issue confronted by the MSSM. When $\lambda$ tends to be zero, the singlet $S$ will decouple from other Higgs fields. Under this limit, the phenomenology of the NMSSM may still be different from the MSSM due to the light singlet and singlino. These particles could affect the features of DM. The decay modes of heavy sparticles produced at the colliders may also change and some new search strategies will be necessary. Therefore, we adopt two scan strategies which allow $\lambda$ variations in the ranges of $[10^{-3}, 0.1]$ and $[0.1, 0.8]$ with a logarithmic and flat distribution, respectively.
    In the NMSSM, $\lambda$ should be smaller than $\sim$ 0.7 when the theory is assumed to be perturbative up to the GUT scale. In this work, we focus on the parameters at the low energy scale and neglect such constraints. For the discussions of perturbation constraints in the NMSSM, see Ref. \cite{King:2012is}.

\item For the gluino and the first two generations of squarks, if their decay products are energetic jets and large MET can be reconstructed, the recent LHC results can put stringent limits on their masses in mSUGRA and phenomenological SUSY, e.g., $M_3 \sim m_{\tilde{q}_{1,2}} > 1.4$TeV \cite{:2012rz}. If the first two generations of squarks are very heavy, the SUSY flavor and CP problems can be solved \cite{Dimopoulos:1995mi}. For the gluino, the naturalness of Higgs mass requires that its mass should not be much larger than $\sim 1$TeV \cite{Papucci:2011wy}. The main decay mode of the gluino may be $\tilde{g} \to t \tilde{t} / b\tilde{b}$. For simplicity, in this work  we focus on the pair production of the third generation of squarks and leave this case for future study. Therefore we fix the soft breaking parameters $M_3=m_{\tilde{q}_{1,2}} = 1.5$TeV. To reduce the number of free parameters, we also assume that the mass parameters of the third generation of right-handed squarks are the same, {\it i.e.} $m_{D_3}=m_{U_3}$.

\item In our scan, we require the mass of the SM-like Higgs to be in the range of $125 \pm 2$GeV and the SM-like Higgs boson can be either $H_1$ or $H_2$. Additionally, several phenomenology and astrophysics experimental limits are also considered. For flavor constraints, we require BR$(B_s \to X_s \gamma)$, BR$(B^+ \to \tau^+ \nu_{\tau})$, BR$(B\to X_s \mu^+ \mu^-)$, $\Delta M_d$ and $\Delta M_s$ satisfying experimental constraints at $2 \sigma$ \cite{Barberio:2006bi,Asner:2010qj,Abulencia:2006ze}. The theoretical uncertainties in these observables are considered as implemented in NMSSMTools. For BR$(B_s \to \mu^+ \mu^-)$, the upper limits have evolved much in these two years. Recently ATLAS, CMS, and LHCb have updated it to $2.2 \times 10^{-8}$ \cite{Aad:2012pn}, $7.7 \times 10^{-9}$ \cite{Chatrchyan:2012rg} and $4.5 \times 10^{-9}$ \cite{Aaij:2012ac} at $95 \%$ confidence level, respectively, which are only a few times above the SM predictions. Here we adopt constraint BR$(B_s \to \mu^+ \mu^-)< 4.5 \times 10^{-9}$ given by LHCb. For the muon anomalous magnetic moment $\alpha_\mu$, we require that SUSY effects explain the discrepancy between the SM prediction and the experimental result at $2 \sigma$. The mass limits for Higgs and charged SUSY particles from LEP and Tevatron are adopted by the NMSSMTools package \cite{Ellwanger:2004xm}. The latest LHC Higgs limits are also taken into account \cite{ATLhgg,CMShll,CMShbb,CMShww,CMShzz}.

\item In our analysis, the lightest neutralino is required to be the lightest supersymmetric particle (LSP) and a candidate of the DM. Considering that there may be several types of DM in our Universe, the LSP in the NMSSM is just one kind of DM, We only require that the thermal abundance of neutralino satisfies a $3\sigma$ upper limit $\Omega_\chi h^2 < 0.1288$ with the correct DM relic density $\Omega h^2=0.112 \pm 0.0056$ reported by the WMAP \cite{Larson:2010gs}.  For this purpose, we define a fraction variable $\xi=\Omega_\chi h^2 / \Omega h^2$, and the neutralino density in halo is $\rho_\chi= \xi \rho_{DM}$. The DM detection limits given by experimental collaborations are obtained by assuming a certain DM density $\rho_{DM}\sim 0.3$ GeV cm$^{-3}$. Thus for the parameter points predicting $\Omega_\chi h^2 \ll \Omega h^2$, these limits need to be rescaled. By doing this, we can examine whether the LSP in the NMSSM can sufficiently accommodate all the data. The dark matter observations are calculated by Micromega \cite{micromega} implemented in NMSSMTools.

\end{itemize}

Because this numerical scan is performed over a multidimensional parameter space, we use a Markov chain Monte Carlo (MCMC) method to increase the scan efficiency in our analysis. The total likelihood function $L_{tot}=\Pi_i L_i$ is evaluated by the likelihood functions based on the phenomenology and astrophysics experimental observables described above. We define $L_i=e^{-\frac{(x_i-\mu_i)^2}{2\sigma_i^2}}$ for two-sided constraints and $L_i=1/(1+e^{\frac{x_i-\mu_i}{\sigma_i'}})$ for the upper limits \cite{Vasquez:2010ru}, where $x_i$ is the observable predicted by the model, $\mu_i \pm \sigma_i$ is the central values and error bars of experimental observables, and $\sigma'$ taken as $\sigma'=0.02 \mu_i$ is the tolerance for the upper limit.

\subsection{125 GeV Higgs Boson}

In the MSSM, the tree-level SM-like Higgs mass is smaller than $M_Z$ which is below the LEP limit $m_h < 114 $ GeV. However, it can be lifted by the loop corrections (say top-stop corrections due the large Yukawa couplings). The one loop formula for $m_h$ is given by
\begin{eqnarray}
m^2_h &=& m^2_{h,tree}+\Delta m^2_{h,loop}  \nonumber \\
&=& M^2_Z \cos^2 2\beta + \frac{3 m^4_t}{4 \pi^2 v^2} \left( \ln\left( \frac{M_{\tilde{t}}^2}{m^2_t}\right)+ \frac{X_t^2}{M^2_{\tilde{t}}} \left( 1- \frac{X^2_t}{12 M^2_{\tilde{t}}} \right) \right) ,
\end{eqnarray}
where $v=174$ is the vacuum expectation value of the SM Higgs, $M_{\tilde{t}}=\sqrt{m_{\tilde{t}_1} m_{\tilde{t}_2}}$ is related to the stop masses, and $X_t\equiv A_t-\mu\cot\beta$ is the stop mixing parameter. The Higgs mass logarithmically depends on stop masses
. This means that the stop bosons should be heavy in order to produce a  large enough correction to the mass of the Higgs boson. We can also see that the Higgs mass depends on stop mixing and is maximal for $X=X_t/M_{\tilde{t}}=\sqrt{6}$.

In the NMSSM, the superpotential $\lambda \hat{S} \hat{H}_u \hat{H}_d $ can induce a new term $\lambda^2 v^2 \sin 2\beta H_u H_d$ in the Higgs potential. After rotating the $2 \times 2$ mass matrix of two CP-even neutral Higgs of doublets $H_U$ and $H_D$ by angle $\beta$, one diagonal element becomes $M_Z^2\cos^2 2\beta + \lambda^2 v^2 \sin^2 2\beta$, which means that compared with the MSSM, the SM-like Higgs mass (mainly the $H_U$ type Higgs boson) obtains a new contribution $\sim \lambda^2 \sin^2 2\beta$ \cite{Ellwanger:2009dp}
\begin{equation}
m^2_h = M_Z^2\cos^2 2\beta + \lambda^2 v^2 \sin^2 2\beta+\Delta m^2_{h,loop}+... ,
\label{nmssmhm}
\end{equation}
where the dots denote the effects from mixing between Higgs doublets and the singlet. For large $\lambda$ with $\lambda v > M_Z$,
the $m_h$ is maximized for $\tan \beta=1$.

\begin{figure}[!htb]
\begin{center}
\includegraphics[width=0.45\columnwidth]{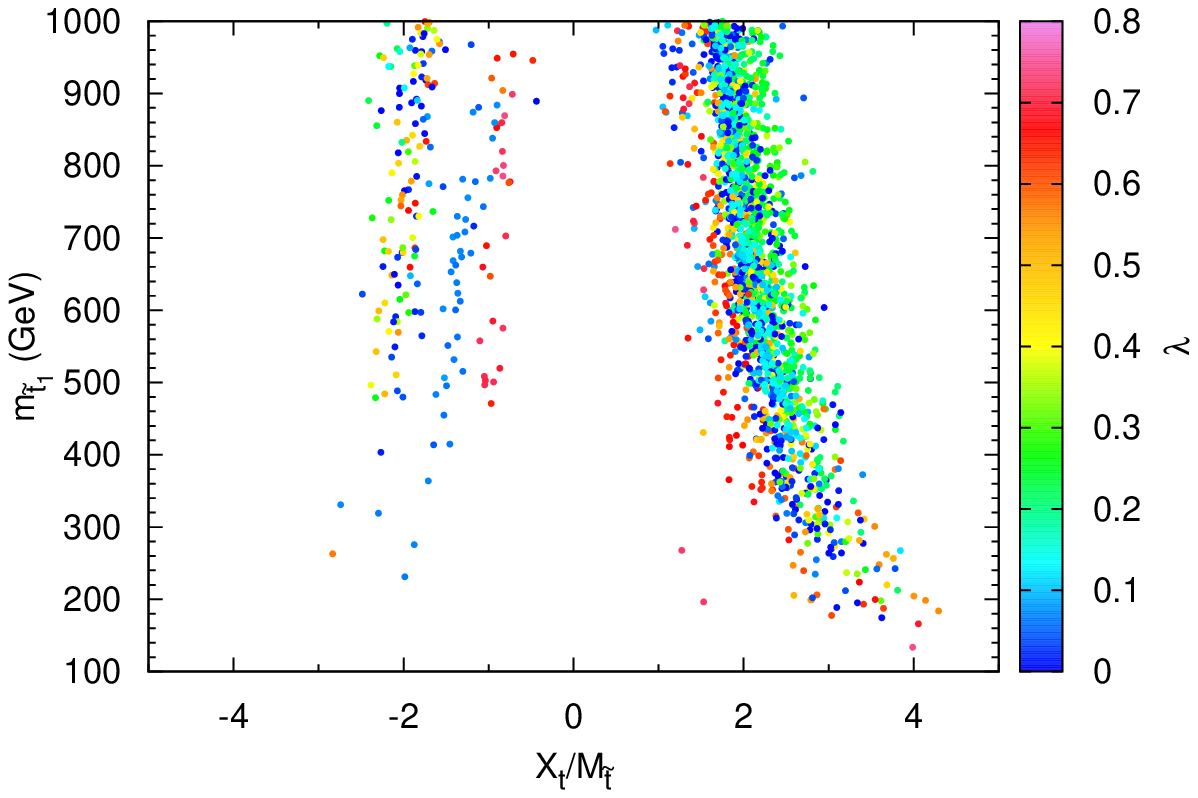}
\includegraphics[width=0.45\columnwidth]{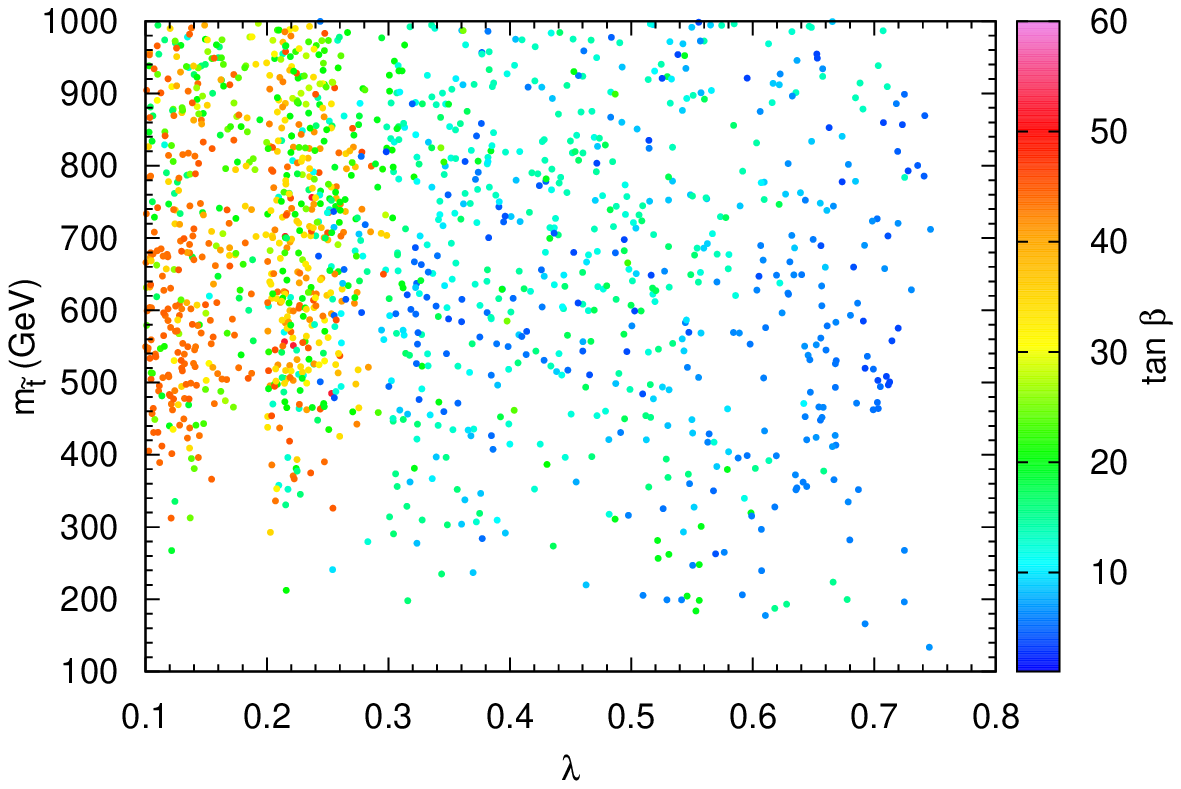}
\caption{Left: $m_{\tilde{t}_1}$ versus $X_{t}/M_{\tilde{t}}$ . Right: $m_{\tilde{t}_1}$ versus $\lambda$. The color scale indicates $\lambda$ (left) and $\tan\beta$ (right), respectively.
\label{hmrl}}
\end{center}
\end{figure}

It is interesting to ask how light the stop can be after taking into account the 125 GeV Higgs boson in the NMSSM. To address such a question, below we show some scattering plots. In these plots, all the points have passed the constraints as described in Sec. \ref{scan}.

In the left panel of Fig. \ref{hmrl}, we show the correlation in the $m_{\tilde{t}_1}-X$ plane; the color bar shows the value of $\lambda$. It is observed that for the large $\lambda \sim 0.6-0.7$, the stop mixing parameter $|X|$ is allowed to be 1.  When the light stop mass decreases, the stop mixing parameter will increase. If $m_{\tilde{t}_1}$ is below 300 GeV, $X$ should be larger than 3. This means there is a large mass splitting between two stop states. Because $\lambda$ can efficiently raise
the mass of the Higgs boson, it is easier to obtain a lighter $\tilde{t}_1$ with a larger $\lambda$.

In the right panel of Fig. \ref{hmrl}, we show the correlations between $m_{\tilde{t}_1}$ and $\lambda$; the color scale indicates $\tan \beta$. We also see that there are many points with small $m_{\tilde{t}_1}$ in the large $\lambda$ regime.
As shown in Eq. (\ref{nmssmhm}), the Higgs mass depends on $\lambda \sin 2\beta$; the values of $\tan \beta$ decrease at large
$\lambda$. We can see $\tan \beta < 30$ for $\lambda > 0.3$ and $\tan \beta < 10$ for $\lambda > 0.6$.

Another interesting question is whether the NMSSM can accommodate the decay modes and the decay branching fractions measured by the LHC collaborations appropriately. Especially whether the NMSSM can explain the diphoton excesses and diboson (mainly $h\to ZZ$) data observed by the LHC collaborations? Below we address this issue.

In the SM, if the Higgs mass is determined, all the Higgs interaction couplings to the SM particles can be obtained. In the new physics model, the Higgs couplings may differ from the SM predictions due to new parameters and particles. Therefore, the determination of the Higgs couplings at the LHC are very important to test the SM and can provide crucial evidence of new physics.

The effective Higgs couplings can be extracted from experimental data and can be compared with theoretical predictions. The modifications to the Higgs couplings to SM particles are denoted by
\begin{equation}
C_{h XX}\equiv \bar{C}^{NP}_{hXX}/\bar{C}^{SM}_{hXX} ,
\end{equation}
where $X$ can denote either heavy fermions, $W$ bosons, $Z$ bosons, photons, or gluons. In the new physics model, both the production cross section and decay width of the Higgs are rescaled by $C^2_h$. The relevant Higgs partial widths would be determined by the ratio
\begin{equation}
R_{hXX}=\frac{\sigma (pp \to h)_{NP} BR(h \to XX)_{NP} }{\sigma (pp \to h)_{SM} BR(h \to XX)_{SM} } .
\end{equation}
For the $\gamma \gamma$, $W^+ W^-$, and $ZZ$ channel, the recorded Higgs events are mainly from the
gluon fusion $gg \to h$ process. If the decay channel of the 125 GeV Higgs is dominated by $h \to b \bar{b}$,
the $R_{hXX}$ ($X=W, Z, \gamma$) is approximated to be $R_{hXX} \sim C^2_{hgg} BR(h\to XX)_{NP}/BR(h \to XX)_{SM}    \sim C^2_{hgg} C^2_{hXX} /C^2_{hb\bar{b}}$. For the $b \bar{b}$ channel, additional electrons or muons are required to suppress the huge QCD background; only the electroweak production
channel $ q \bar{q} \to hV $ is used to search for the Higgs signal. Therefore, the $R_{hXX}$ can be given by
$R_{hb\bar{b}} \sim C^2_{hVV} BR(h\to b\bar{b})_{NP}/BR(h \to b\bar{b})_{SM}$.

In the NMSSM, the Higgs mass basis $H^{mass}_i=\{H_1, H_2, H_3 \}$ and interaction basis $H^{int}_a=\{H_d, H_u, S\}$ are related by $H^{mass}_i=S_{ia} H^{int}_a$. The reduced Higgs couplings to fermions and heavy gauge bosons can be given by
\begin{equation}
C_{hb \bar{b}}=C_{h\tau \bar{\tau}}= \frac{S_{i1}}{\cos \beta},\;\;\; C_{ht \bar{t}}= \frac{S_{i2}}{\sin \beta},\;\;\; C_{hV V}= S_{i1}\cos \beta+ S_{i2} \sin \beta.
\label{Chffvv}
\end{equation}
In the SM, effective Higgs coupling to gluons $C_{h gg}$ is dominantly determined by the triangle top loop. In the SUSY model, the stop loop would also contribute to $C_{h gg}$; it can be written as
\begin{equation}
C_{hgg} = \frac{\bar{C}^{SUSY}_{h gg,\; t} + \bar{C}^{SUSY}_{h gg,\; \tilde {t}} }{ \bar{C}^{SM}_{h gg,\; t} } \sim C_{ht \bar{t}}+ C_{\tilde {t}} ,
\label{Chgg}
\end{equation}
where $\bar{C}_{hXX, \; A}$ is the loop contribution from particle $A$ to the effective Higgs coupling $C_{hXX}$, and $C_{\tilde {t}}$ is defined as $C_{\tilde {t}}=\bar{C}^{SUSY}_{h gg,\; \tilde {t}}/\bar{C}^{SM}_{h gg,\; t} $. Here, by using the NMSSMTools package, the higher order QCD corrections to the gluon fusion cross section have not been taken into account. These effects are important for calculating the SM Higgs production cross section \cite{Dittmaier:2011ti}. The SUSY-QCD corrections from light stops can also modify the Higgs production cross section depending on the squark masses and mixing angle \cite{Dawson:1996xz,Harlander:2003bb,Degrassi:2010eu,Anastasiou:2006hc,Muhlleitner:2006wx,Anastasiou:2008rm,Pak:2010cu}. Since the Higgs production cross section including next-to-next-to leading order (NNLO) QCD corrections has been adopted by the experimental collaborations \cite{:2012gk,:2012gu}, it is also necessary to take into account such important corrections in the NMSSM. More detailed discussions and treatments can be found in Ref. \cite{King:2012is} where the NNLO QCD corrections are included by using the package HIGLU \cite{Spira:1995mt}.

For the $C_{h \gamma \gamma}$, the main contributions arise from the W loop and top loop, which are related by $\bar{C}^{SM}_{h \gamma \gamma,\; W} / \bar{C}^{SM}_{h \gamma \gamma,\; t} \sim  -8.3/1.8 $ in the SM. In the SUSY model, light charged particles \cite{Djouadi:1996pb} such as light charginos, charged Higgs, stops, sbottoms, and staus would provide additional contributions. We can also get $C_{h \gamma \gamma}$ approximately
\begin{equation}
C_{h\gamma \gamma} = \frac{\bar{C}^{SUSY}_{h \gamma \gamma,\; t} + \bar{C}^{SUSY}_{h \gamma \gamma,\; W} +  \bar{C}^{SUSY}_{h \gamma \gamma} }{ \bar{C}^{SM}_{h \gamma \gamma,\; t} + \bar{C}^{SM}_{h \gamma \gamma,\; W } } \sim
1.28 C_{h VV} -0.28 (C_{h t \bar{t}} + C_{\tilde {t}}) + C_{\tilde {\tau}} + C_{\tilde {\chi}^+},
\label{Chgaga}
\end{equation}
where we have used the relations $\bar{C}^{SUSY}_{h \gamma \gamma,\; W}/\bar{C}^{SM}_{h \gamma \gamma,\; W} \sim C_{h VV} $, $\bar{C}^{SUSY}_{h \gamma \gamma,\; t}/\bar{C}^{SM}_{h \gamma \gamma,\; t} \sim C_{h t \bar{t}} $, and $\bar{C}^{SUSY}_{h \gamma \gamma,\; \tilde {t}}/\bar{C}^{SM}_{h \gamma \gamma,\; t} \sim \bar{C}^{SUSY}_{h gg,\; \tilde {t}}/\bar{C}^{SM}_{h gg,\; t}$
(neglecting the high order QCD corrections induced by the stop), and $C_{\tilde {\tau}}$ and $C_{\tilde {\chi}^+}$ are defined as $C_{\tilde {\tau}}=\bar{C}^{SUSY}_{h \gamma \gamma,\; \tilde {\tau}}/\bar{C}^{SM}_{h \gamma \gamma}$ and $C_{\tilde {\chi}^+}=\bar{C}^{SUSY}_{h \gamma \gamma,\; \tilde {\chi}^+}/\bar{C}^{SM}_{h \gamma \gamma}$, respectively.. Note that $C_{\tilde {\tau}}$ is proportional to the Yukawa coupling $y_{h\tau \bar{\tau}}$ and inversely proportional to the stau mass $m_{\tau}$. This means $C_{\tilde {\tau}}$ is important for large $\tan \beta > 50\sim 60$ and light stau \cite{Carena:2011aa}.

An important feature of Higgs phenomenology in the NMSSM is the exotic Higgs decay modes to light neutralinos or scalars \cite{Cao:2012im,Vasquez:2012hn}. If $H_1$ is SM-like, $H_2$ might be much heavier than $125$ GeV. If $H_2$ is SM-like, $H_1$ and $A_1$ would be light singlets.
In this case, the branching ratios of $H_2 \to \tilde{\chi}^0_1 \tilde{\chi}^0_1$, $H_2 \to H_1 H_1$ and $H_2 \to A_1 A_1$ could be large due to Higgs mixing and kinematics. For the invisible Higgs decay $H_2 \to \tilde{\chi}^0_1 \tilde{\chi}^0_1$,
the possible search channels are $f \bar{f} \to Vh \to$ leptons+MET or $gg \to hj \to monojet+MET$. For the decay channels $H_2 \to H_1 H_1$ and $H_2 \to A_1 A_1$, the light singlets would decay into $\tau \bar{\tau}$ or $b \bar{b}$. The final states are four fermions which can be significant for some parameter points. However, these exotic decay modes might suppress
the standard Higgs decay modes to heavy gauge bosons, photons, and $b \bar{b}$, and could be tested in the future by the global-fitting of Higgs decay partial widths.

\begin{figure}[!htb]
\begin{center}
\includegraphics[width=0.32\columnwidth]{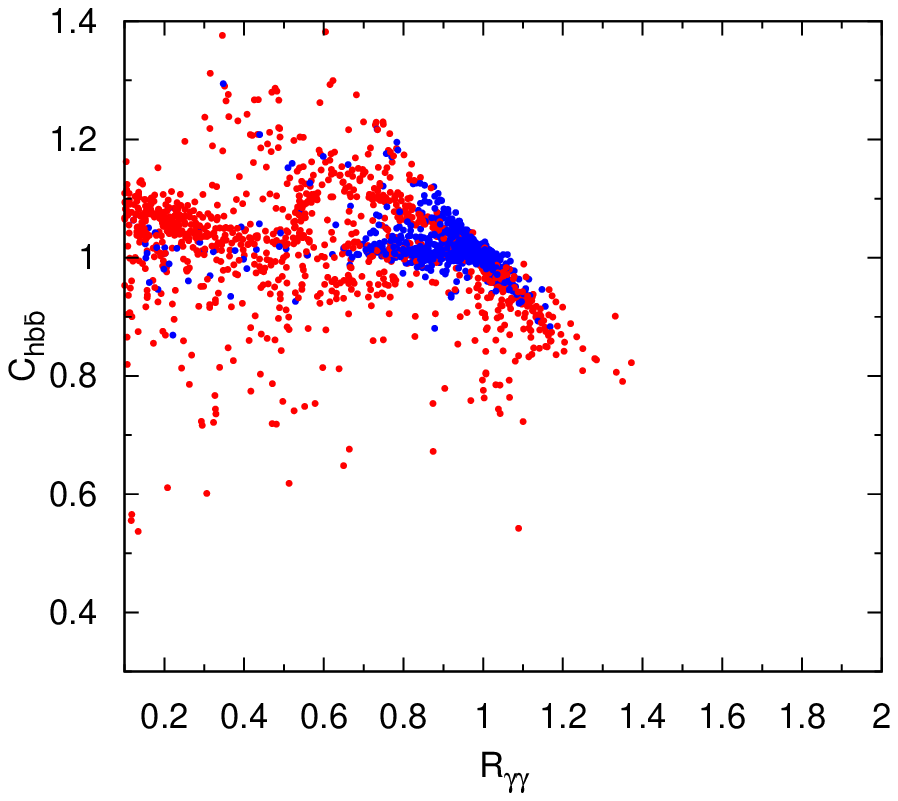}
\includegraphics[width=0.32\columnwidth]{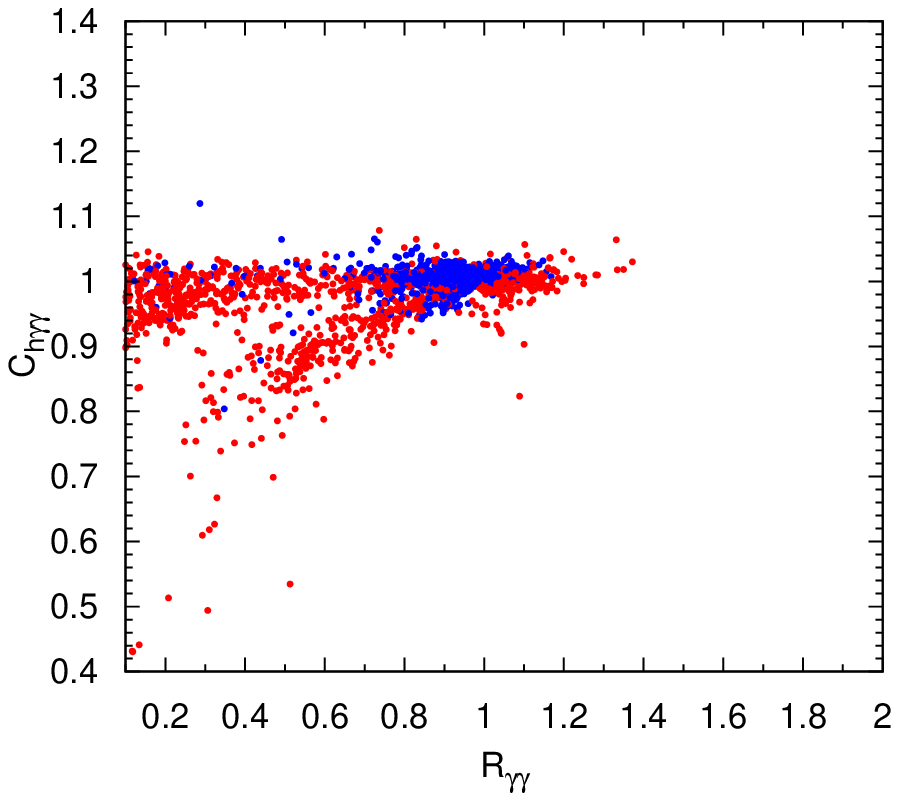}
\includegraphics[width=0.32\columnwidth]{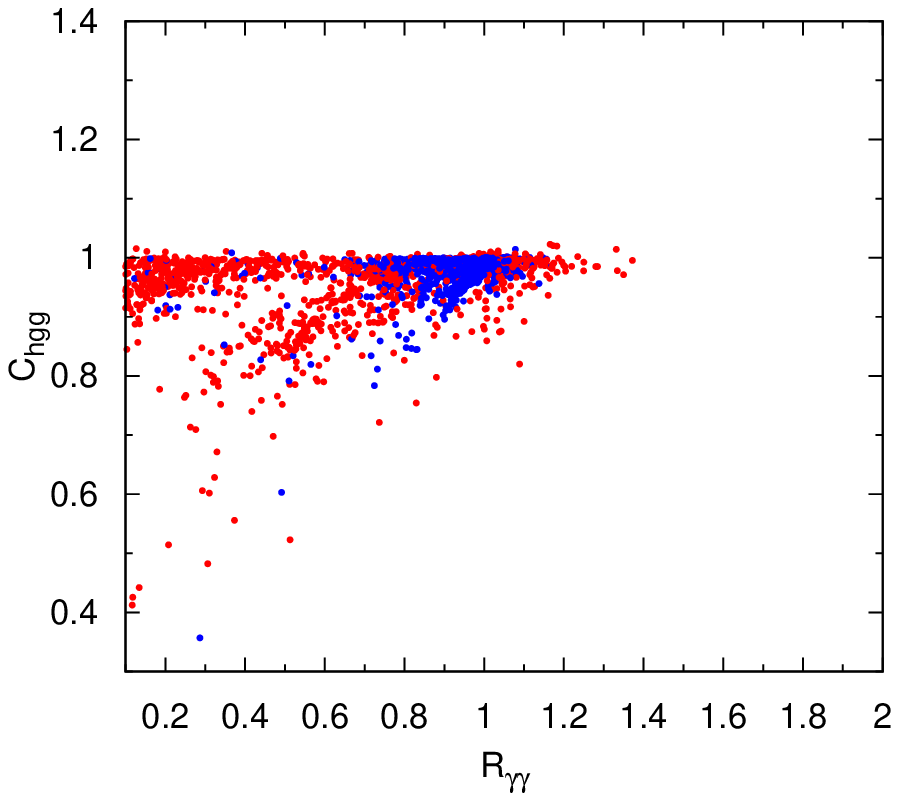}
\caption{The correlations between $R_{\gamma \gamma}$ and $C_{h b \bar{b}}$/$C_{h \gamma \gamma}$/$C_{h gg}$ are demonstrated. The blue/red points denote that the SM-like Higgs boson is $H_1$/$H_2$.
\label{RggC}}
\end{center}
\end{figure}

To examine the question whether the NMSSM can accommodate the diphoton excess, we show scattering plots to demonstrate the correlations between the effective couplings and $R_{\gamma \gamma}$. In the left/middle/right panel of Fig. \ref{RggC}, we demonstrate the correlations between $R_{\gamma \gamma}$ with $C_{h b \bar{b}}$/$C_{h \gamma \gamma}$/$C_{h gg}$; the blue/red points denote that the SM-like Higgs is $H_1$/$H_2$. We can see $C_{h \gamma \gamma}$ and $C_{hgg}$ always vary in the range of $\sim 0.8 -1.1$, and $R_{\gamma \gamma}$ is sensitive to $C_{h b \bar{b}}$. We also find $R_{\gamma \gamma}$ is inversely proportional to $R_{b \bar{b}}$ and $R_{b \bar{b}}$ is larger than 1 for $R_{\gamma \gamma}> 1$.

These results can be understood in terms of the explanation given above. In the decoupling limit $M_A \gg M_Z$ ($M_A$ is the mass of the heavy CP-odd Higgs), the main component of the SM-like Higgs is $H_u$ due to $S_{i2}\sim \sin\beta \sim 1 $ for $\tan \beta \gg 1$. From Eq. (\ref{Chffvv}), we can get $C_{ht \bar{t}}$ and $C_{hWW}$. If the stop mass parameter $\sqrt{m_{\tilde{t}_1} m_{\tilde{t}_2}}$ is much larger than $m_t$, the stop loop would not provide larger contributions to $C_{h gg}$. Then $C_{hgg} \sim 1$ and $C_{h \gamma \gamma} \sim 1$ can be obtained from Eqs. (\ref{Chgg}) and (\ref{Chgaga}), respectively. Depending on the parameters in the Higgs sector, $C_{h b \bar{b}}$ can be significantly reduced by the mixing effect. As we have mentioned, if the main SM-like Higgs decay mode is $h \to b \bar{b}$, and $R_{hXX}$($X= W,Z,\gamma$) is approximately to be $R_{hXX} \sim C^2_{hgg} C^2_{hXX} /C^2_{hb\bar{b}} \sim 1/C^2_{hb\bar{b}}$. This relation explains the  inversely proportional correlation between $R_{\gamma \gamma}$ and $C_{h b \bar{b}}$ for $R_{\gamma \gamma} > 0.8$, as shown in Fig. \ref{RggC}. This is the case for the SM-like Higgs $H_1$.

If $H_2$ is SM-like, then the branching ratios of exotic decay modes might be significant, and the $R_{\gamma \gamma}$ would be much suppressed even if the $C_{h b \bar{b}}$ is still $\sim 1$. For all the points, small $R_{b \bar{b}}$ and large $R_{\gamma \gamma}$ can be acquired by tuning the singlet-doublet mixing parameters.

\begin{figure}[!htb]
\begin{center}
\includegraphics[width=0.45\columnwidth]{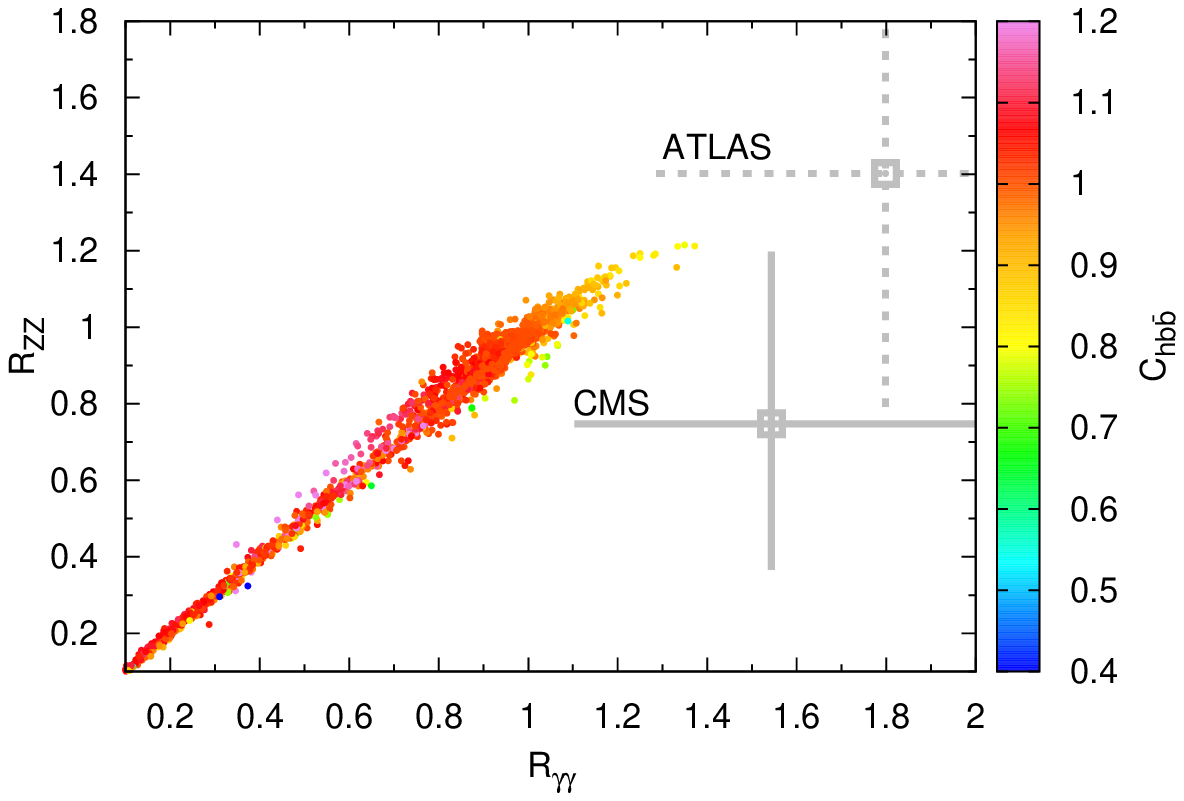}
\includegraphics[width=0.45\columnwidth]{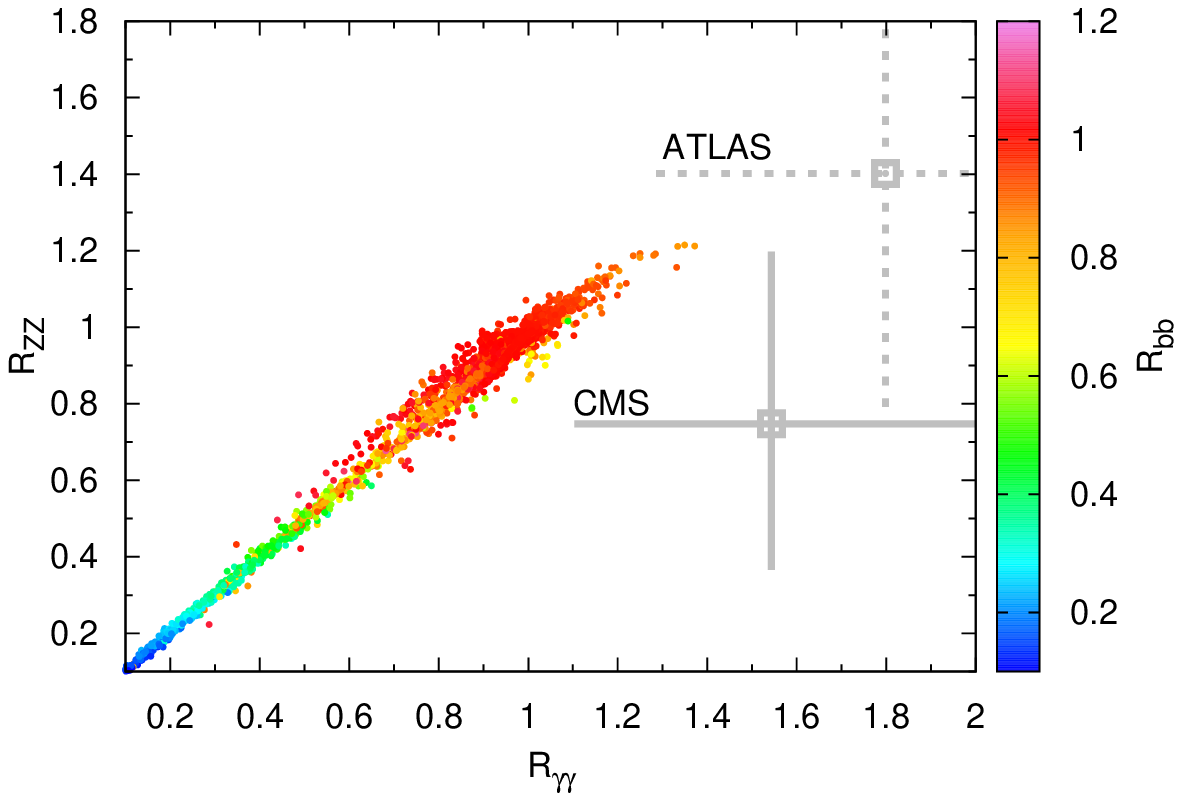}
\caption{The correlations between $R_{ZZ}$, $R_{\gamma \gamma}$ and $C_{b {\bar b}}$/$R_{b{\bar b}}$ are displayed in the left/right panel. The observed values given by ATLAS and CMS are also shown.
\label{RggRvv}}
\end{center}
\end{figure}

We also show the correlations between $R_{\gamma \gamma}$ and $R_{ZZ}$ in Fig. \ref{RggRvv}, the color bar indicates $C_{h b \bar{b}}$ ($R_{b \bar{b}}$) in the left (right) panel of Fig. \ref{RggRvv}. From the left panel, we can also see that $R_{b \bar{b}}$ is smaller than 1 for $R_{\gamma \gamma}> 1$. Moreover, the $R_{\gamma \gamma}$ is always proportional to $R_{VV}$ due to the approximations $R_{hVV}/R_{\gamma \gamma} \sim C^2_{hVV}/C^2_{h \gamma \gamma} $ and $C_{\gamma \gamma} \sim 1.28 C_{VV} - 0.28 C_{t \bar{t}} + C_{SUSY}$. As we mentioned above, $R_{hb\bar{b}}$ depends on $C_{VV}$ due to the production process. As $C_{VV}$ is almost $\sim 0.8 \sim 1.1$ shown in Fig. \ref{RggC}, $R_{hb\bar{b}}$ is mainly determined by BR$(h \to b\bar{b})$. For the large $R_{\gamma \gamma} > 1$, BR$(h \to b\bar{b})$ is suppressed due to small $C_{h b \bar{b}}$. We also find the $R_{\gamma \gamma}$ can be smaller than $0.5$ due to the significant Higgs exotic decays; both the $R_{VV}$ and $R_{b \bar{b}}$ are suppressed in this case, too.

According to the analysis given by the CMS collaboration, the ratio of the couplings of Higgs to fermions is around $0.5 \pm 0.3$. In the NMSSM, this suppression can be accommodated by the mixing between the singlet and doublet Higgs bosons, while keep the couplings to vector weak bosons close to one. In Fig. \ref{RggRvv}, we also mark out two current global values of $R_{ZZ} \sim$ ($0.7 \pm 0.4$), $R_{\gamma\gamma} \sim$ ($1.6 \pm 0.4$) and $R_{ZZ} \sim$ ($1.4 \pm 0.5$), $R_{\gamma\gamma} \sim$ ($1.8 \pm 0.4$) given by CMS \cite{:2012gu} and ATLAS \cite{:2012gk}, respectively.
For the light stau region, which may be helpful to ease this tension, we find the $\delta (g-2)_{\mu}$ and flavor physics put stringent bounds to the stau mass and $\tan \beta$ in our scan. The parameter region providing the enhancement by the light chargino/charged Higgs boson to $R_{\gamma \gamma}$ has not been reached since the parameter $\lambda$ is confined to be less than 1 in our scan.

The light stop can affect the $C_{hgg}$ and $C_{h\gamma\gamma}$ simultaneously. For small $A_t$ and stop mixing terms,
the stop and top loop contributions interfere constructively. In this case, the Higgs coupling to gluons and the production cross section are enhanced. Since the top and W loop contributions interferer destructively, small $A_t$ would suppress the Higgs coupling to photons. If $C_{hgg}$ is large enough, $R_{\gamma\gamma}$ can still be enhanced. For large $A_t$, the stop contribution can suppress the Higgs coupling to gluons and enhance the Higgs coupling to photons. In our MCMC scan, we have chosen starting points with large $A_t=$1.5 TeV; thus, we do not get parameter points with enhanced $C_{hgg}$. Another possible reason for the lack of points with enhanced $C_{hgg}$ is attributed to the omission of high order QCD corrections to the Higgs production cross section. Since our MCMC scan does not lead to points with significantly enhanced $C_{hgg}$ and $C_{h\gamma\gamma}$, there are no points with simultaneous enhancement of $C_{hb\bar{b}}$ and $R_{\gamma\gamma}$ in Fig. \ref{RggC}. Detailed studies on the light stop effects to Higgs couplings can be found in Ref. \cite{Blum:2012ii,King:2012tr}.

It should be noticed that with the current experimental error bars and statistics, it is too early to conclude that the decay patterns of the Higgs boson have confirmed the existence of new physics. As pointed out in Ref. \cite{Baglio:2012et}, the large uncertainty in the parton distribution function can also affect these results. Future data and reduction in the uncertainty are needed to make sure whether the new physics has been indicated in the Higgs decay modes already.

\subsection{Dark Matter Bounds}

In this section, we will discuss the constraints from the DM detections. We assume the LSP and DM candidate is the lightest neutralino. Because the neutralinos have an additional singlino component in the NMSSM, the phenomenology of DM is different from that in the MSSM, especially because the LSP can be a pure singlino. In this case, the LSP can be lighter than $100$ GeV and can easily escape the constraints from the invisible Z decay measurements due to its almost vanishing coupling to the Z boson. Below we will address the issue whether the singlino in the NMSSM can help ease the tension between the theories and the experiments.

\begin{figure}[!htb]
\begin{center}
\includegraphics[width=0.32\columnwidth]{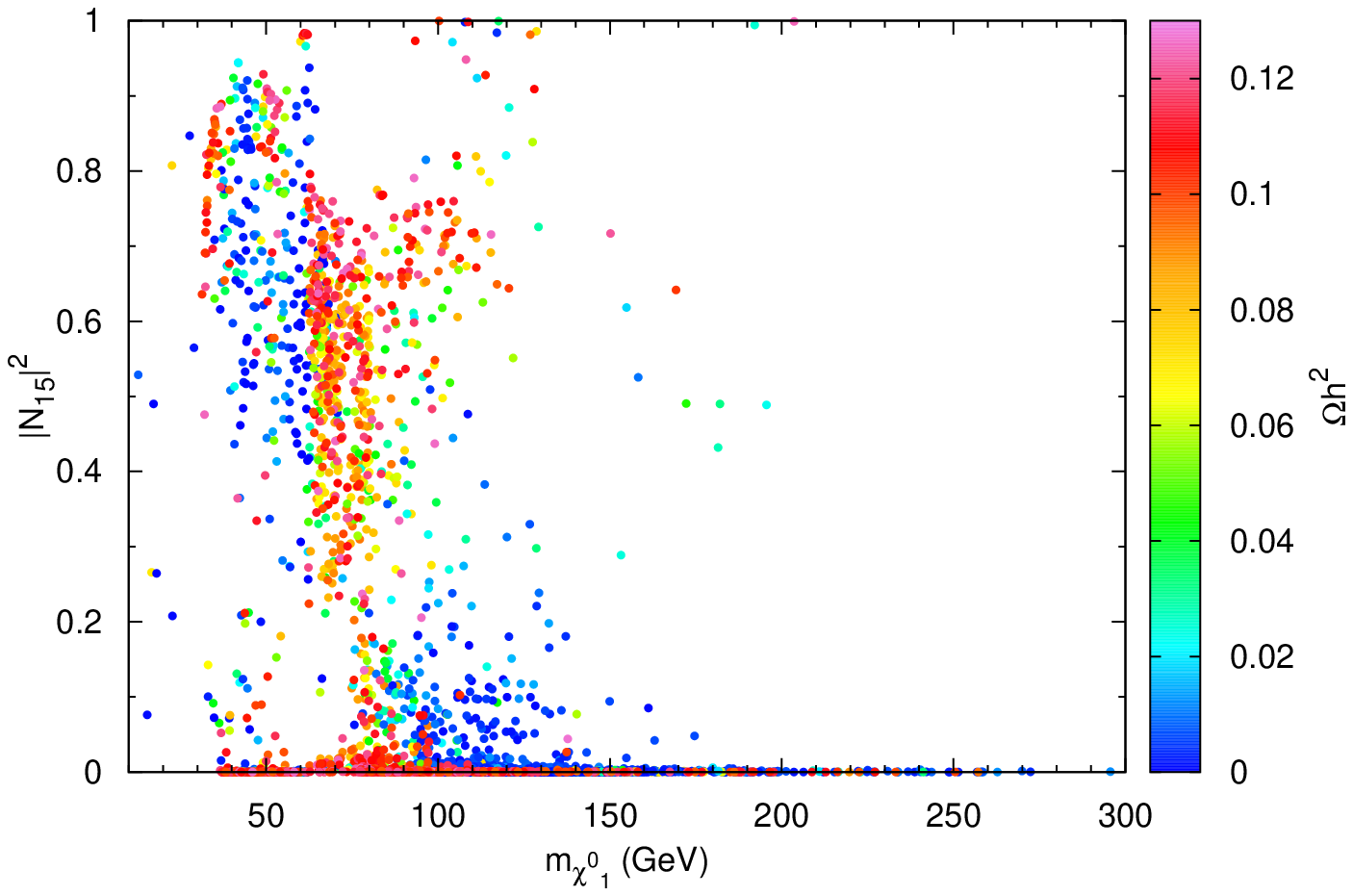}
\includegraphics[width=0.32\columnwidth]{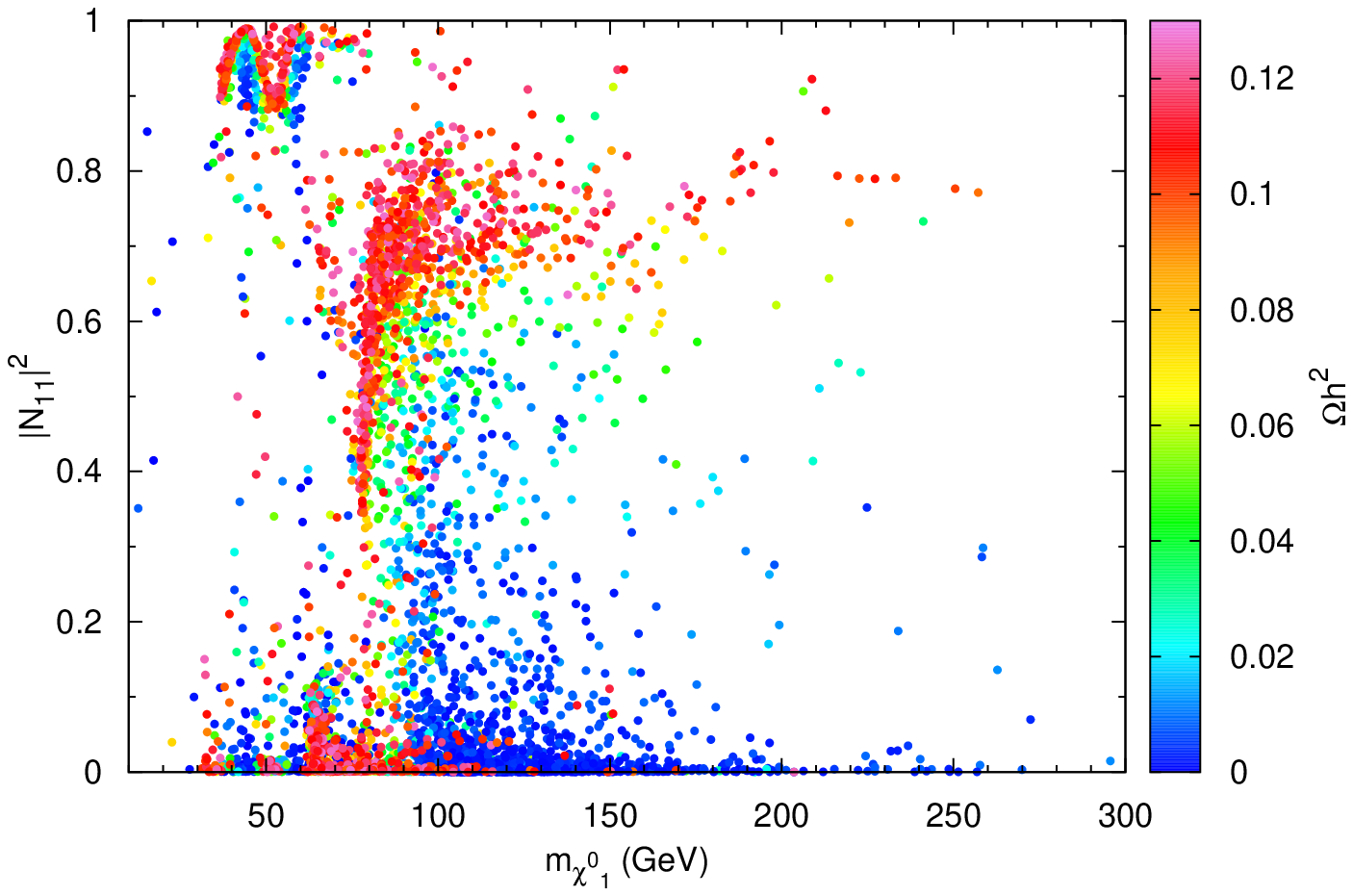}
\includegraphics[width=0.32\columnwidth]{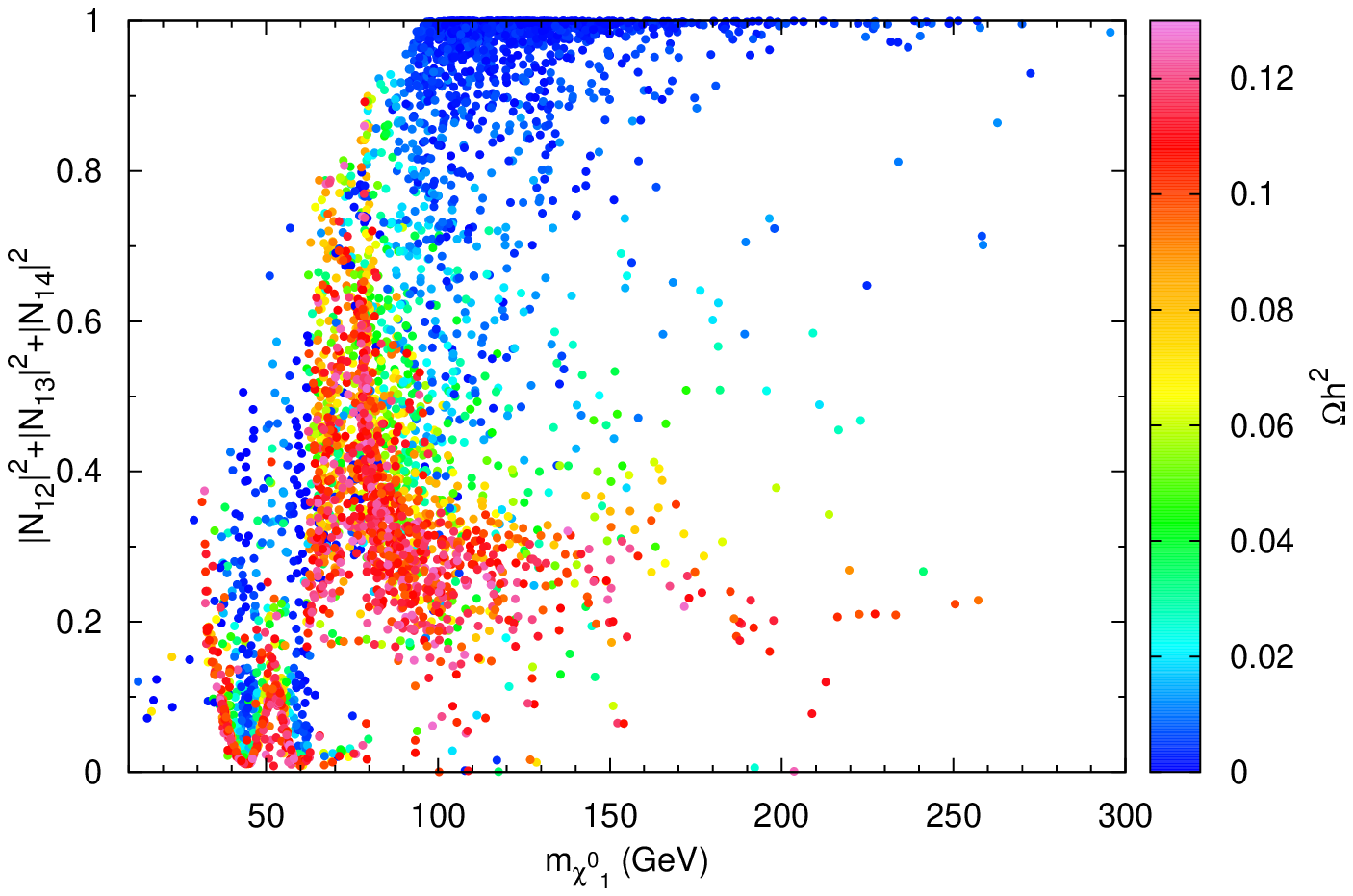}
\caption{The singlino/bino/(wino+Higgsino) content of the LSP versus LSP mass, represented in the left, middle, and right panels, respectively. The color scale indicates the LSP relic density.
\label{chicompo}}
\end{center}
\end{figure}

First, Let us examine the constraints from the dark matter relics density. As we know, the neutralino mass basis $\tilde{\chi}^0_{i=1-5}$ and interaction basis $\psi^0_i=\{ \tilde{B}, \tilde{W}, \tilde{H}_d, \tilde{H}_u, \tilde{S} \}$ are related by $\tilde{\chi}^0_i=N_{ij} \psi^0_j$. In the left/middle/right panel of Fig. \ref{chicompo}, we show the singlino/bino/Higgsino+wino content ($|N_{15}|^2$/$|N_{11}|^2$ /$|N_{12}|^2+|N_{13}|^2+|N_{14}|^2$) of the lightest neutralino. The color scale indicates the neutralino thermal relic density $\Omega_\chi h^2$. To obtain the suitable relic density $\Omega_\chi h^2 < 0.1388$, the neutralinos should have enough annihilation cross section $\langle \sigma v \rangle  > 3\times 10^{-26}$cm$^{3}$s$^{-1}$. In the left panel, we can see most of the lightest neutralinos with a significant singlino content are lighter than 100 GeV.
Because the mixing terms between the singlino and Higgsinos in the mass matrix are proportional to $\lambda$, the LSP as a pure singlino means $\lambda$ is small. In this case, the singlino mass is approximately $2\kappa \mu/\lambda$; the lightest CP-even Higgs and CP-odd Higgs are also singletlike with small masses. The main neutralino annihilation process is via the s-channel Z resonance or CP-odd/CP-even Higgs resonance .\footnote{For the s-channel Z exchange annihilation, small Higgsino content is still needed because $Z \tilde{\chi}^0_1 \tilde{\chi}^0_1$ coupling is proportional to $|N_{13}|^2-|N_{14}|^2$.} This is the reason why there are many points condensing around the range of $\sim 40-70$ GeV. Moreover, if the $\lambda$ is not very small, neutralinos could annihilate into light scalar pairs $H_1 H_1$, $A_1 A_1$ or $H_1 A_1$ via the t channel by the $\tilde{\chi}^0$ exchange or the sufficient large singlet-singlino interaction.

If the lightest neutralino is bino dominated, the annihilation cross section is often too small to produce the correct neutralino relic density. From the middle panel of Fig. \ref{chicompo}, we find that the Z resonance or Higgs resonance can enhance the annihilation cross section, and avoid the overproduction of neutralinos. We can also see that the correct DM relic density is easily acquired when the lightest neutralino is the bino-wino/Higgsino mixture.

If the lightest neutralino has sufficient Higgsino or wino content, the neutralino annihilation cross section can be large via
t-channel chargino exchange to $W^+W^-$. This is the case for points with $\Omega_\chi h^2 \ll 0.1$ and $m_\chi > 80$ GeV in the right panel of Fig. \ref{chicompo}. Because the lightest chargino $\tilde {\chi}^+_1$ can be pure wino or Higgsino, it is also possible to find the parameter points with almost degenerate neutralino $\tilde {\chi}^0_1$ and chargino $\tilde {\chi}^+_1$, which means neutralino can obtain suitable relic density via large coannihilation $\tilde {\chi}^0_1\tilde {\chi}^+_1$ and $\tilde {\chi}^+_1 \tilde {\chi}^-_1$.

\begin{figure}[!htb]
\begin{center}
\includegraphics[width=0.5\columnwidth]{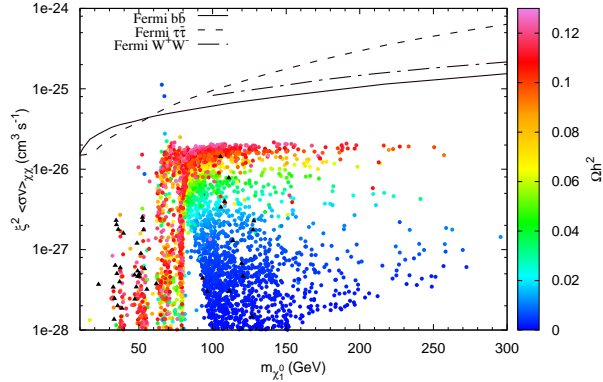}
\caption{$\xi^2 \sigma v$ versus $m_{\tilde{\chi}^0_1}$. The color scale indicates the neutralino relic density. The black triangles are the LSP with singlino component dominated ($|N_{15}|^2>0.8$). Also shown are the upper limits on three annihilation channels $\chi \chi \to b\bar{b}$, $\tau \bar{\tau}$, and $W^+W^-$ given by Fermi-LAT dwarf galaxies observations \cite{Ackermann:2011wa}.
\label{csvgammar}}
\end{center}
\end{figure}

Next we consider the constraints from the indirect astrophysics search. If the DM particles annihilate into heavy quarks, charged fermions and gauge bosons at the present time, these annihilation final states can induce significant gamma-ray flux which can be detected by air shower Cherenkov detectors or satellite detectors. In Fig. \ref{csvgammar}, we plot parameter points in the $m_{\tilde {\chi}^0_1}$ vs. $\xi^2 \langle \sigma v \rangle$ plane. The color scale indicates $\Omega_{\chi} h^2$. Note that the DM thermal averaged annihilation cross section is rescaled by $\xi^2$ because the gamma-ray flux depends on the DM density square, and the actual neutralino density may be $\xi \rho_{DM}$. In Fig. \ref{csvgammar}, the upper limits derived from Fermi gamma-ray observations towards dwarf spheroidal galaxies are also shown \cite{Ackermann:2011wa}. These limits are combined by null results from ten dwarf galaxies, and only a few times above the ``natural value" $\langle \sigma v \rangle =3 \times 10^{-26}$ cm$^3$ s$^{-1}$ for DM with $m_{DM} \sim O(10^2)$ GeV. For parameter points with large wino and Higgsino contents and then large $\langle \sigma v \rangle \gg 3\times 10^{-26}$ cm$^3$ s$^{-1}$, the thermal neutralino density $\xi \rho_{DM}$ is very small. Thus the reduced annihilation cross section rescaled by $\xi^2$ can escape the constraints easily. The gamma-ray limits become more stringent when DM mass decreases. However, for light neutralino $m_{\tilde {\chi}^0_1}< 70$ GeV with significant singlino content, the annihilation process via s-channel Z or CP-even Higgs exchange is p wave which is much suppressed at the present universe. Therefore we can see these limits do not exclude many parameter points.\footnote{The $\langle \sigma v \rangle$ is often estimated at $v \sim 10^{-3}$ which is the typical DM velocity in the present Galactic halo. However, the velocities of DM particles in the dwarf galaxies are about an order of magnitude smaller. When the gamma-ray limits from dwarf galaxies are taken into account, this effect needs to be considered for the velocity dependent annihilation cross section \cite{AlbornozVasquez:2011js}. This effect may enhance the cross section significantly when the main annihilation process is an s channel CP-odd Higgs exchange. The large enhancement often requires a very narrow Breit-Wigner resonance with tiny mass splitting parameter $|1-m^2_A/4m_\chi^2| \ll 1$ and decay width $\Gamma_A/m_\chi \ll 1$. This effect would not change our results very much and is neglected here.} The more stringent limits can be derived by Fermi gamma-ray observations towards Galactic center because DM particles are more condensate in this regime. The main problem is how to precisely subtract the complicated astrophysical backgrounds. However, it is possible to improve the constraints to be below the ``nature value" for $O(10^2)$ GeV DM in the future.

\begin{figure}[!htb]
\begin{center}
\includegraphics[width=0.5\columnwidth]{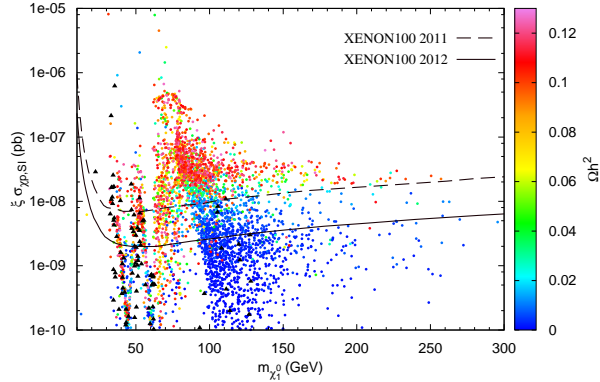}
\caption{$\xi \sigma_{SI}$ versus $m_{\tilde{\chi}^0_1}$. The color scale indicates the neutralino relic density. The black triangles are the LSP with singlino component dominated ($|N_{15}|^2>0.8$). The upper limits set by XENON100 in 2011 \cite{Aprile:2011hi} and 2012 \cite{:2012nq} are also shown.
\label{sicschipr}}
\end{center}
\end{figure}

Then we examine the constraints from the direct searches. We consider both spin-independent and spin-dependent cases. For the spin-independent constraints, we focus on the constraint from the XENON100, which is the most stringent bound for dark matter direct searches. For the spin-dependent constraints, we include bounds from the neutrino flux measurements. In Fig. \ref{sicschipr}, we plot parameter points in the $m_{\tilde {\chi}^0_1}$ vs. $\xi \sigma_{SI}$ plane. The color scale indicates $\Omega_{\chi} h^2$. Here, $\xi \sigma_{SI}$ is the reduced spin-independent neutralino-nucleon elastic scattering cross section. We also show the most stringent constraints set by the XENON100 in 2011 \cite{Aprile:2011hi} and 2012 \cite{:2012nq}. We can see the XENON limits have excluded many parameter points with the correct DM relic density $\Omega_\chi h^2 \sim 0.11$. The spin-independent scattering processes via squark exchanges are strongly suppressed due to the heavy squarks that are assumed to avoid collider constraints. The main process is the exchange of Higgs through the t channel. The cross section of such process depends on the wino and Higgsino contents of the lightest neutralino and the Higgs masses. Therefore, the neutralinos with masses of $m_\chi > 80$ GeV and intermediate wino and Higgsino contents are strongly disfavored by direct detections unless the $\sigma_{SI}$ is reduced by small $\xi$. Moreover, if the neutralinos have significant singlino contents, and the lightest CP-even Higgs is singletlike, then the neutralino-nucleon scattering can be enhanced by additional Higgs-neutralino couplings and small Higgs mass in the propagator. These parameter points might also be easily excluded by XENON limits.

\begin{figure}[!htb]
\begin{center}
\includegraphics[width=0.45\columnwidth]{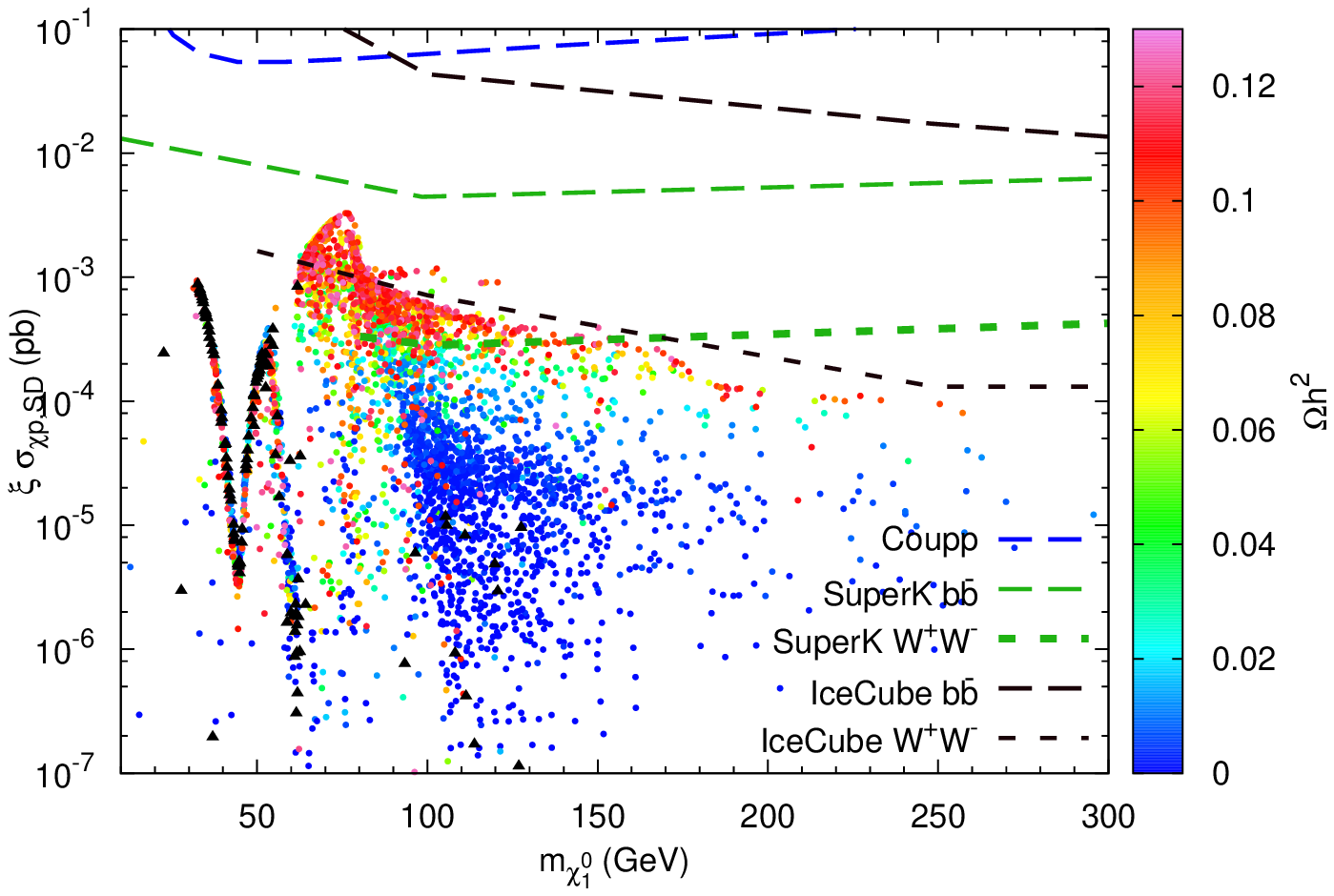}
\includegraphics[width=0.45\columnwidth]{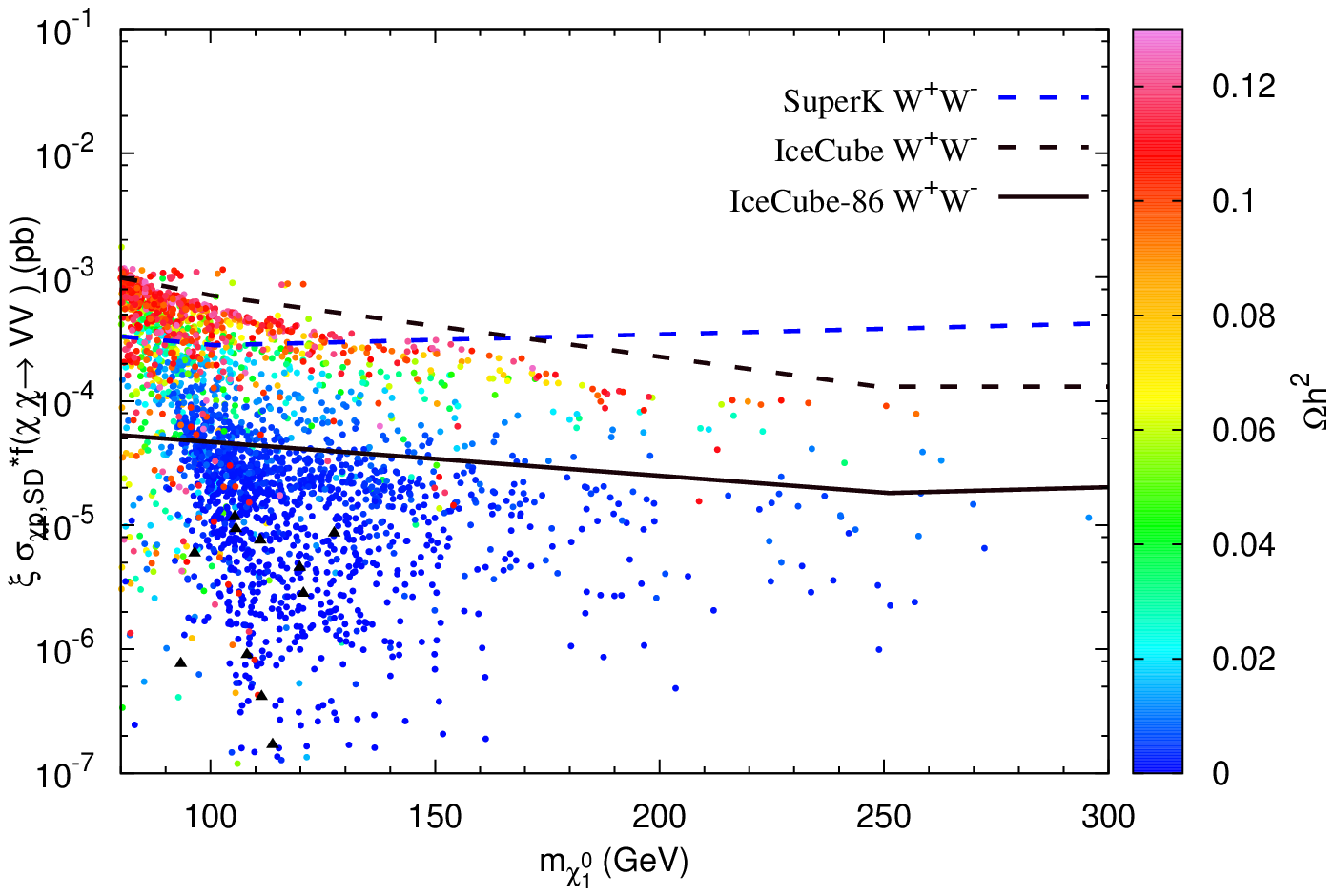}
\caption{Left: $\xi \sigma_{SD}$ versus $m_{\tilde{\chi}^0_1}$. Right: $\xi \sigma_{SD} f(\tilde{\chi}^0_1 \tilde{\chi}^0_1 \to VV)$ versus $m_{\tilde{\chi}^0_1}$. The color scale indicates the neutralino relic density. The black triangles are the LSP with singlino component dominated ($|N_{15}|^2>0.8$). The limits given by Coupp \cite{Behnke:2010xt}, Super-kamionkande \cite{Tanaka:2011uf}, and IceCube \cite{IceCube:2011aj} are also shown.
\label{sdcschip}}
\end{center}
\end{figure}

For the spin-dependent neutralino-nucleon scattering, the constraints established by direct detections are very weak. The most strong direct constraints given by the COUPP \cite{Behnke:2010xt} and KIMS collaborations \cite{Lee.:2007qn} are of the order of $O(10^{-1})$ pb. More stringent limits can be set by high energy neutrino telescopes. If the DM particles lose their energies by scattering with solar nucleons and are trapped in the center of the Sun, they could annihilate into heavy fermions or gauge bosons and produce detectable high energy neutrino signatures. The signature flux can be determined by the DM-nucleon spin-dependent scattering rate. Moreover, it also depends on the fractions of certain DM annihilation channels $f_i=\sigma(\chi\chi \to X_i X_i)/\sigma_{\chi\chi}$ which could produce neutrinos. In the left (right) panel of Fig. \ref{sdcschip}, we present the parameter points in the $m_{\tilde {\chi}^0_1}$ vs. $\xi \sigma_{SD}$ ($m_{\tilde {\chi}^0_1}$ vs. $\xi \sigma_{SD} f_{VV}$) plane. The limits given by Coupp \cite{Behnke:2010xt}, Super-Kamiocande \cite{Tanaka:2011uf}, and IceCube \cite{IceCube:2011aj} are also shown. Note that the neutrino energy spectra from different DM annihilation channels are different, the limits from neutrino telescopes are derived from experimental results for different channels. Because the neutrino energy spectra from $b \bar{b}$ ($q \bar{q}$, $\tau \bar{\tau}$) are much softer than those from $W^+W^-$ ($ZZ$, $t \bar{t}$); the limits from the $W^+W^-$ channel are more stringent. From the left panel of Fig. \ref{sdcschip}, we can see that recent direct detections and neutrino signatures from soft channels do not constrain the models very much. However, if the neutralinos have intermediate Higgsino contents, the spin-dependent cross section might be large due to $Z\tilde{\chi}^0_1 \tilde{\chi}^0_1$ coupling which is proportional to $|N_{13}|^2-|N_{14}|^2$. In this case, some parameter points with $\xi \sigma_{SD} > O(10^{-4})$pb and $f_{VV}\sim 1$ have been excluded. The expected limit which can be set by IceCube 86 strings are also shown. We can see that IceCube has the capability to exclude most of the parameter points which predict neutralinos with the significant Higgsino contents and correct DM relic density $\Omega_\chi h^2 \sim 0.11$.

To examine the impact of a singlino to the dark matter searches, in Figs. \ref{csvgammar}-\ref{sdcschip}, we have marked out (with black triangles) those allowed points with a dominant singlino constituent. It is observed that when the LSP is singlino dominant, it can easily escape the constraint from Fermi-LAT, as demonstrated in Fig. \ref{csvgammar}. While the bounds from XENON100 2011 and 2012 are really impressive and can exclude a certain fraction of those allowed points even the LSP is singlino dominant. For the neutrino bounds, since many singlino dominant LSPs are lighter than W bosons, they can be still consistent with the data due to weak constraints fromthe $b \bar{b}$ channel. For the LSP heavier than the W boson, the annihilation cross section of $\tilde{\chi}^0 \tilde{\chi}^0 \to W^+ W^-$ is always small enough and can be safe. From the analysis shown above, we can see that the singlino/bino in the NMSSM can help to ease the tension between theories and experiments, while the wino-like and Higgsino-like dark matter candidates are more constrained.

\section{LHC SUSY Search Bounds}

\subsection{The production and decay of the light stop/sbottom at the LHC}

In this section we study the SUSY search bounds on the light
stop/sbottom pair signatures from the LHC. For our purpose below we
make two additional requirements for the parameter points that passed all of the
constraints in Sec. II. First, we require that the stop mass is
lighter than 500 GeV in order to have a large enough production
rate. Since we have assumed $m_{U_3}=m_{D_3}$, the lightest
sbottom is light in our analysis. Then, the largest color sparticle
signatures are the stop pair and sbottom pair productions. Second,
to accommodate the SM-like Higgs in the NMSSM, we require
$R_{ZZ} > 0.8$ and $R_{\gamma \gamma} > 0.8$ in the parameter space.
Implicitly, this condition also means that the branching ratios of
the exotic Higgs decay modes should not be very large.
Consequently, it also suggests that the lightest neutralino,
CP-odd and CP-even Higgs cannot be very light. This feature will
also affect the decays of other sparticles. After taking into
account these two extra requirements, we choose $552$
parameter points that survived all our criteria for our
simulations. Below we study the constraints from direct SUSY
searches at the LHC to these points.

\begin{figure}[!htb]
\begin{center}
\includegraphics[width=0.45\columnwidth]{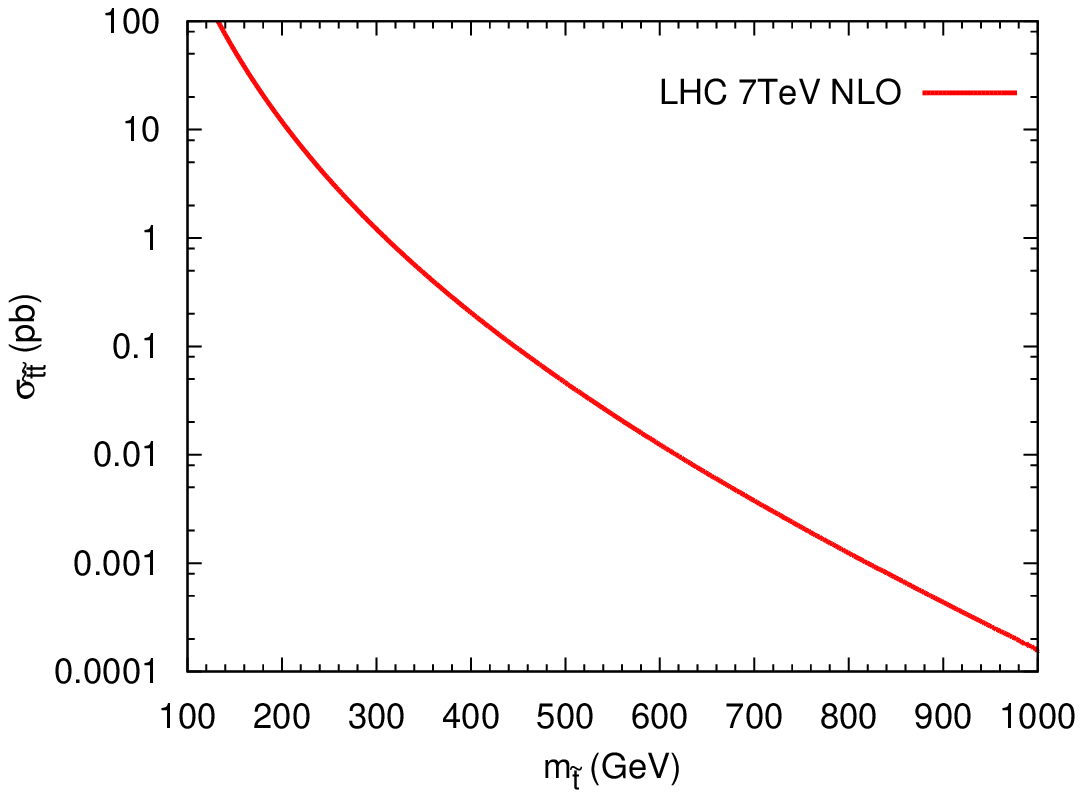}
\includegraphics[width=0.45\columnwidth]{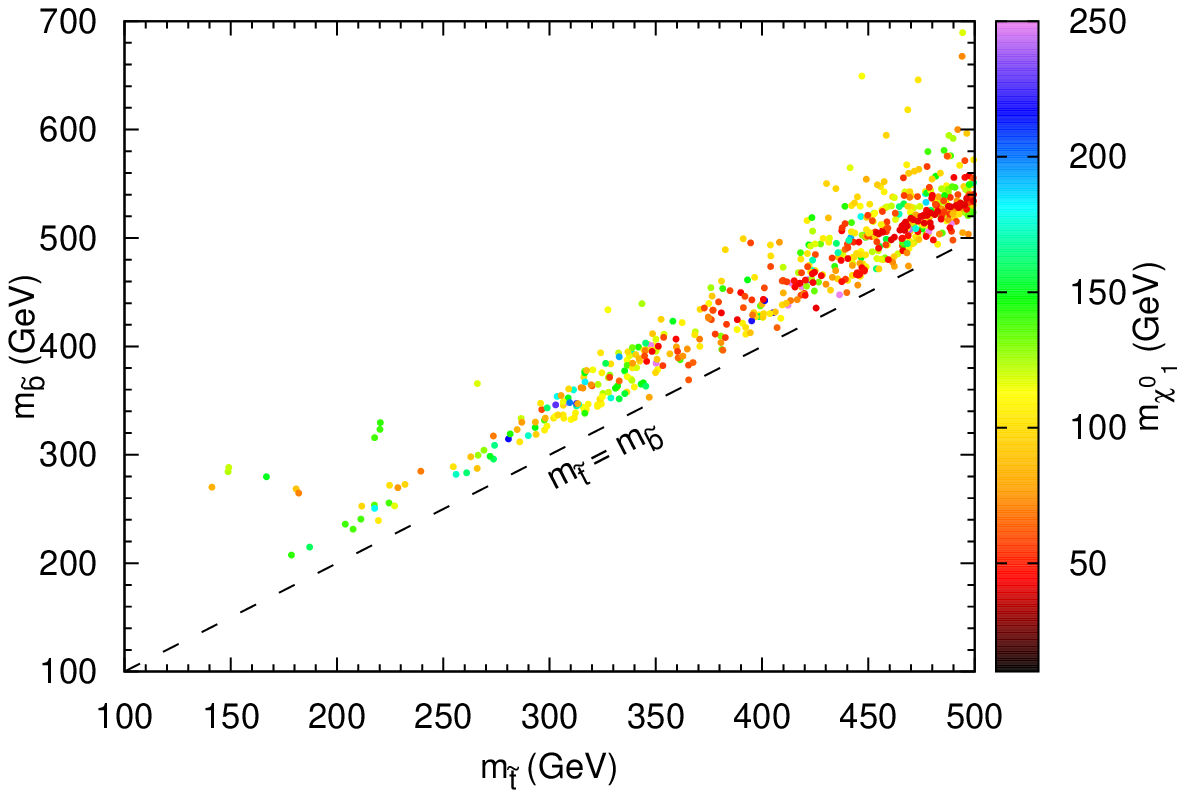}
\caption{Left: the NLO cross section of stop pair production as a function of stop mass. Right: sbottom mass versus stop mass, where the color scale indicates neutralino mass.
\label{stopmass}}
\end{center}
\end{figure}

In the left panel of Fig. \ref{stopmass}, we show the cross section of stop pair production $\sigma_{\tilde{t} \tilde{t}}$ at the LHC with $\sqrt{s}=7$ TeV. Here we use the package Prospino2 \cite{Beenakker:1996ed} to calculate $\sigma_{\tilde{t} \tilde{t}}$ including the NLO corrections. Because the main production channel of stop pair at the LHC is $gg\to \tilde{t} \tilde{t}$, we can see $\sigma_{\tilde{t} \tilde{t}}$ is uniquely determined by stop mass. For the stop with mass of $m_{\tilde{t}} \leq 500$ GeV, the $\sigma_{\tilde{t} \tilde{t}}$ is larger than 45 fb, and there would be more than $\sim 200$ stop pair events at the LHC with 5 fb$^{-1}$ of data.

The masses of the stop and sbottom are shown in the
right panel of Fig. \ref{stopmass} where the color scale indicates
the lightest neutralino mass. We can see that the mass splitting
between the stops and sbottoms is small due to our assumption
$m_{U_3}=m_{D_3}$ and the lighter stop and sbottom quarks are
either dominantly left handed or right handed. It is supposed that this parameter configuration can easily  pass the electroweak
precision tests \cite{King:2012is,Maniatis:2012ka}.

\begin{figure}[!htb]
\begin{center}
\includegraphics[width=0.45\columnwidth]{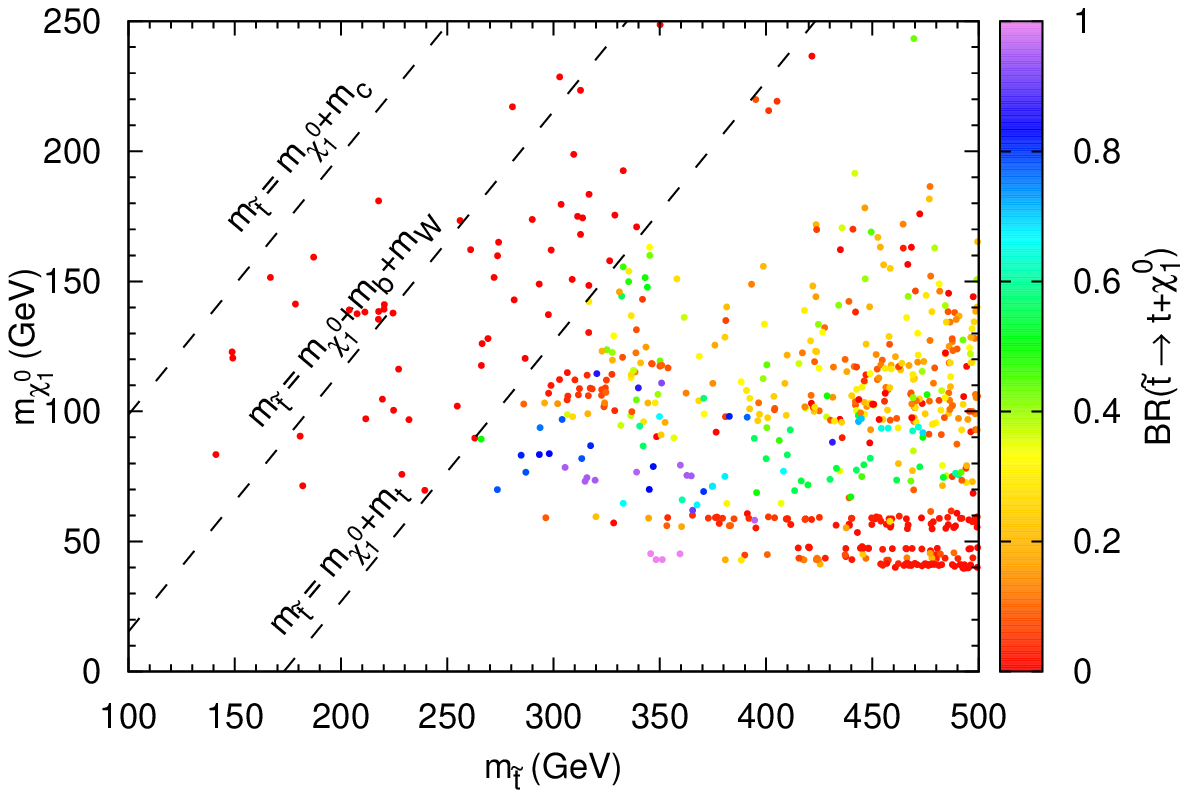}
\includegraphics[width=0.45\columnwidth]{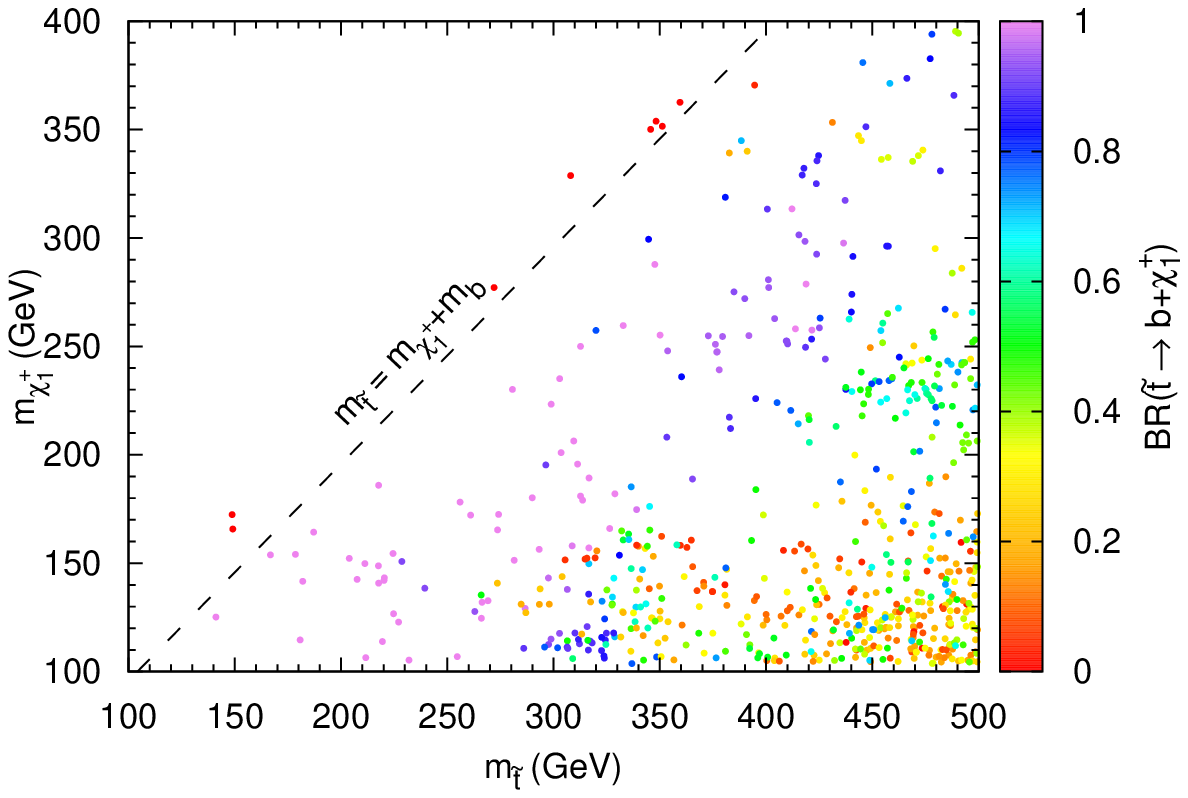}
\caption{The lightest neutralino/chargino mass versus the stop mass in the left/right panel. The color scale indicates the branching ratio of stop BR($\tilde{t}\to t \tilde{\chi}^0_1 $)/BR($\tilde{t}\to b \tilde{\chi}^+_1 $) in the left/right panel.
\label{stopatln}}
\end{center}
\end{figure}

The stop decay pattern is dominantly determined by the mass splitting between the stop and light neutralinos/charginos. We use NMSDECAY \cite{Das:2011dg} based on SDECAY \cite{Muhlleitner:2003vg} to calculate the decay branching ratios of the sparticles. If the stop is much heavier than the lightest neutralino and chargino, the main decay channels are two-body decays $\tilde{t}\to t \tilde{\chi}^0_1 $ and $\tilde{t}\to b \tilde{\chi}^+_1 $. In Fig. \ref{stopatln}, we show the relations between the sparticle mass spectra and the branching ratios of $\tilde{t}\to t \tilde{\chi}^0_1 $ and $\tilde{t}\to b \tilde{\chi}^+_1 $. We can see the decay modes depend on the neutralino and chargino mass spectra. The processes of the stop decay into $t\tilde{\chi}^0_2$, $t\tilde{\chi}^0_3$, and $b\tilde{\chi}^+_2$ might also be significant for the heavy stop  (as demonstrated in first two benchmark points given in Table. \ref{benchmarkps}). In this case, even the LSP is very light $< 100$GeV, and the decay mode $\tilde{t}\to t \tilde{\chi}^0_1 $ may also be suppressed. These processes have longer decay chains than $\tilde{t}\to t \tilde{\chi}^0_1 $ and produce softer final states. When the mass splitting is too small to forbid above two-body decays, the three-body decay channels  $\tilde{t}\to b W \tilde{\chi}^0_1 $ and $\tilde{t}\to b \nu \tilde{l} / b l \tilde{\nu} $ become important. If these processes are also kinematic forbidden, the loop induced two-body FCNC decay $\tilde{t} \to c  \tilde{\chi}^0_1$ \cite{Hikasa:1987db,Muhlleitner:2011ww} would be dominant. Moreover, the four final state decay modes $\tilde{t}\to b j_1 j_2 \tilde{\chi}^0_1 $ and $\tilde{t}\to b \ell \nu_{\ell} \tilde{\chi}^0_1 $ \cite{Boehm:1999tr} are also possible if the $W$ boson in the three body decay mode is not on shell. Because the NMSDECAY has not included the stop four-body decay process, we do not calculate its branching ratio in this work.

The main two-decay modes of the sbottom include ${\tilde b}_1 \to b \tilde{\chi}^0_i$ and ${\tilde b}_1 \to t \tilde{\chi}^-_i$. The three-body decay modes include ${\tilde b}_1 \to t^* \tilde{\chi}^-_i \to b W^+ \tilde{\chi}^-_i$. The decay chain of the sbottom can be quite long, and all the final state from its decay can be soft as demonstrated in the third benchmark point in Table. \ref{benchmarkps}.

\subsection{Constraints on stop/sbottom pair signatures}

The general SUSY search bounds for squarks and gluino have been provided by ATLAS and CMS, corresponding to an integrated luminosity of $2 \sim 5$ fb$^{-1}$. In this subsection, we investigate the constraints on light stop pair and sbottom pair productions based on these results. In our studies, parton-level events $pp \to \tilde{t}\tilde{t}$ and $pp \to \tilde{t}\tilde{t}+jets$ are generated by MadGraph5 \cite{Alwall:2011uj}. PYTHIA \cite{Sjostrand:2006za} is used to perform the parton shower, decay, final state radiation, and hadronization processes. The detector effects are simulated by PGS 4 \cite{pgs}. To avoid the double-counting issue, we adopt the MLM matching scheme and choose $Q_{cut}=80$ GeV in our simulation. Jet candidates are reconstructed by using the anti-kt jet algorithm (which is infrared and collinear safe) with a distance parameter of 0.4/0.5 for the ATLAS/CMS searches.\footnote{The basic selected conditions for jets and leptons are slightly changed in different searches. In general, these conditions are $p_{t}> 20-40$ GeV, $|\eta| < 2.5 - 3$ for jets, $p_{t}> 10-25$ GeV, $|\eta| < 2.0-2.5$ for electrons and muons. Moreover, the electron candidates in the barrel-endcap transition region, with $1.44<\eta<1.57$, are rejected. Here we used the basic selections as adopted by ATLAS and CMS according to different research, and do not list them in the following discussions.}

Currently, most of experimental groups from both CMS and ATLAS collaborations work in the simplified model. The signals are assumed to be $pp \to {\tilde b}_1 {\tilde b}_1^* \to b {\bar b} \chi^0_1 \chi^0_1$ or $pp \to {\tilde t}_1 {\tilde t}_1^* \to t {\bar t} \chi^0_1 \chi^0_1$. It is interesting to examine what might happen in a concrete model like the NMSSM and how new decay modes can be affected by these direct searches. We will consider two categories of constraints from the direct search: 1) the searches for the final states without b-jets; 2) the searches for the final states with b-jets.

\subsubsection{Constraints for Final states without b-jets}

If the dominated decay modes of squarks and gluino are $\tilde{q} \to q \tilde{\chi}^0_1$ and $\tilde{g} \to q \bar{q} \tilde{\chi}^0_1$, the main features of events are energetic jets and large MET. This signature channel can set the most stringent constraints on the CMSSM and simplified model without a light stop/sbottom. For the light squarks of the third generation, such signatures would be suppressed due to smaller production cross section and different decay modes. Because such analysis requires very hard jets and large MET, the events of stop pair with many soft jets can easily escape the constraints. Below we investigate how the $m_{eff}$ and $ \sl{E}$ cuts as well as the associated monojet searches can affect our selected points.

\begin{table}[th]
\begin{center}%
\begin{tabular}
[c]{|c|c|c|c|c|c|c|c|}\hline
Requirements & A & A' & B  & C & D & E  \\\hline\hline
$\sl{E}_T$ [GeV] $>$  &  \multicolumn{6}{|c|}{160} \\\hline
$N_{Jet}(p_T>130$ GeV$) \geq $ &  \multicolumn{6}{|c|}{1}  \\\hline
$N_{Jet}(p_T>60$ GeV$) \geq$ & 2 & 2 & 3  & 4 & 4 & 4  \\\hline
$N_{Jet}(p_T>40$ GeV$) \geq$ & - & - & -  & - & 5 & 6  \\\hline
$\Delta \phi(\vec{j}_i,\vec{\sl{E}}_T)_{min} >$ & \multicolumn{3}{|c|}{0.4 (i=1,2,(3))}   &  \multicolumn{3}{|c|} {0.4 (i=1,2,3), 0.2 ($p_T^j >$ 40 GeV)}  \\ \hline
$ \sl{E}_T/m_{eff}(N_{Jet}) >$ & 0.3 (2j) & 0.4 (2j) & 0.25 (3j) & 0.25 (4j) & 0.2 (5j) & 0.15 (6j)  \\\hline
$ m_{eff}$(incl.) [GeV] $>$ & 1900/1400/- & -/1200/- & 1900/-/- & 1500/1200/900 & 1500/-/- & 1400/1200/900  \\\hline\hline
$ N_{lim}^{obs} $ & 2.9/25/- & -/29/- & 3.1/-/- & 16/18/58 & 10/-/- & 12/12/84  \\\hline
\end{tabular}
\end{center}
\caption{Summary of cuts and observed 95\% CL upper limits on the excess event number, following the ATLAS jets+MET analysis for 4.7 fb$^{-1}$\cite{:2012rz}. }%
\label{ATLjet5}%
\end{table}

The first constraint is from the jets plus missing energy searches. We list the cut conditions adopted by the ATLAS collaboration in Table. \ref{ATLjet5} \cite{:2012rz}. This analysis is based on $4.7$ fb$^{-1}$ of data, and all the events with isolated electrons or muons are rejected. The azimuthal angle $\Delta \phi(\vec{j}_i,\vec{\sl{E}}_T)$ is defined as the azimuthal angle separation between the $\sl{E}_T$ and the jets. The effective mass is defined as
\begin{equation}
m_{eff}= \sl{E}_T + \sum_{i=1}^{N_j} p_T^j + \sum_{i=1}^{N_l} p_T^l .
\end{equation}
It is obvious to see the effective mass characterizes the mass scale of SUSY particles directly produced by pp collisions.
Large $m_{eff} \sim 1$TeV is helpful to reduce the SM backgrounds such as $W+jets$, $Z+jets$, $t\bar{t}$, and single top, but it also severely suppresses light stop/sbottom pair events with soft jets.

\begin{figure}[!htb]
\begin{center}
\includegraphics[width=0.40\columnwidth]{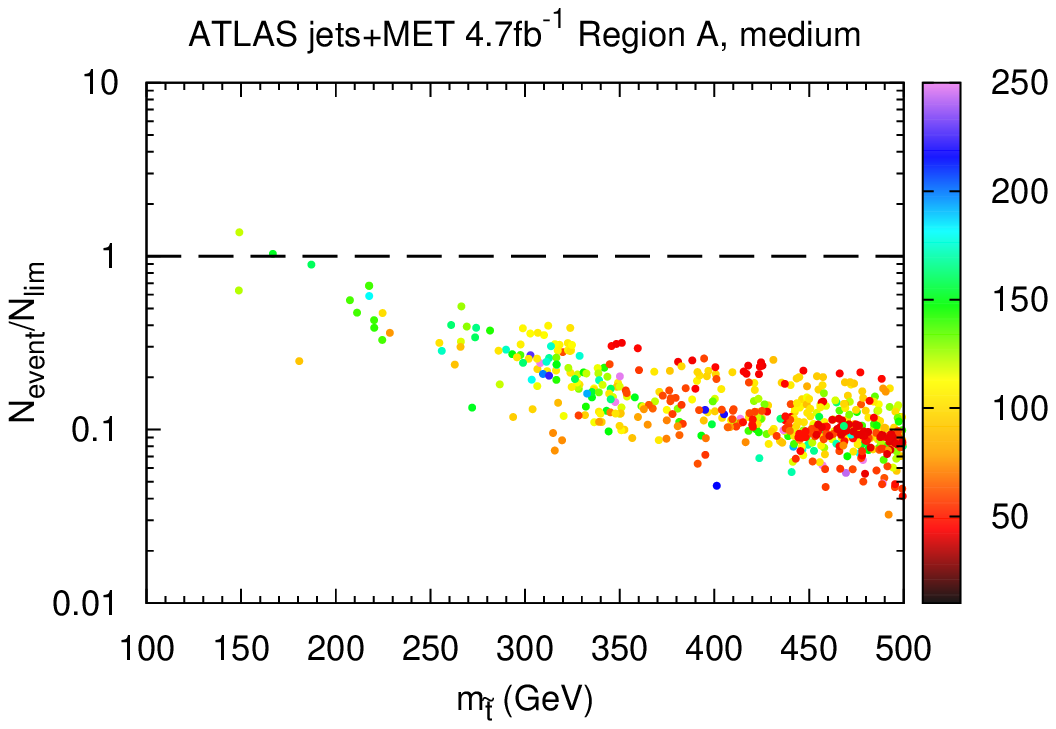}
\includegraphics[width=0.40\columnwidth]{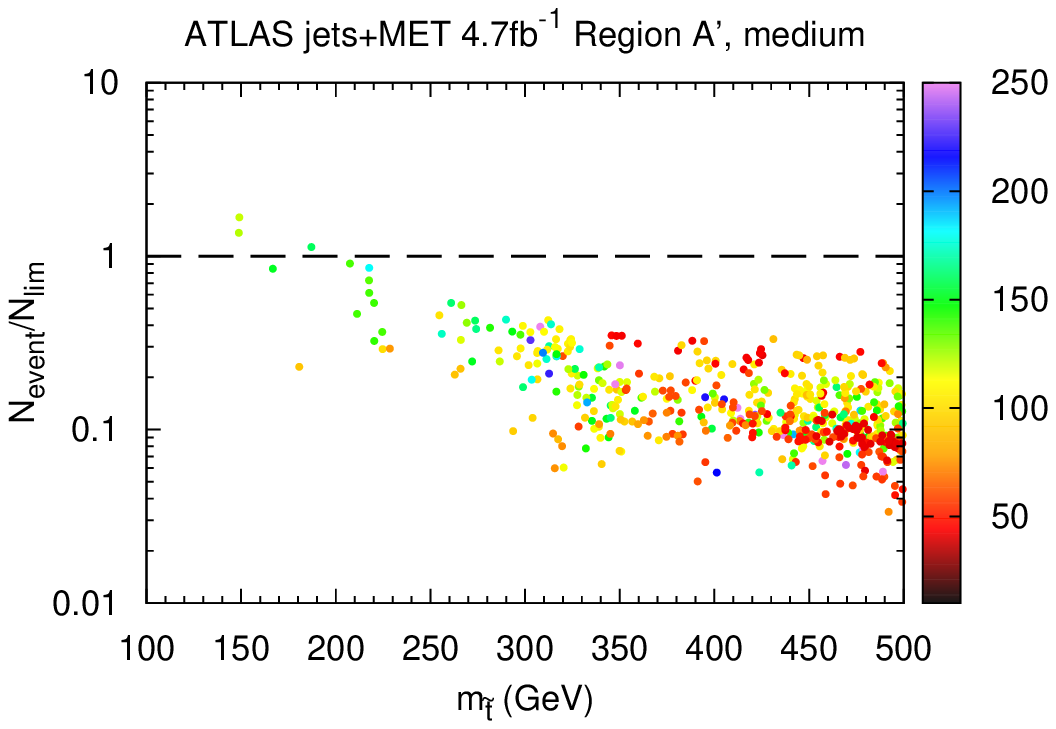}
\includegraphics[width=0.40\columnwidth]{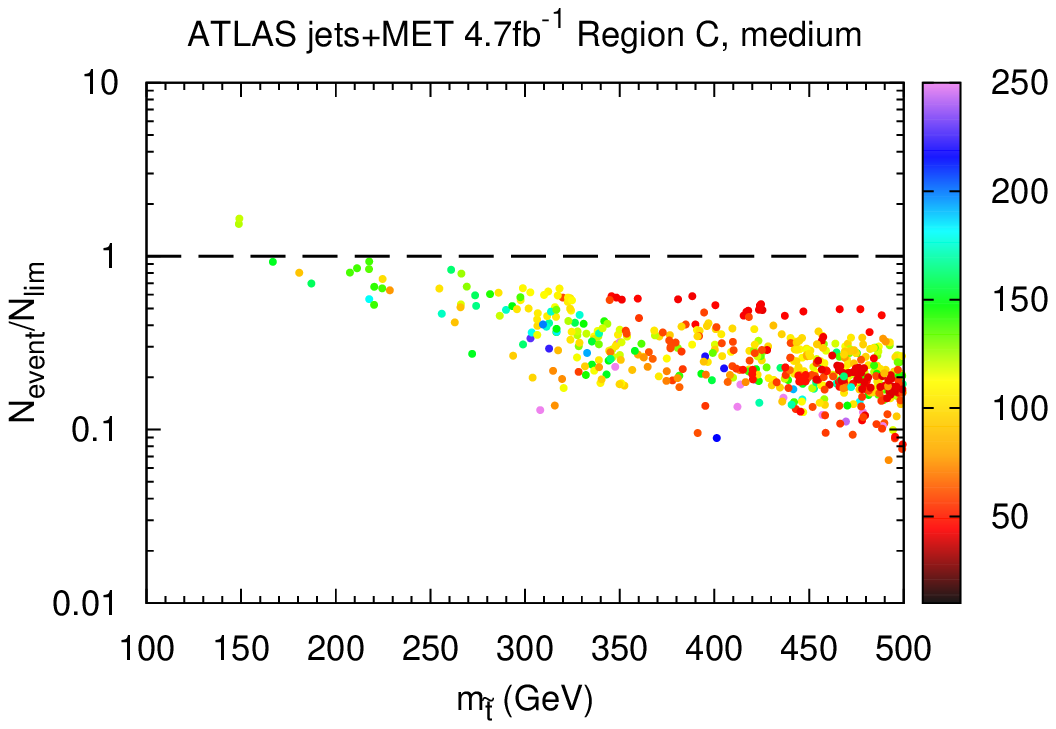}
\includegraphics[width=0.40\columnwidth]{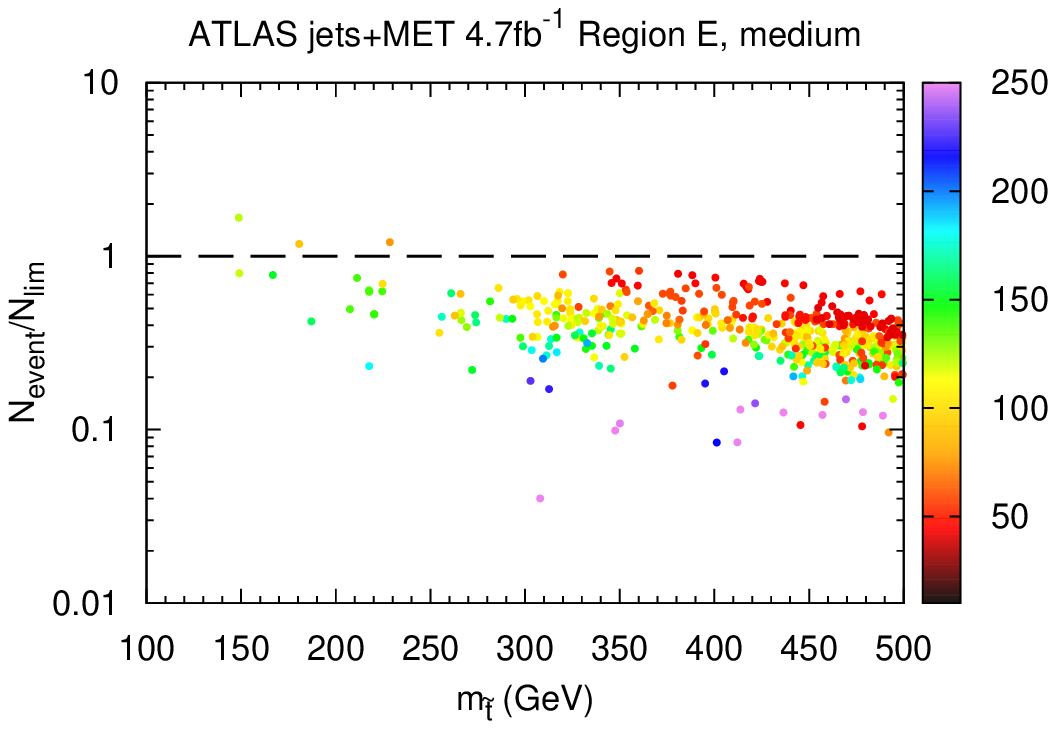}
\caption{The ratio of ${\tilde t} {\tilde t}^*$+${\tilde b} {\tilde b}^*$ event number and observed 95\% C.L. upper limit versus stop mass in the signal regions A, A', C, E in the upper-left, upper-right, lower-left, and lower-right panels, respectively. The color scale indicates neutralino mass.
\label{ATLjet}}
\end{center}
\end{figure}

For each signal region, three $m_{eff}$ cut conditions (denoted by
``tight, medium and loose") are taken into account. In Table.
\ref{ATLjet5}, the $95\%$ C.L. observed upper limits $N_{lim}$ on the number of
new physics events given by the experimental collaboration are also listed
\cite{:2012rz}. For comparison, we show the ratio of predicted
event number $N$ to upper limit $N_{lim}$ in Fig. \ref{ATLjet}
where the color scale denotes the LSP mass.\footnote{It should be
noticed that the bounds shown in our analysis have not included
theoretical and experimental errors. Although our
simulated results are close to the experimental results, it should be
remembered that when these uncertainties are taken into account,
the lines should become bands.} Here we have summed the
signatures of the stop pair and sbottom pair together. If the $m_{eff}$
cut is chosen to be too large, most of the signals would be
rejected; otherwise, there are too many background events which
lead to much weaker upper-limits for new physics. Better constraints can
be demonstrated by the four medium channels A, A', C and E in Fig.
\ref{ATLjet}.

We have also checked the results for all the channels and find that the jets+MET channel can set marginal constraints on most of the parameter points when the stop mass is below $\sim 200$ GeV and the main decay mode is $\tilde{t} \to c \tilde{\chi}^0_1 $. The large cross section of stop pair production $\sim$ O(10) pb can induce event excess, as indicated in the A and A' regions in Fig. \ref{ATLjet}. Interestingly, it is noticed that some parameter points with light LSP and stop cannot be excluded by this search. These parameter points also predict light charginos, sleptons, or second neutralino; therefore, the decay chain can be long and the final states contain fewer hard jets, which can hide into the background events, as shown in all regions of Fig. \ref{ATLjet}.

We can see there are quite a fraction of points
in A and A' situated below $N/N_{lim}=0.1$, while all those points
are above $N/N_{lim}=0.1$ in C and E. For the heavier stop with dominated decay mode $\tilde{t} \to t \tilde{\chi}^0_1$, the channels requiring high jet multiplicity should be more efficient. For instance, from Fig. \ref{ATLjet} we can see that the values of $N/N_{lim}$ in the region E medium for parameter points with heavy stops and light neutralinos are close to $1.0$.

We also check the constraints from ATLAS jets+MET research based on 1 fb$^{-1}$ of data \cite{Aad:2011ib}. In this analysis, the $m_{eff}$ cut is required to be
O(100) GeV. Considering that the huge SM backgrounds prevent us from setting a better constraint to stop/sbottom mass, we find the $N/N_{lim}$ is much lower, e.g. $<$O$(10^{-1})$.

\begin{table}[th]
\begin{center}%
\begin{tabular}
[c]{|c|c|c|c|c|}\hline
$N_{Jet}(p_T > 100)$ & $N_{Jet}(p_T > 30)$ & $\Delta\phi(j_1,j_2) $ & $\sl{E}_T$ & $N_{lim}^{obs}$ \\\hline\hline
$\geq 1$ & $\leq 2$ & $\leq 2.5$ & $>$ 250/ 300/ 350/ 400   & \; 600/ 368/ 158 /95  \\\hline
\end{tabular}
\end{center}
\caption{Summary of cuts and observed 95\% C.L. upper limits on the excess event number, following the CMS monojet+MET analysis for 4.98 fb$^{-1}$ \cite{Chatrchyan:2012pa}. }
\label{CMSmjett}
\end{table}

The second constraint that will be considered here is from the associated monojet production. As is well known, when the dark matter particles are directly produced by pp collisions, one possible search channel is the monojet + MET \cite{Fox:2011pm}. The monojet is produced by the initial state radiation and can be energetic. If the stop is nearly degenerate with the LSP, the soft jets from the stop decay might not be reconstructed by the detectors. In this case, the stop production $\tilde{t} \tilde{t} j$ can be constrained by monojet + MET research \cite{Carena:2008mj,Drees:2012dd,Ajaib:2011hs}. The cut conditions and upper limits given by CMS are summarized in Table. \ref{CMSmjett} \cite{Chatrchyan:2012pa}. This analysis is based on 4.98 fb$^{-1}$ of data, and all the events with isolated electrons or muons are rejected.

\begin{figure}[!htb]
\begin{center}
\includegraphics[width=0.40\columnwidth]{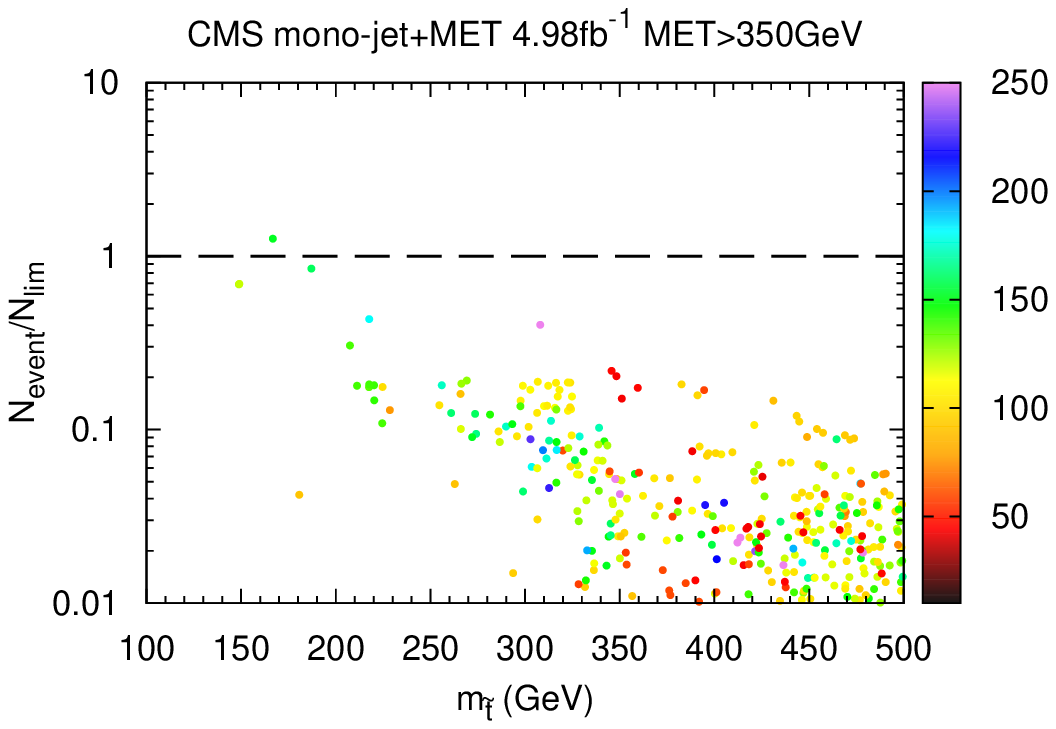}
\includegraphics[width=0.40\columnwidth]{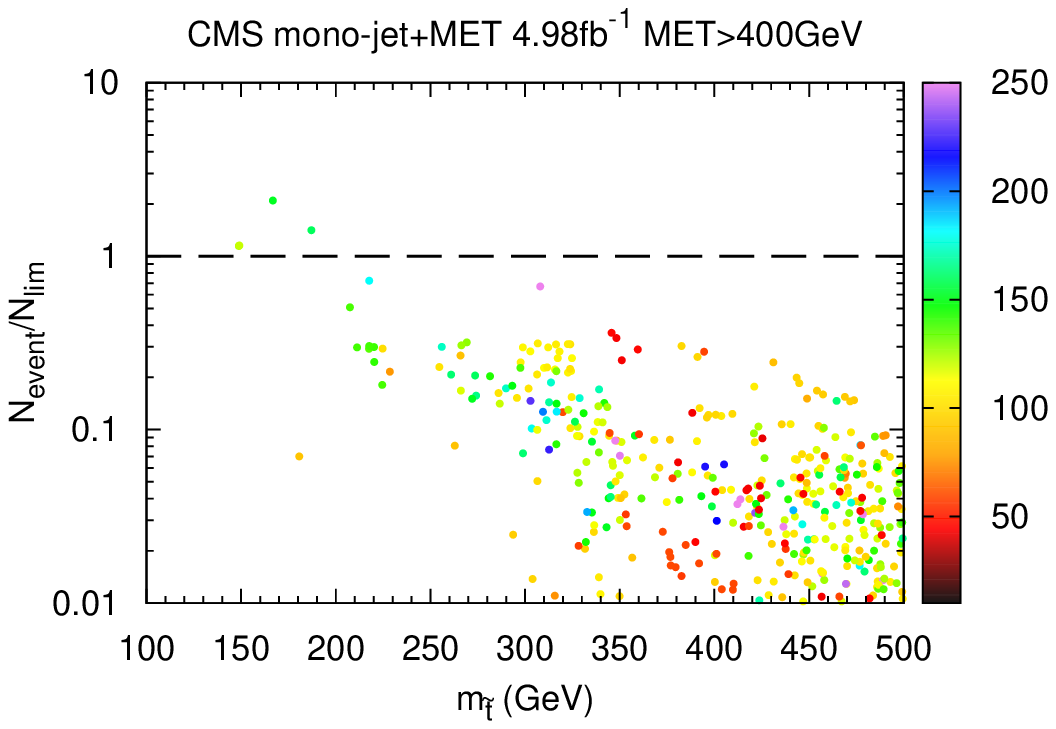}
\caption{The ratio of ${\tilde t} {\tilde t}^*$+${\tilde b} {\tilde b}^*$ event number and observed 95\% C.L. upper limit versus stop mass in the signal regions for $\sl{E}_T>350$ GeV and $\sl{E}_T>400$ GeV in the left and right panels, respectively. The color scale indicates neutralino mass.
\label{CMSmjetf}}
\end{center}
\end{figure}

In Fig. \ref{CMSmjetf}, we show the ratio $N/N_{lim}$ with both the $\sl{E}>350$ GeV and $\sl{E}>400$ GeV cases. For these large $\sl{E}_T$ cut conditions, the jet from the initial state radiation is required to be very energetic $p_j^T \sim \sl{E}_T$. It is obvious to see that only the light stop with $m_{\tilde{t}}<200$ GeV and the large production cross section can have a large event rate, as demonstrated in Fig. \ref{CMSmjetf}. For heavier stops with $\tilde{t} \to t \tilde{\chi}^0_1 $, the cut condition on the third jet will suppress the events with high jet multiplicities and leads to a weaker constraint. For the sbottom which is not degenerate with the LSP, the dominated decay mode can be $\tilde{b} \to b\tilde{\chi}^0_1$. Therefore, the cuts on the transverse momentum of jets can be satisfied easily. But the sbottom pair cannot induce a large $\sl{E}_T$ signal as required in this analysis which also leads to a weaker constraint.

\subsubsection{Constraints for the final states with b-jets}

In this subsection, we explore the impact of b-jets+MET searches on stop/sbottom pair production signatures. The identification of a b-jet is  helpful to reduce the huge QCD backgrounds. In this case, the dominated SM backgrounds are the top pair production and associated production of W/Z  with heavy flavor jets. The dibosons production $WW$, $ZZ$ and $WZ$ are sub-dominated due to a smaller electro-weak cross section.

As discussed in Ref. \cite{Papucci:2011wy}, the ``naturalness" of
SM-like Higgs mass suggests a light stop $\leq 700$ GeV and a not
very heavy gluino $\sim 1$ TeV in the SUSY model. In this case,
the gluino pair production has a moderate cross section, and the
cascade decay productions would contain many top and bottom
quarks. For example, the typical SUSY search channels are
$\tilde{g} \tilde{g} \to t\bar{t}\tilde{t}\tilde{t}^*  \to
tt\bar{t}\bar{t}+MET$ or $\tilde{g} \tilde{g} \to
b\bar{b}\tilde{b}\tilde{b}^* \to bb\bar{b}\bar{b}+MET$. The
multi-b jets in the final states are very powerful to suppress the
SM backgrounds. Therefore, the LHC has strong capability to test
or exclude such scenario.

In this work, we assume the gluino is very heavy $> 1.5$ TeV.
Therefore, the main production signatures are $pp \to {\tilde t}_1
{\tilde t}_1^*$ and $pp \to {\tilde b}_1 {\tilde b}_1^*$. For the
light sbottom pair production, the b-jets+MET search can constrain
the channel $\tilde{b} \tilde{b} \to b\bar{b}+MET$ if the $\Delta
m_{\tilde b} = m_{\tilde{b}}-m_{\tilde{\chi}^0_1}$ is large enough
(say larger than 50 GeV). For the stop pair production, if the
dominated stop decay mode is $\tilde{t} \to bW\tilde{\chi}$ or
$\tilde{t} \to t\tilde{\chi} \to b W \tilde{\chi}$, the b-jets in
final states are less energetic. To pass the $p_{_T}$ cut on the
leading b-jet, the $\Delta m_{\tilde t} =
m_{\tilde{t}}-m_{\tilde{\chi}^0_1}$ is required to be large, and
the detectable capability is lower than that of the sbottom pair. Here
we point out that if the chargino is light and the light stop has a large left-handed component then the $\tilde{t} \to b
\tilde{\chi}^+_1$ can become significant. In particular, if the
chargino mass is nearly degenerate with the LSP which can be used to
obtain the required DM relic density through the coannihilation
mechanism, the kinematics of $\tilde{t} \to b \tilde{\chi}^+_1$ is
very similar to the $\tilde{b} \to b \tilde{\chi}^0_1$. In this
case, the b-jets+MET search is also useful to test or exclude stop
pair signatures. This feature is also addressed in the first
benchmark point in Ref. \cite{Baer:2012uy}.

\begin{table}[th]
\begin{center}%
\begin{tabular}
[c]{|c|c|c|c|c|c|c|c|c|c|}\hline
$p_T^{bjet_1}$ & $p_T^{bjet_2}$ & $p_T^{j_3}$ &   $\Delta \phi(\vec{j}_i,\vec{\sl{E}}_T) > $ & $\sl{E}_T >$ & $\sl{E}_T/m_{eff}^{N_j=2} >$ & $m_{CT} >$ & $N_{lim}^{obs}$ \\\hline\hline
$> 130$ & $> 50$ & $< 50$ &  0.4(0.2) (i=1,2,(3)) & 130 & 0.25 & 100/150/200 & 27.5/19.7/11.5 \\\hline
\end{tabular}
\end{center}
\caption{Summary of cuts and observed 95\% C.L. upper limits on the excess event number following the ATLAS 2b-jets+MET analysis for 2.05 fb$^{-1}$\cite{Aad:2011cw}. }
\label{ATLjet2bt}
\end{table}

\begin{figure}[!htb]
\begin{center}
\includegraphics[width=0.40\columnwidth]{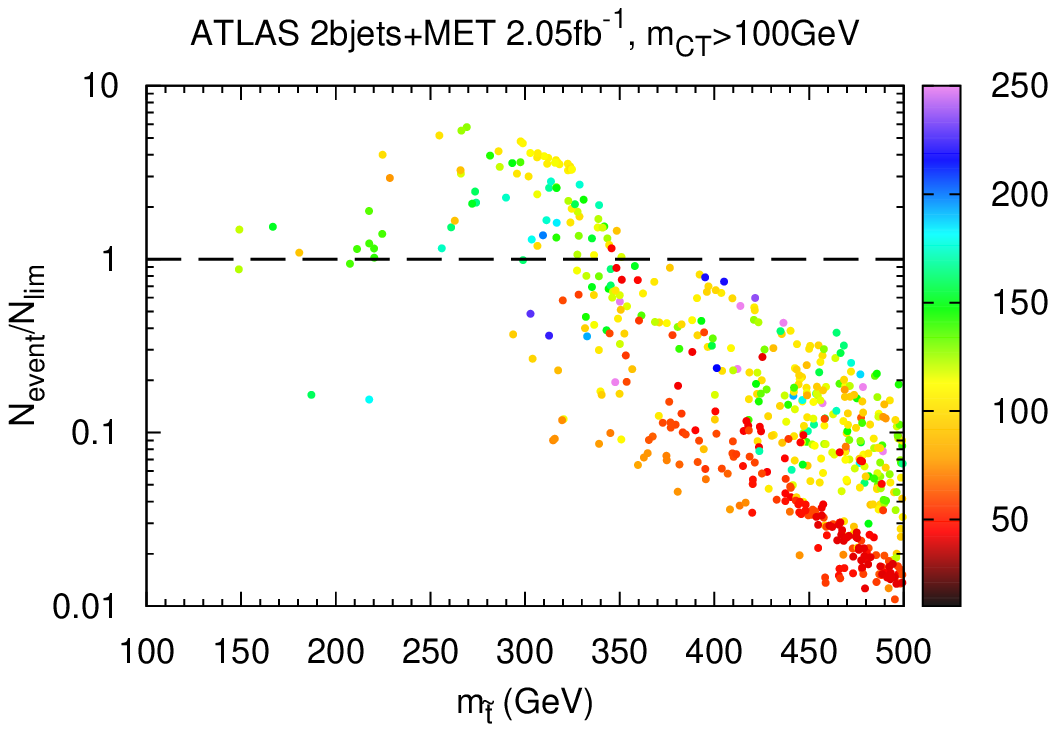}
\includegraphics[width=0.40\columnwidth]{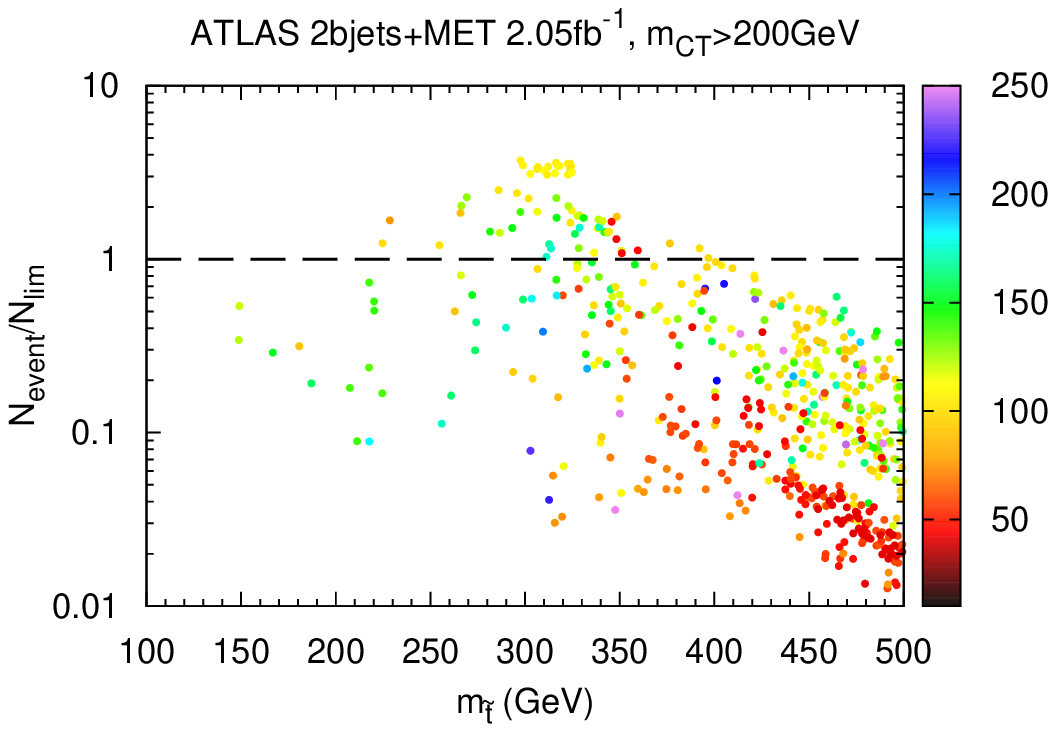}
\caption{The ratio of ${\tilde t} {\tilde t}^*$+${\tilde b} {\tilde b}^*$ event number and observed 95\% C.L. upper limit versus stop mass in the signal regions for $m_{CT}>100$ GeV and $m_{CT}>200$ GeV in the left and right panels, respectively. The color scale indicates neutralino mass.
\label{ATLjet2bf}}
\end{center}
\end{figure}

\begin{figure}[!htb]
\begin{center}
\includegraphics[width=0.40\columnwidth]{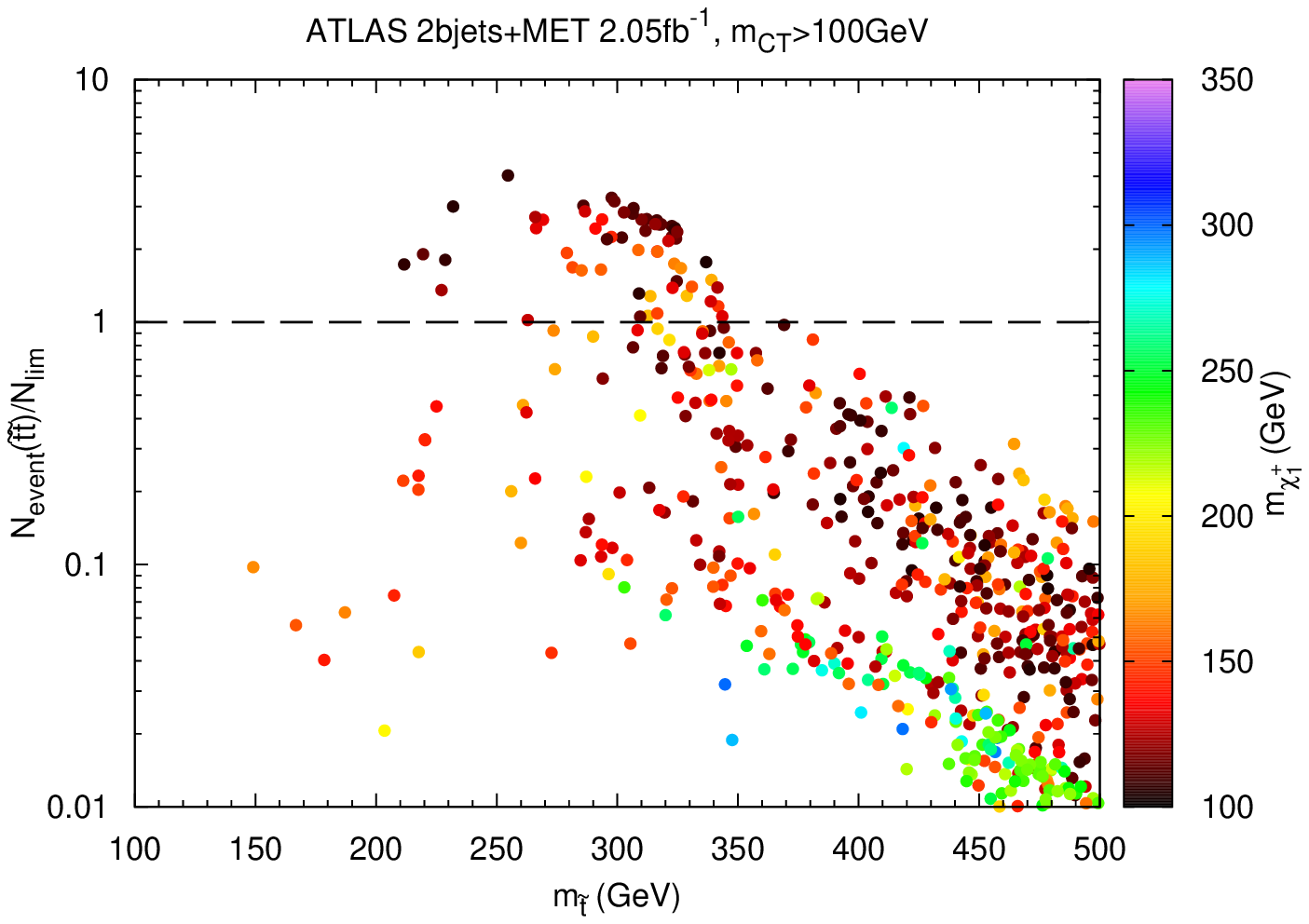}
\includegraphics[width=0.40\columnwidth]{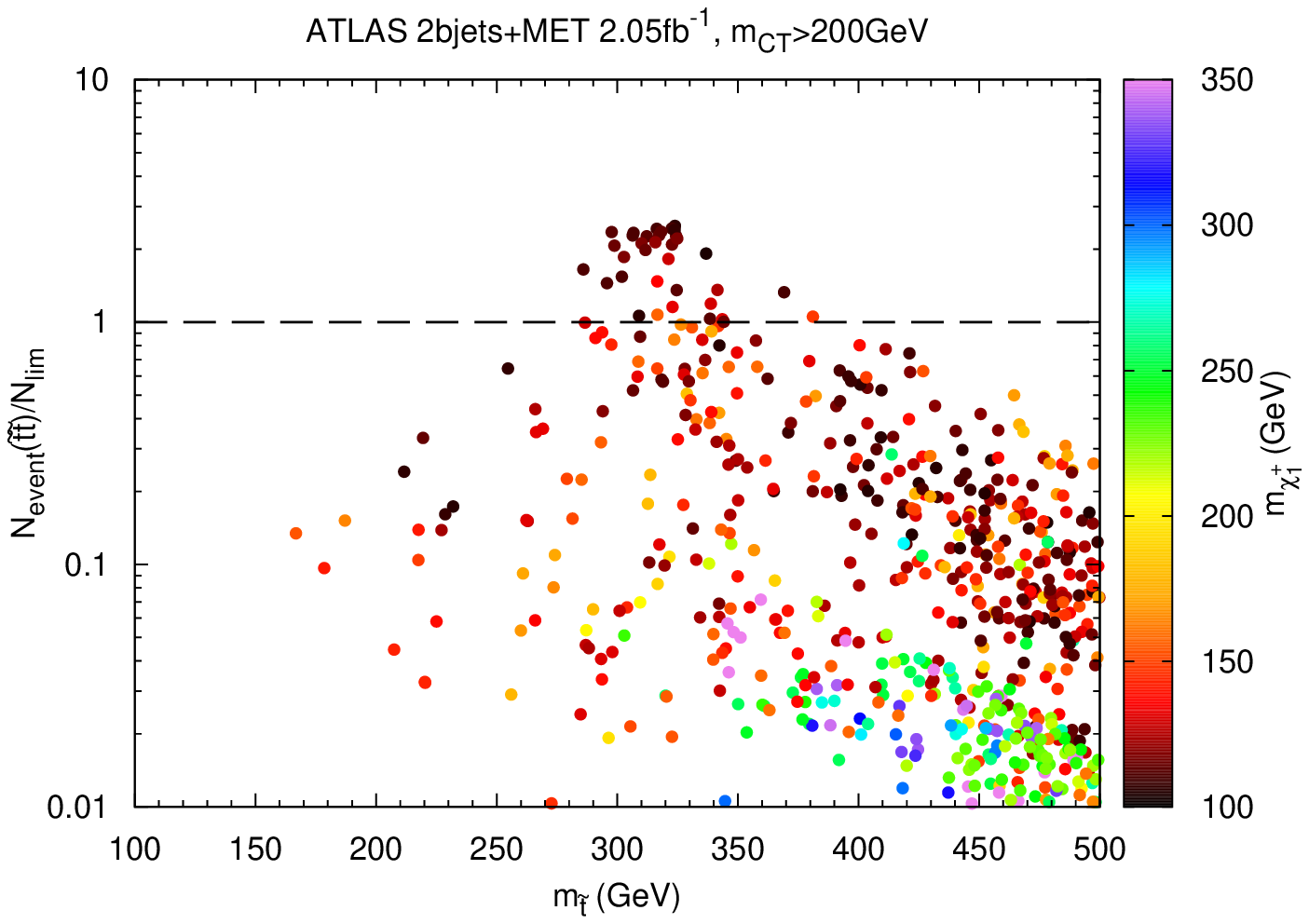}
\caption{The ratio of ${\tilde t} {\tilde t}^*$ event number and observed 95\% C.L. upper limit versus stop mass in the signal regions for $m_{CT}>100$ GeV and $m_{CT}>200$ GeV in the left and right panels, respectively. The color scale indicates chargino mass.
\label{ATLjet2bfst}}
\end{center}
\end{figure}

First, we consider the ATLAS 2b-jets+MET search based on 2.05 fb$^{-1}$ of data \cite{Aad:2011cw}. This search is optimized for sbottom pair production with the sbottom branching ratio of $BR(\tilde{b} \to b\tilde{\chi}^0_1)=100\%$. The corresponding cut conditions and limits are summarized in Table. \ref{ATLjet2bt}. In this analysis, the number of jets with $p_t>50$ GeV is required to be exactly two. No $m_{eff}$ cut is imposed because the sbottom should be light to provide large production cross section. A boost-corrected contransverse mass $m_{CT}$ is introduced as \cite{Tovey:2008ui}
\begin{equation}
m_{CT}=\sqrt{(E_T^{j_1}+E_T^{j_2})^2-(\vec{p}_T^{j_1}-\vec{p}_T^{j_1})^2} ,
\label{mct}
\end{equation}
where $m_{CT}$ is invariant under contralinear equal magnitude boosts. It can be expected that the distribution of $m_{CT}$ display an end point at $(m_{\tilde{b}}^2-m_{\tilde{\chi}^0_1}^2)/m_{\tilde{b}}$ when the two b-jets are colinear. In Fig. \ref{ATLjet2bf}, the ratios $N/N_{lim}$ are shown . We find that this search can exclude many selected parameter points, especially when the mass splitting $\Delta m_{\tilde b} = m_{\tilde{t}}-m_{\tilde{\chi}^0_1}$ is large enough with $m_{{\tilde t}_1}$ in the range of 250 GeV to 350 GeV.  When comparing the left and right panels of Fig. \ref{ATLjet2bf}, it is easy to read out the fact that the smaller $m_{CT}$ is more sensitive to smaller stop/sbottom mass region. It might be interesting to notice that the maximum excluded $m_{{\tilde t}_1}$ can reach to 380 GeV by this observable.

Moreover, we show the ratios $N/N_{lim}$ only for stop pair production in Fig. \ref{ATLjet2bfst} where the color scale indicates the mass of chargino $\tilde{\chi}^+_1$. It is noticed that even if the sbottom is very heavy, the 2b jets search can be useful to put constraints on stop pair production. This can occur for the points with significant decay mode $BR(\tilde{t} \to b \tilde{\chi}^+_1)$ with very small $m_{\tilde{\chi}^+_1}-m_{\tilde{\chi}^0_1}$, as we have mentioned before.

It should be noted that the $BR(\tilde{t} \to b \tilde{\chi}^+_1)$ cannot simply be determined by $m_{\tilde{\chi}^+_1}$. When the mass of the LSP $m_{\tilde{\chi}^0_1}$ is much smaller than that of lighter chargino $m_{\tilde{\chi}^+_1}$, the decay mode $\tilde{t} \to t\tilde{\chi}^0_1$ is still significant. Furthermore, if the kinematics is allowed, other decay modes $\tilde{t} \to t\tilde{\chi}^0_2$ and $\tilde{t} \to b\tilde{\chi}^+_2$ are open and can reduce the decay mode $BR(\tilde{t} \to b \tilde{\chi}^+_1)$. For both cases, the constraints from this search become weaker or even invalid.

\begin{table}[th]
\begin{center}%
\begin{tabular}
[c]{|c|c|c|c|c|c|c|c|c|c|c|c|c|}\hline
Requirements   &$N_{l}$ & $N_{bjet}\geq$ & $p_T^{j_1}$ & $p_T^{j_{2,3}}$ &$p_T^{j_4}$  & $m_{T}>$& $\sl{E}_T$ $>$ & $\sl{E}_T/m_{eff} >$ & $\Delta \phi > $ & $m_{eff} >$ &  $N_{lim}^{obs}$\\\hline\hline
SR0-A1/B1/C1 &0 &1  & $>$ 130 & $>$ 50 &- &- & 130 & 0.25 & 0.4 & 500/700/900 & 580/133/31.6  \\\hline
SR0-A2/B2/C2 &0 &2  & $>$ 130 & $>$ 50 & -& -& 130 & 0.25 & 0.4 & 500/700/900 & 124/29.6/8.9  \\\hline
SR1-D &1 &1  & $>$ 60 & $>$ 50 & $>$50& 100&80 & - &-& 700 & 45.5 \\\hline
SR1-E &1 &1  & $>$ 60 & $>$ 50 & $>$50& 100& 200 & - &-& 700 & 17.5  \\\hline
\end{tabular}
\end{center}
\caption{Summary of cuts and observed 95\% C.L. upper limits on the excess event number following the ATLAS b-jets+MET analysis for 2.05 fb$^{-1}$ \cite{ATLAS:2012ah}. }%
\label{ATL1lepblept}%
\end{table}

\begin{figure}[!htb]
\begin{center}
\includegraphics[width=0.40\columnwidth]{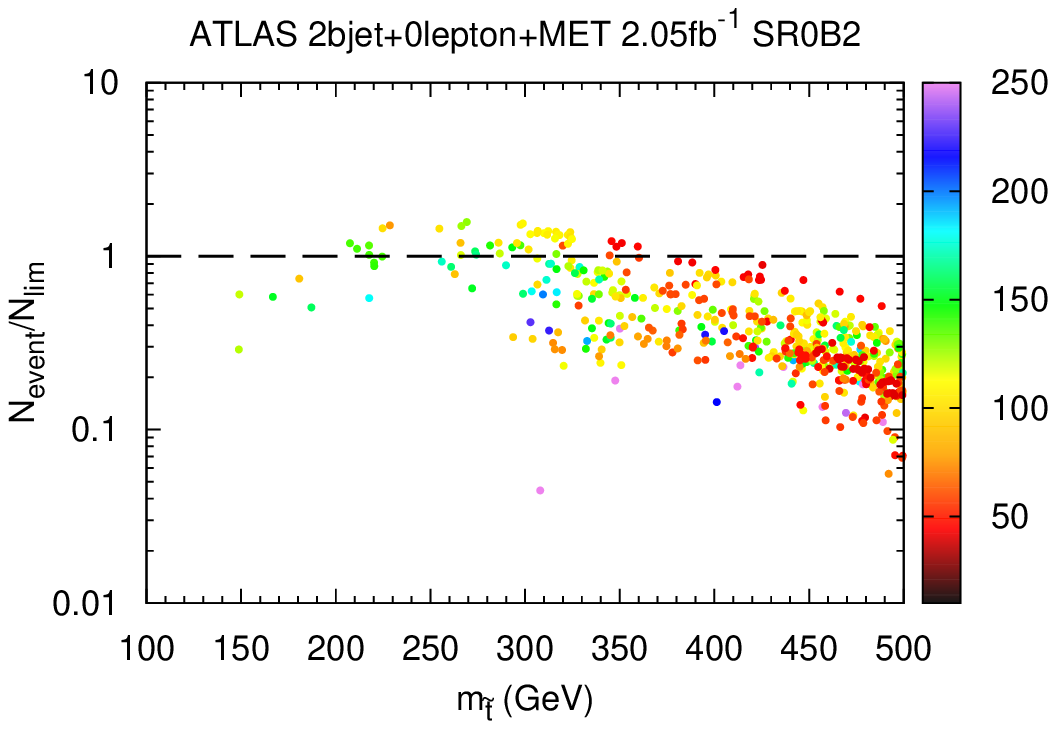}
\includegraphics[width=0.40\columnwidth]{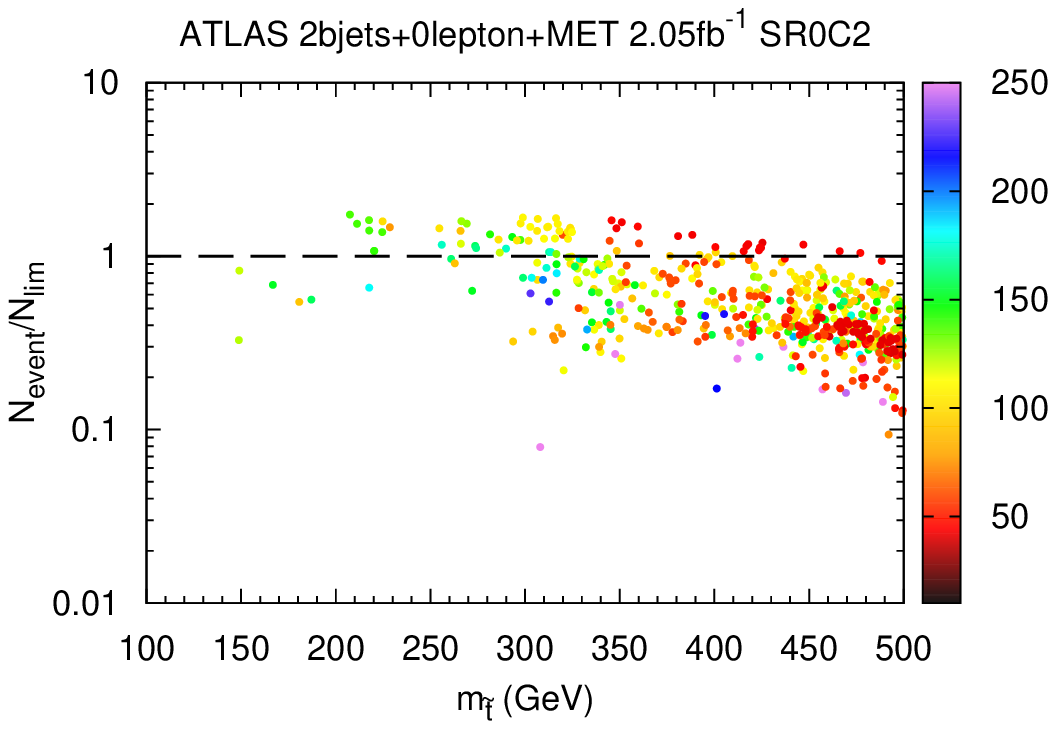}
\caption{The ratio of ${\tilde t} {\tilde t}^*$+${\tilde b} {\tilde b}^*$ event number and observed 95\% C.L. upper limit versus stop mass in the signal regions SR0-B2 and SR0-C2 in the left and right panels, respectively. The color scale indicates neutralino mass.
\label{ATL1lepblepf}}
\end{center}
\end{figure}

Next, we consider the ATLAS b-jets+MET search based on 2.05 fb$^{-1}$ of data \cite{ATLAS:2012ah}. In this analysis, the number of b-jets is required to be at least one or two. Moreover, two signal regions allow one lepton in the final states. This search can be supposed to constrain the the signal $p p \to {\tilde t}_1 {\tilde t}_1^*$. The corresponding cut conditions and limits are summarized in Table. \ref{ATL1lepblept}. The ratios $N/N_{lim}$ from the two most stringent channels SR0B2 and SR0C2 are shown in Fig. \ref{ATL1lepblepf}. We can see even these channels require a large $m_{eff}$ cut as $m_{eff}> 700-900$ GeV, they can still exclude many parameter points. It is remarkable that the channel SR0B2 (SR0C2) can exclude signals with maximum $m_{{\tilde t}_1}$ up to 380 GeV(480 GeV). The constraints from 1b-jet+MET signal regions are weaker than 2b-jets+MET due to the large backgrounds. The lepton + b-jets + MET in the same analysis also cannot achieve better constraints and is omitted here.

\begin{table}[th]
\begin{center}%
\begin{tabular}
[c]{|c|c|c|c|c|c|c|c|c|c|c|c|c|}\hline
Requirements   & $p_T^{j_{1,2,3}}$ & $\Delta\phi_{norm}$ & $N_{bjet} \geq$  & $H_T$ & $\sl{E}_T$\\\hline\hline
1BL   & $>$50 & 4.0 & 1 & 400 & 250  \\\hline
1BT   & $>$50 & 4.0 & 1 & 500 & 500 \\\hline
2BL   & $>$50 & 4.0 & 2 & 400 & 250 \\\hline
2BT   & $>$50 & 4.0 & 2 & 600 & 300 \\\hline
\end{tabular}
\end{center}
\caption{Summary of cut conditions, following the CMS b-jets+MET analysis for 4.98 fb$^{-1}$. The capital ``L" and ``T" mean ``loose" and  ``tight", respectively.}%
\label{CMSjet1bt}%
\end{table}

\begin{figure}[!htb]
\begin{center}
\includegraphics[width=0.40\columnwidth]{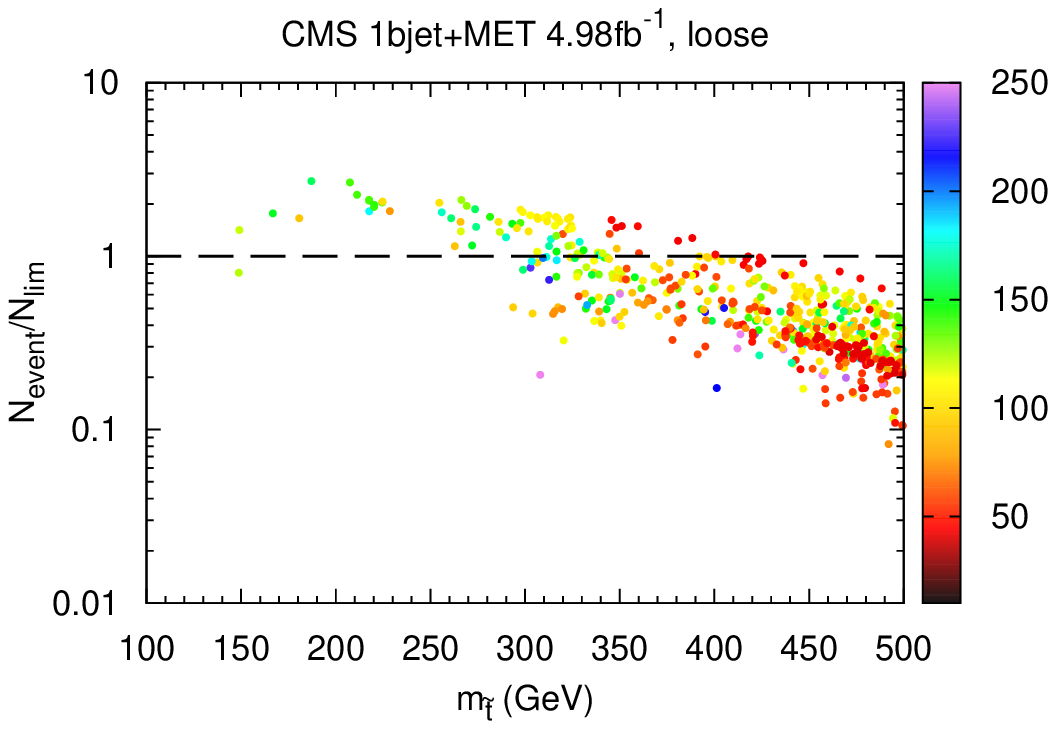}
\includegraphics[width=0.40\columnwidth]{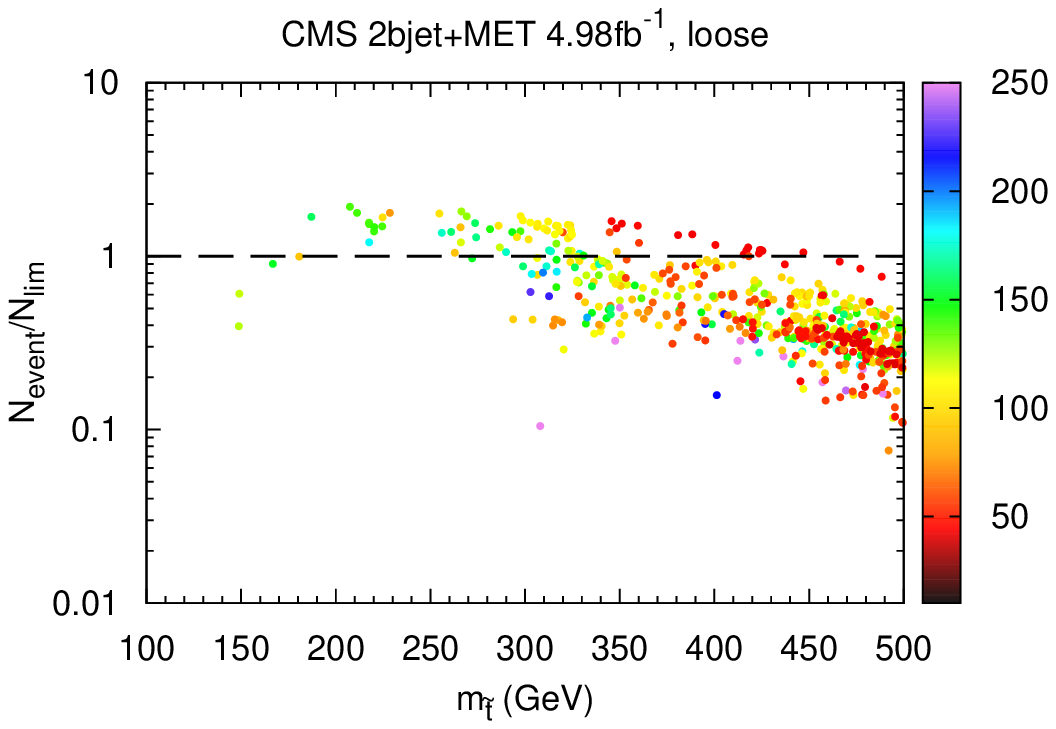}
\caption{The ratio of ${\tilde t} {\tilde t}^*$+${\tilde b} {\tilde b}^*$ event number and observed 95\% C.L. upper limit versus stop mass in the signal regions 1BL and 2BL in the left and right panels, respectively. The color scale indicates neutralino mass.
\label{CMSjet1bf}}
\end{center}
\end{figure}

Then we consider the CMS b-jets+MET search based on 4.98 fb$^{-1}$ of data \cite{CMSbjets}. The corresponding cut conditions are summarized in Table. \ref{CMSjet1bt}. In this analysis, $H_T$ defined as the scalar sum of momenta of all the energetic jets are used to set cut conditions. $\Delta \phi_{norm}$ is the normalized azimuthal separation between the $\sl{E}$ direction and jets. The collaboration has not yet provided limits on the number of new physics explicitly. Here we use the formula given in Ref. \cite{Fox:2011pm} to roughly estimate the upper limits,
\begin{equation}
\chi^2=\frac{\left[N_{obs}-N_{SM}-N_{BSM} \right]^2}{N_{BSM}+N_{SM}+\sigma^2_{SM}}
\end{equation}
where $N_{obs}$ is the number of observed events, $N_{BSM}$ and $N_{SM}$
are the numbers of predicted beyond standard model (BSM) events and SM backgrounds, respectively, and $\sigma_{SM}$ is the
uncertainty due to statistic and systematic reasons.
Requiring $\chi^2 < 3.84$, we can get the 95\% C.L. upper bounds on the numbers of events contributing from new physics. For signal regions 1BL, 1BT, 2BL, and 2BT, these upper bounds can be translated as the numbers of 124, 14.7, 58, and 41, respectively. Because this analysis dose not require a very hard leading jet, the numbers of events passing all the cut conditions are larger than the last
analysis. The ratios $N/N_{lim}$ are given in Fig.
\ref{CMSjet1bf}. Because the ``tight'' cut conditions require very
large $\sl{E}$, the signals are significantly reduced in these
signal regions. Here we only show the constraints from ``loose"
searches.

\begin{figure}[!htb]
\begin{center}
\includegraphics[width=0.40\columnwidth]{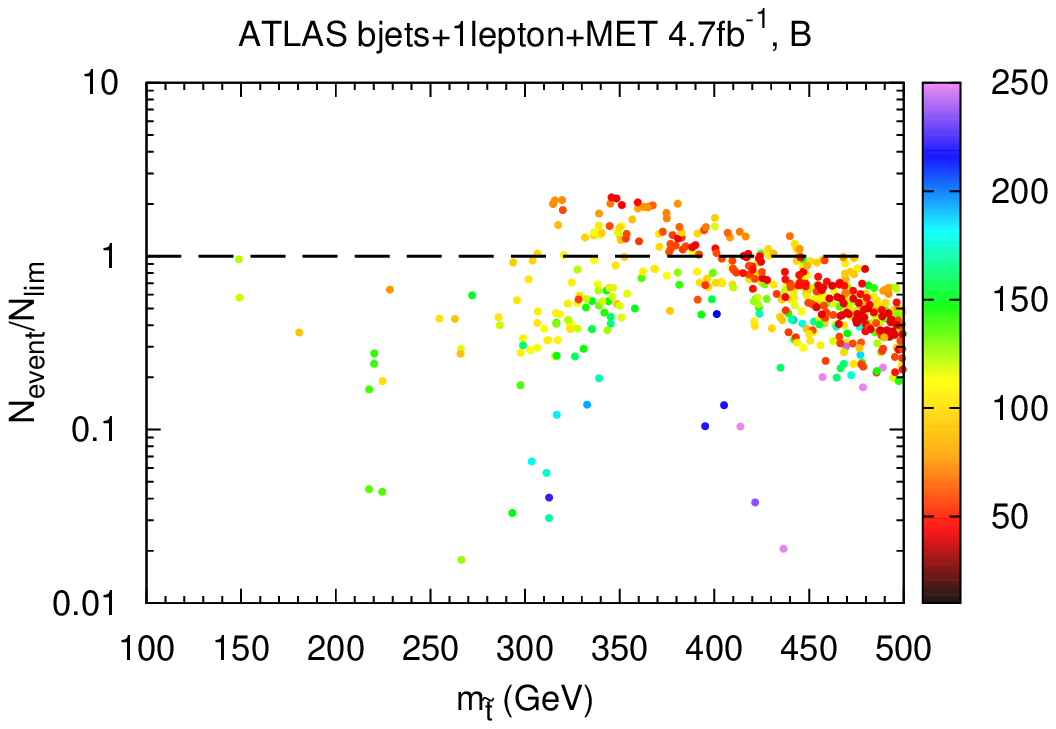}
\includegraphics[width=0.40\columnwidth]{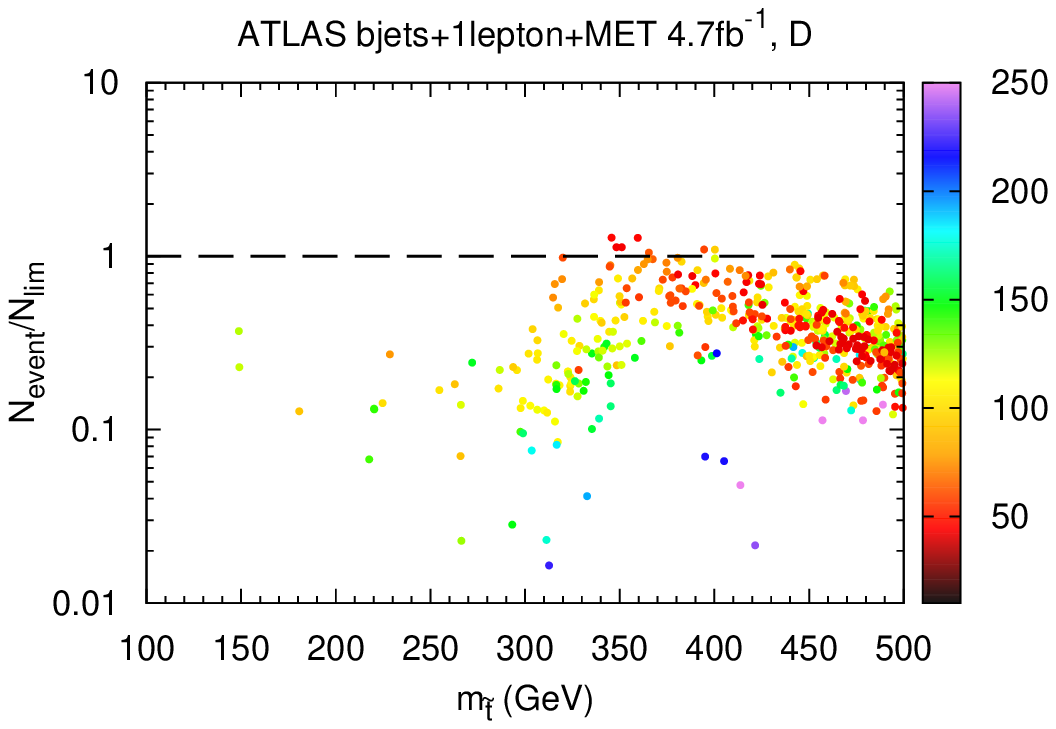}
\caption{The ratio of ${\tilde t} {\tilde t}^*$+${\tilde b} {\tilde b}^*$ event number and observed 95\% C.L. upper limit versus stop mass in the signal regions B and D in the left and right panels respectively. The color scale indicates neutralino mass.
\label{ATL1lepf}}
\end{center}
\end{figure}

The bounds obtained from inclusive and exclusive 2bjets+MET searches can be compared from Figs. \ref{ATLjet2bf}, \ref{ATL1lepf} and \ref{CMSjet1bf}.
Roughly speaking, the bounds from inclusive 2bjets+MET searches seem to be more stringent, as demonstrated by the fact the selected points are squeezed in a smaller range of $N/N_{lim}$ between 0.1 and 1 and the maximum excluded $m_{{\tilde t}_1}$ can reach to $\sim 400$ GeV.

\begin{table}[th]
\begin{center}%
\begin{tabular}
[c]{|c|c|c|c|c|c|c|c|c|c|c|c|c|}\hline
Requirements   & $p_T^{j_1} >$ & $p_T^{j_2} >$ & $p_T^{j_3} >$ & $p_T^{j_4} >$ & $\phi(\vec{j}_{1,2},\vec{\sl{E}}_T) >$ & $m_{jjj}$ &  $\sl{E}_T >$ & $\sl{E}_T/\sqrt{H_T} >$ & $m_T >$ & $N_{lim}$\\\hline\hline
A/B/C   & 80 & 60 & 40 & 25 & 0.8 & [130, 205] & 150 & 7/9/11 &  120 & 15.1/10.1/10.8 \\\hline
D   & 80 & 60 & 40 & 25 & 0.8 & [130, 205] & 225 & 11 &  130 & 8.4 \\\hline
E   & 80 & 60 & 40 & 25 & 0.8 & [130, 205] & 275 & 11 &  140 & 8.2  \\\hline
\end{tabular}
\end{center}
\caption{Summary of cuts and observed 95\% C.L. upper limits on the excess event number following the ATLAS ``heavy top" 1lepton+b-jets+MET analysis for 4.7 fb$^{-1}$ \cite{CMSbjets}.}%
\label{ATL1lepbt}%
\end{table}

Since the semileptonic mode of $p p \to {\tilde t}_1 {\tilde t}_1^* \to t {\bar t} \tilde{\chi}^0_1 \tilde{\chi}^0_1$ can have a large branching fraction and enjoys a smaller SM background, so it is expected the search for this mode should be stringent. Then we consider the constraints from the ATLAS 1lepton+b-jets+MET searches based on 4.7 fb$^{-1}$ of data \cite{Collaboration:2012ar}.\footnote{The ATLAS analysis of 0 lepton+b-jets+MET in Ref. \cite{Collaboration:2012si} is also optimized for stop pair production with decay mode $\tilde{t} \to t\tilde{\chi}^0_1$, while tops are assumed to decay hadronically. The constraints from this channel are weaker than the semileptonic channel over most of the parameter space.} The corresponding cut conditions and limits are summarized in Table. (\ref{ATL1lepbt}). It requires that the number of isolated lepton is exactly one. Obviously, this analysis is optimized for searching stop pair production with decay mode $\tilde{t} \to t\tilde{\chi}^0_1$. One top from stop decay is required to decay hadronically and the other semileptonically. $m_T$ is the transverse mass defined as
\begin{equation}
m_T=\sqrt{2p_T^l\sl{E}\;(1-\cos\Delta\phi(l,\vec{\sl{E}}_T))}.
\end{equation}
$m_T$ denotes the mass scale of mother particles which decay into charged leptons, and $m_T$ cut can be used to reduce $W+jets$ backgrounds.
To suppress the backgrounds from dileptonically decaying top pair, a specific cut on three-jet invariant mass $m_{jjj}$ is required. Two jets
with $m_{jj}> 60$ GeV and smallest $\Delta R$ are assumed to be originated from a hadronically decaying $W$ boson, and a third jet which is closest to
the reconstructed $W$ boson is selected. These three jets may be the decay products of a hadronically decaying top, and the invariant mass $m_{jjj}$ is required to be around top mass 130 GeV$< m_{jjj}< 205$ GeV.

The ratio $N/N_{lim}$ is shown in Fig. \ref{ATL1lepf}. It is obvious that this channel is sensitive to the mass range of stops from 270 GeV to 400 GeV when the main decay mode of the stops can be $\tilde{t} \to t\tilde{\chi}^0_1$. On the other hand, for the parameter points with heavy stops $> 400$ GeV and light LSPs $< 150$ GeV in our scan, the main decay modes of stops $\tilde{t} \to t\tilde{\chi}^0_2$, $\tilde{t} \to t\tilde{\chi}^0_3$ or $\tilde{t} \to b\tilde{\chi}^+_2$ are open. In this case, although the mass difference between the stop and LSP is large, the limits become weaker due to the absence of energetic top quarks in the final states. From Fig. \ref{ATL1lepf} we can also see the constraints from signal region B are stronger than region E due to the smaller $\sl{E}$ cut condition.

\begin{figure}[!htb]
\begin{center}
\includegraphics[width=0.40\columnwidth]{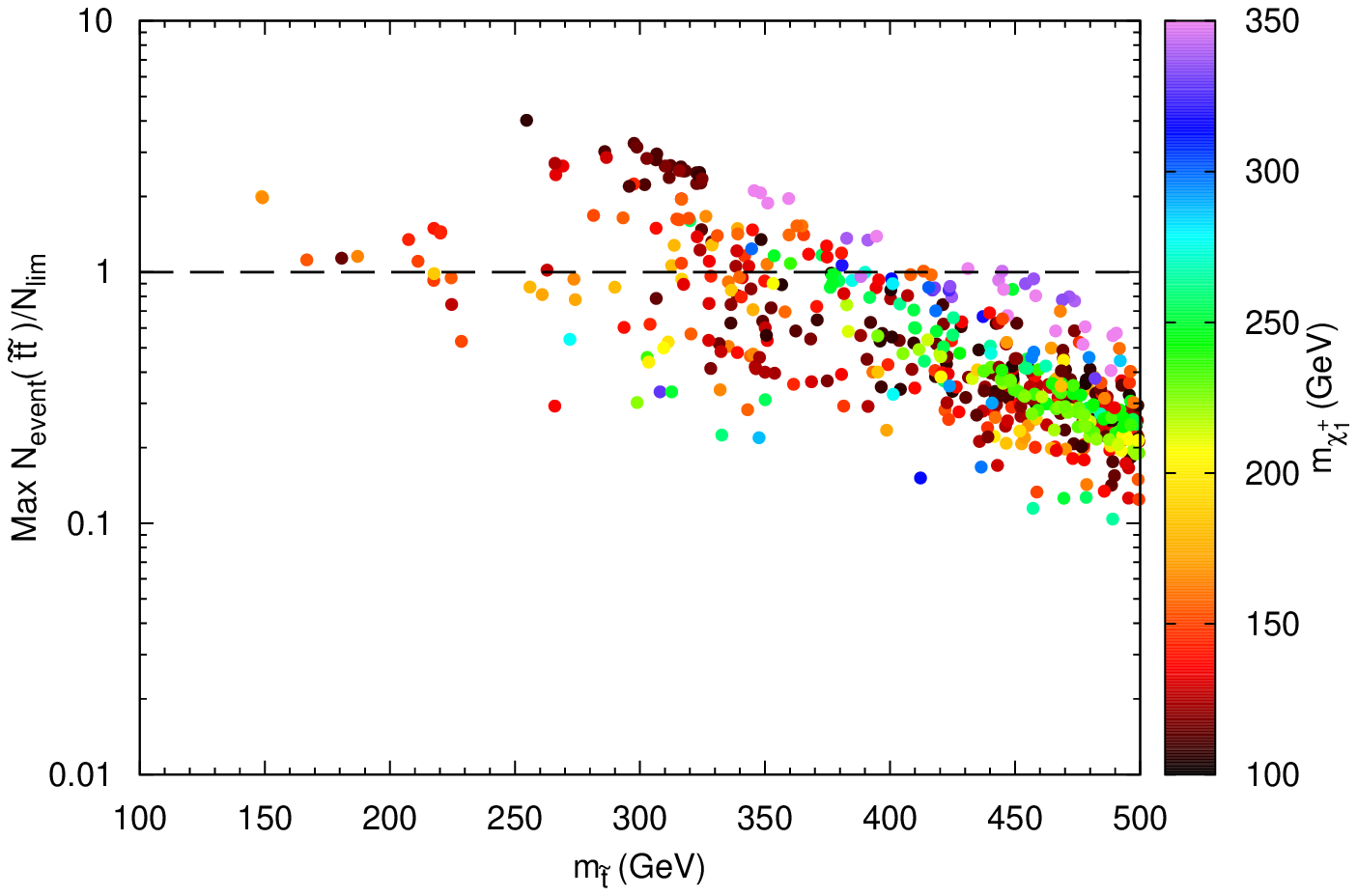}
\includegraphics[width=0.40\columnwidth]{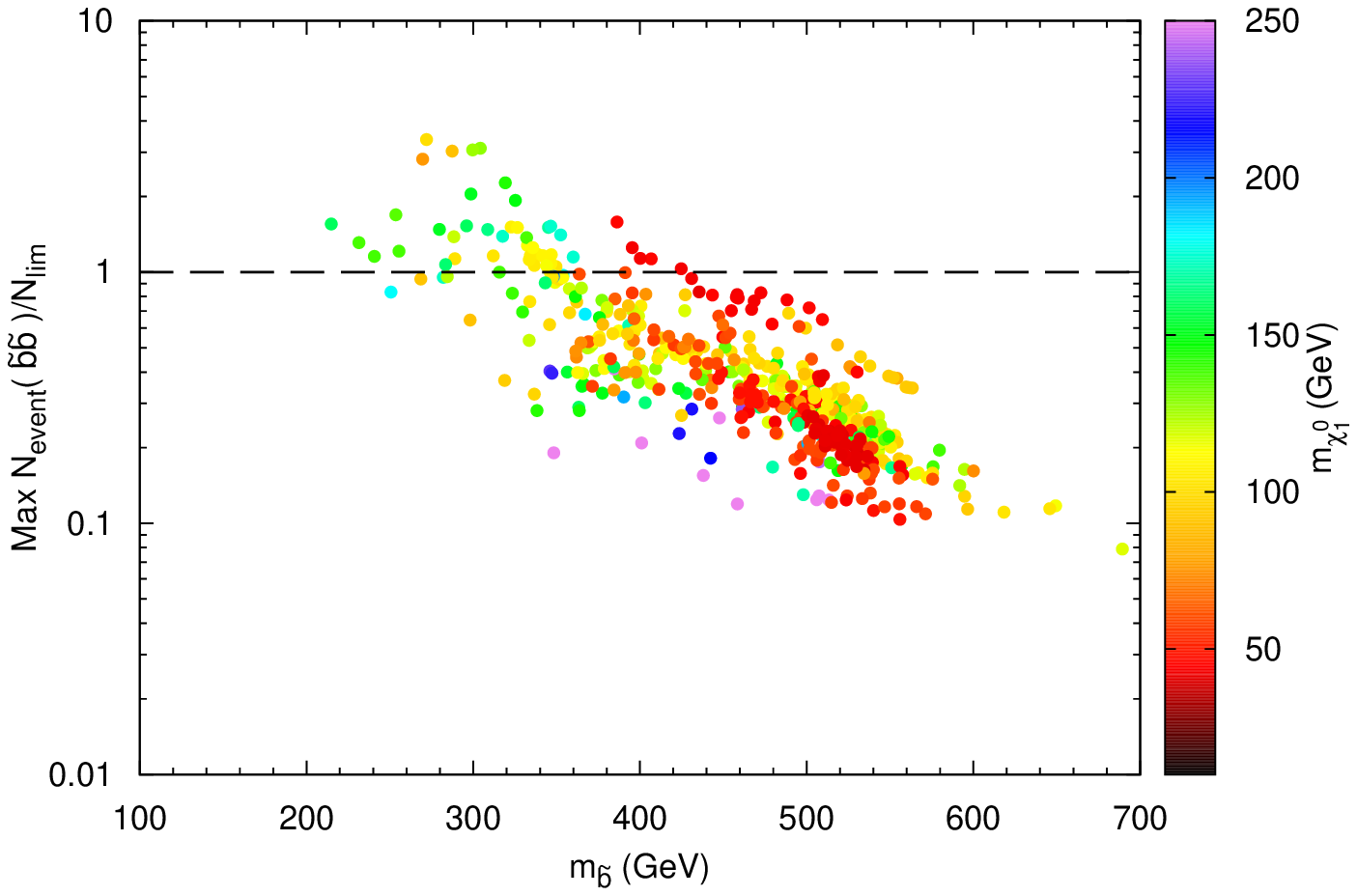}
\includegraphics[width=0.40\columnwidth]{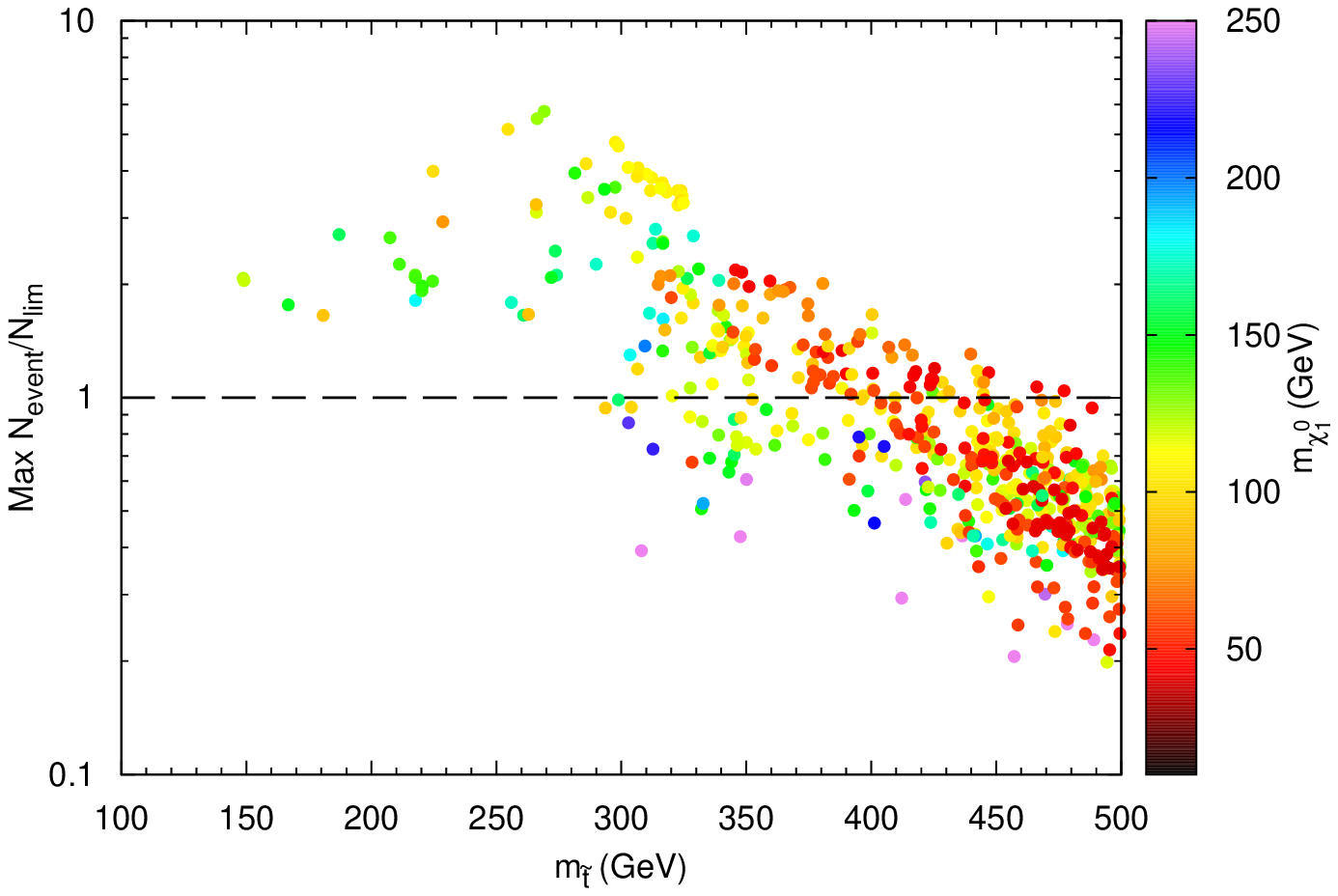}
\includegraphics[width=0.40\columnwidth]{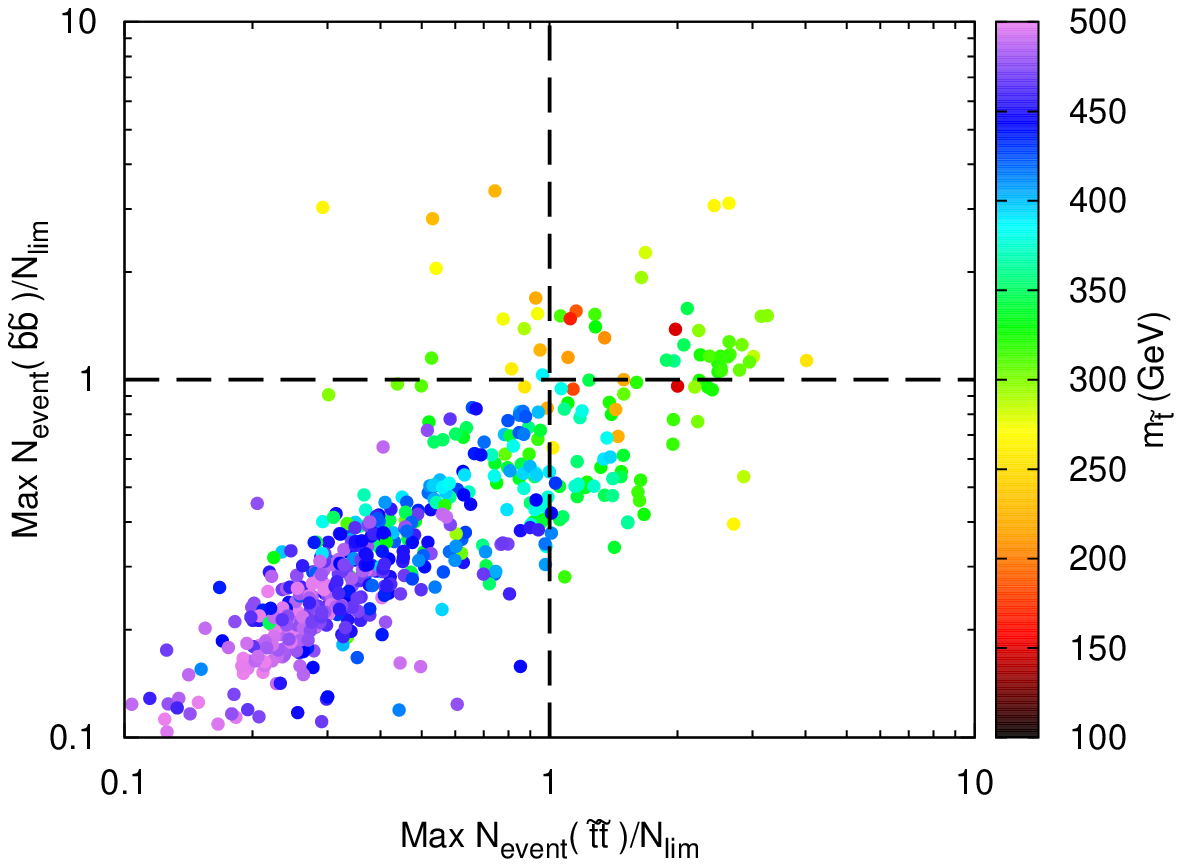}
\caption{Top left: the maximum value of $N(\tilde{t}\tilde{t}^*)/N_{lim}$  (which is determined by considering all $N(\tilde{t}\tilde{t}^*)/N_{lim}$  from different channels) versus stop mass, where the color scale indicates chargino mass. Top right: the maximum value of $N(\tilde{b}\tilde{b}^*)/N_{lim}$ versus sbottom mass, where the color scale indicates neutralino. Bottom left: the maximum value of $(N(\tilde{t}\tilde{t}^*)+N(\tilde{b}\tilde{b}^*))/N_{lim}$ versus stop mass, where the color scale indicates neutralino mass. Bottom right: the maximum value of $N(\tilde{b}\tilde{b}^*)/N_{lim}$ versus the maximum value of $N(\tilde{t}\tilde{t}^*)/N_{lim}$, where the color scale indicates stop mass.
\label{lastplot}}
\end{center}
\end{figure}

In Fig. \ref{lastplot}, we summarize the most stringent bounds from direct SUSY searches for both stop and sbottom pair productions. It is observed that for the stop pair production, the current LHC collaborations can exclude the stop(sbottom) up to 400 GeV or so when only the exclusive signals are considered. When the inclusive signals are considered, the direct searches can exclude the mass of the stop up to 500 GeV, as shown by the bottom-right plot.

Finally, it is should be mentioned that CMS and ATLAS have released a series of results of light stop/sbottom searches with 5 fb$^{-1}$ of data \cite{CMSbjetsalphat,CMSbjetsrazor,:2012tx,ATLAS2bjet5}. Some results based on $\sqrt{s}=8$ TeV have also been provided \cite{ATLsamelep,CMSbjets8TeV}. In these searches, some powerful methods based on additional kinematic variables such as ``$\alpha_T$" \cite{CMSbjetsalphat} and ``$Razor$" \cite{CMSbjetsrazor} are performed. These analyses can set very stringent constraints on the sbottom and stop pair production with large $BR(\tilde{b} \to b \tilde{\chi}^0_1)$ and $BR(\tilde{t} \to t \tilde{\chi}^0_1)$. Some parameter points with $m_{\tilde{t}}> 500$ GeV may also be excluded.
Here we do not perform these analyses and leave them for future studies.

\subsection{Benchmark points}

\begin{table}[th]
\begin{center}%
\begin{tabular}
[c]{|c|c|c|c|c|}\hline
Point   & BMP1 & BMP2 & BMP3 & BMP4\\\hline \hline
$\tan\beta$   & 5.84 & 11.8 & 16.7 & 3.66\\\hline
$\lambda$   & 0.66 &  0.68 & 0.41 & 0.71\\\hline
$\kappa$   & 0.18 &  0.34 & 0.47 & 0.16\\\hline
$\mu$ (GeV)  & 183 & 152 & 223 & 221\\\hline \hline
$A_\lambda$ (GeV)&1110& 1742 & 2903 & 826\\\hline
$A_\kappa$ (GeV)& 13.7 & -2.83 & -122 & - 136\\\hline
$A_t=A_b=A_\tau$ (GeV)& 1370& 1813 & 2233 & -785\\\hline \hline
$M_{\tilde {Q}_3}$ (GeV) &556& 514 & 1968 & 1189\\ \hline
$M_{\tilde {t}_R }=M_{\tilde {b}_R }$ (GeV) &998 & 1348 & 397 & 484\\ \hline
$M_{\tilde {\ell}_L}=M_{\tilde {\ell}_R}$ (GeV) & 200 & 261 & 520 & 140 \\ \hline
\hline
$M_1 $ (GeV) &977  & 118 & 191 & 530\\ \hline
$M_2$ (GeV) & 332 & 490 & 141 & 160\\ \hline
\end{tabular}
\end{center}
\caption{Input parameters of four benchmark points.}%
\label{benchmarkps}%
\end{table}

\begin{table}[th]
\begin{center}%
\begin{tabular}
[c]{|c|c|c|c|c|}\hline
Point   & BMP1 & BMP2 & BMP3 & BMP4\\\hline \hline
$m_{_{H_1}}$ (GeV) & 108 & 125 & 125 & 84\\ \hline
$m_{_{H_2}}$ (GeV) & 126 &  151& 476 & 124\\ \hline
$m_{_{A_1}}$ (GeV) & 85 &  100 & 290 & 180\\ \hline
$m_{\tilde{\chi}^0_1}$ & 78.5 & 66.7 & 118 & 80\\ \hline
$m_{\tilde{\chi}^0_2}$ & 211 & 135 & 182 & 163\\ \hline
$m_{\tilde{\chi}^\pm_1}$ & 165 & 149 & 125 & 124\\ \hline
$m_{\tilde{\chi}^\pm_2}$ & 370 & 516 & 267 & 274\\ \hline
$m_{\tilde t_1}$ & 497 & 475 & 346 & 462\\ \hline
$m_{\tilde b_1}$ & 534 & 504 & 402 & 474\\ \hline \hline
$R(H_{SM} \to \gamma \gamma)$ & 1.17 & 1.29 & 1.01 & 1.23\\ \hline \hline
$R(H_{SM} \to V V)$ & 1.11 & 1.20 & 0.98 & 1.16\\ \hline
$R(H_{SM} \to b {\bar b})$ & 0.89 & 0.83 & 1.00 & 0.91\\ \hline \hline
$BR(\chi^\pm_1 \to W^\pm  \tilde{\chi}^0_1)$ & 100\% &100\% & ($W^*$) 100\%  & ($W^*$) 100\% \\ \hline \hline
$BR({\tilde t_1} \to t \tilde{\chi}^0_1 )$ & 41.7\% & 38.7\% & 8.0\% & 22\% \\ \hline
$BR({\tilde t_1} \to t \tilde{\chi}^0_2 )$ & 9.9\% & 17.9\%& - & -\\ \hline
$BR({\tilde t_1} \to t \tilde{\chi}^0_3 )$ & 26.1\% & 32.8\% & - & 12.3\%\\ \hline
$BR({\tilde t_1} \to t \tilde{\chi}^0_4 )$ & - & 5.4\% & - & 1.9\%\\ \hline
$BR({\tilde t_1} \to b \tilde{\chi}^+_1 )$ & 1.8\% & 5.2 \% & 28.5\% & 30.0\% \\ \hline
$BR({\tilde t_1} \to b \tilde{\chi}^+_2 )$ & 20.4\% & -& 63.5\% & 34.3 \% \\ \hline \hline
$BR({\tilde b_1} \to b \tilde{\chi}^0_1)$ & 1.8\% &  2.8\% & 20.2\% & 10.4\%\\ \hline
$BR({\tilde b_1} \to b \tilde{\chi}^0_2 )$ & 3.5\% &  0.3\% & 33\% & 8.5 \% \\ \hline
$BR({\tilde b_1} \to b \tilde{\chi}^0_3 )$ & 0.7\% &  2.3\% & 19.0\% & 23.4\% \\ \hline
$BR({\tilde b_1} \to b \tilde{\chi}^0_4 )$ & 11.1\% &  1.3\% & 10.4\% & 20.7\% \\ \hline
$BR({\tilde b_1} \to t \tilde{\chi}^-_1 )$ & 82.9\% & 93.2\% & 17.2\% & 28.4\% \\ \hline
$BR({\tilde b_1} \to t \tilde{\chi}^-_2 )$ & - & - & -  & 8.4\% \\ \hline
\end{tabular}
\end{center}
\caption{Mass spectra as well as main decay modes of stops and sbottoms in four benchmark points, where the label $W^*$ in the row marked by $Br(\chi^\pm_1 \to W^\pm  \tilde{\chi}^0_1)$ means the off-shell W boson.}%
\label{benchmarkps1}%
\end{table}

\begin{figure}[!htb]
\begin{center}
\includegraphics[width=0.46\columnwidth]{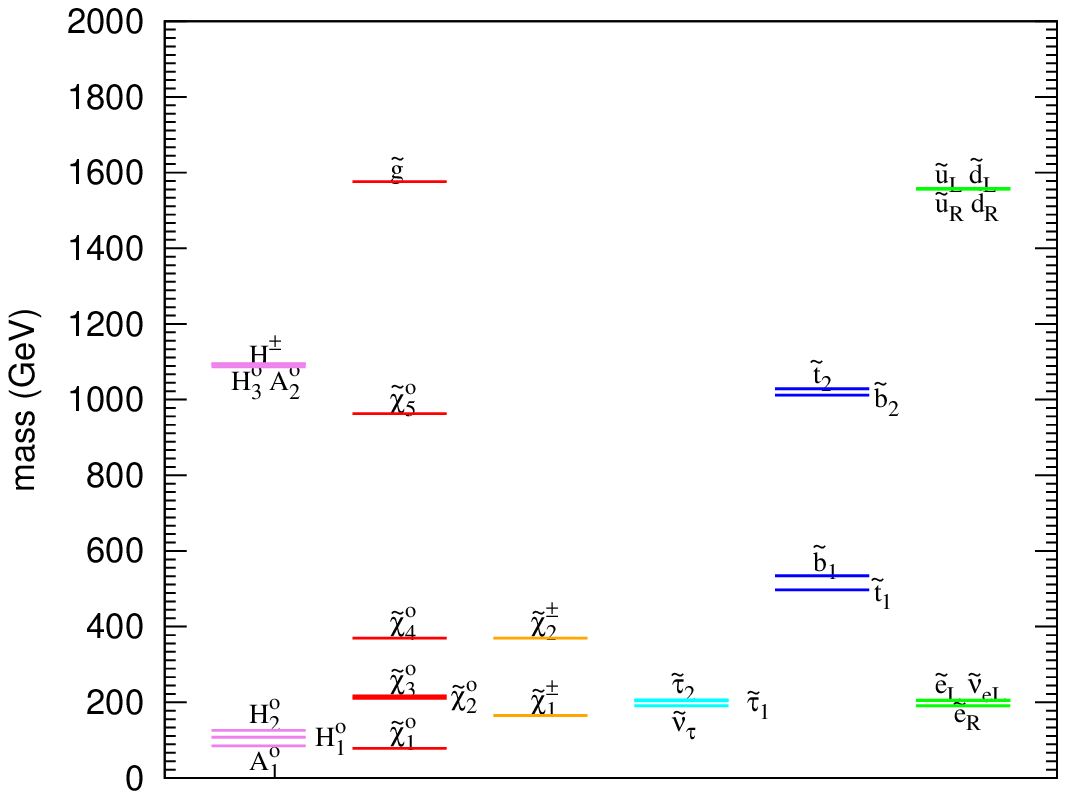}
\includegraphics[width=0.46\columnwidth]{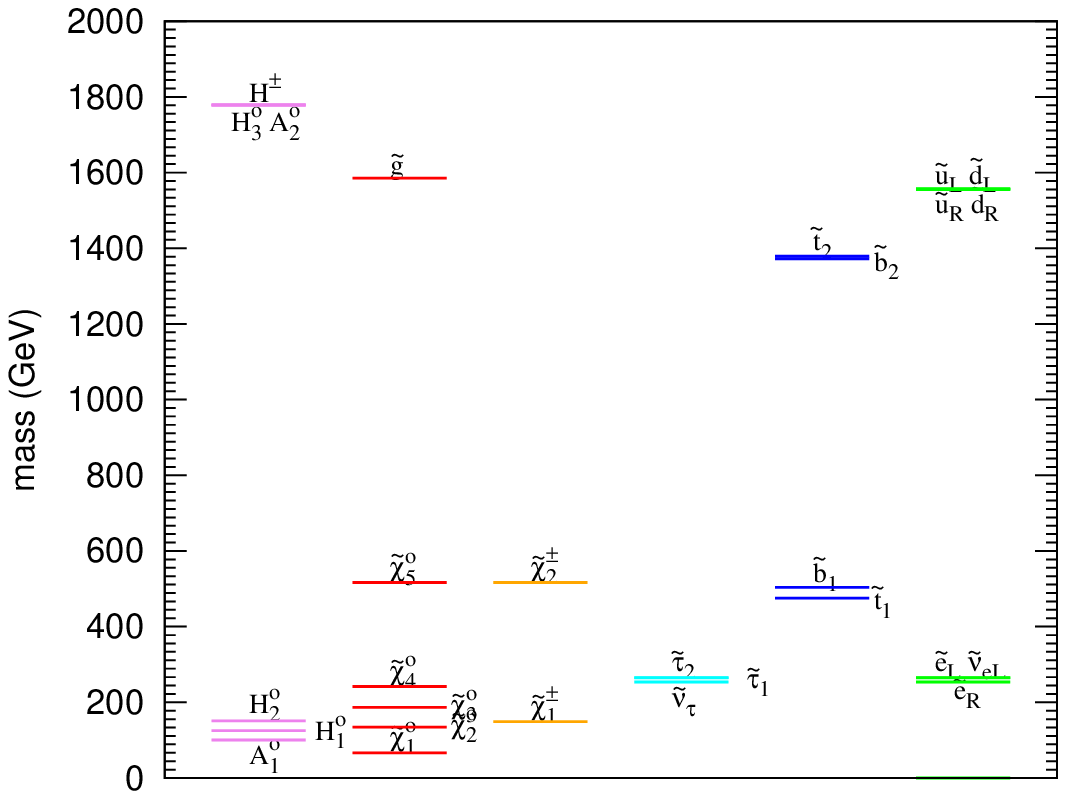}\\
\includegraphics[width=0.46\columnwidth]{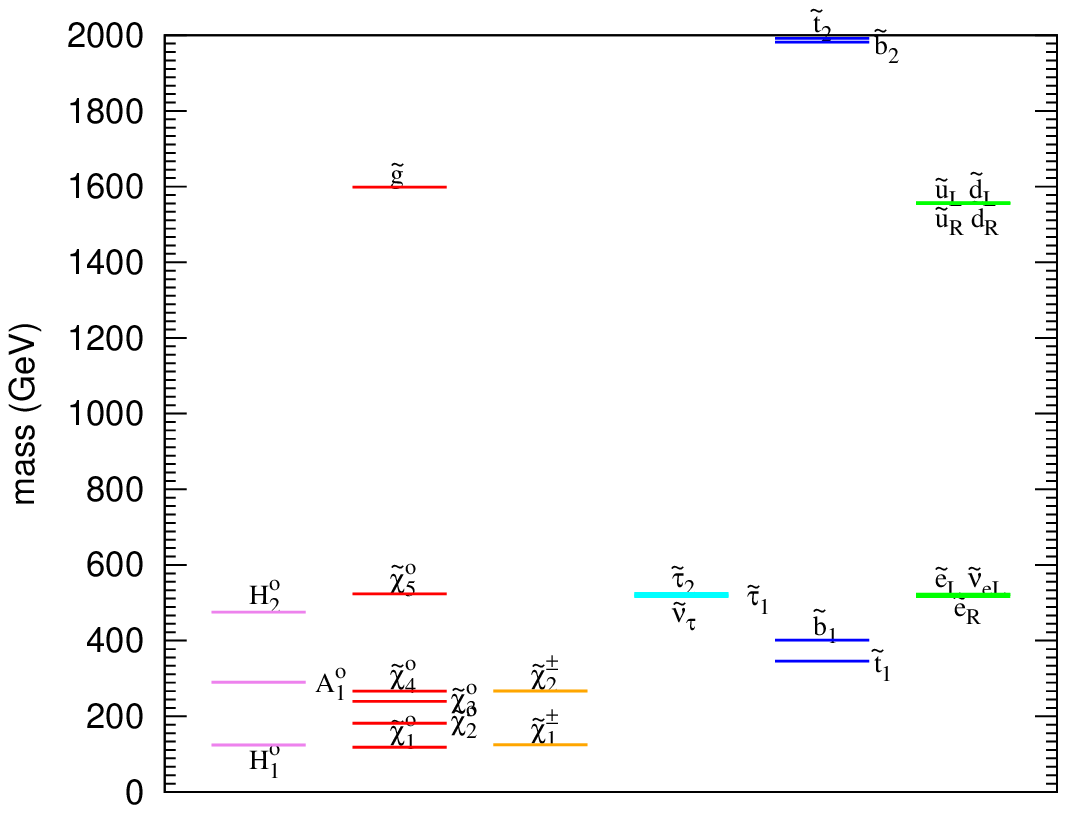}
\includegraphics[width=0.46\columnwidth]{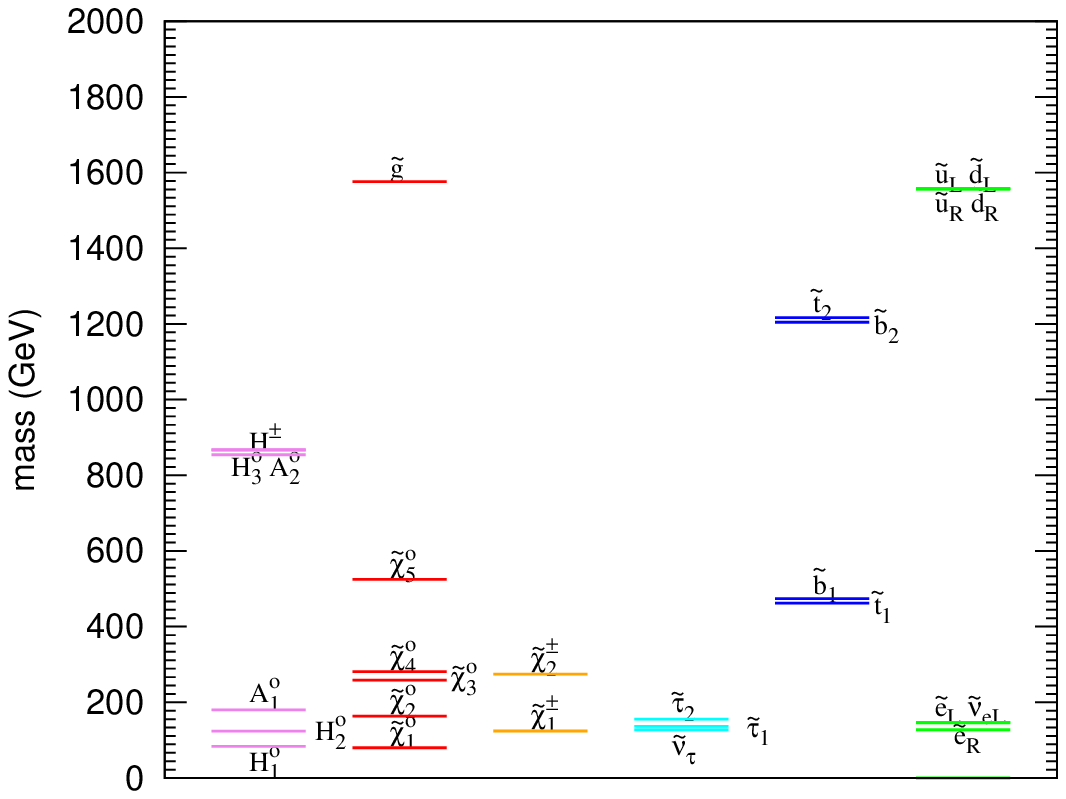}
\caption{Mass spectra of benchmark point 1 (top left), benchmark point 2 (top right), benchmark point 3 (bottom left ), and benchmark point 4 (bottom right).
\label{massspectra}}
\end{center}
\end{figure}

In this subsection we choose four benchmark points that have passed
all the constraints we considered above and discuss their features at the LHC.
In Tables. \ref{benchmarkps}, \ref{benchmarkps1}, and Fig.
\ref{massspectra}, we tabulate the mass spectra as well as main
decay modes of the stops and sbottoms of four benchmark points. \footnote{After taking into account the latest LHC Higgs measurements (after November 2012) and the superposition effect of the two almost degenerate Higgs bosons, it is noticed the first two benchmark points are on the edge of exclusion from the Higgs boson measurements. However, our discussions on the signatures of the stop and sbottom are not affected by such results.}
In the first and fourth benchmark points, the SM-like Higgs boson is $H_2$; while in the second and third benchmark point, the SM-like Higgs boson is $H_1$. For all four benchmark points, the $H_d$ dominated Higgs bosons, including $H_3$, $A_2$ and $H^\pm$ are quite heavy and are decoupled.

It is also noticed that the lightest stop and sbottom are left-handed dominant in the first two benchmark points, while they are right-handed dominant in the other two benchmark points. Due to the branching fractions $BR({\tilde t_1} \to t \tilde{\chi}^0_1)$ of these benchmark points are less than $50\%$ and also partially due to the mass splitting between $\delta m_{\tilde t} = m_{\tilde t_1} - m_{\tilde{\chi}^0_1}$, such four benchmark points have not been excluded by the LHC searches.

There are some interesting phenomenologies for these benchmark points. It is noticed that the branching fraction of ${\tilde t_1} \to t \tilde{\chi}^0_2$ and ${\tilde t_1} \to t \tilde{\chi}^0_3$ can be quite large for the first two benchmark points. For the first benchmark point, the $\tilde{\chi}^0_2$ can dominantly decay to $H_2 \tilde{\chi}^0_1 $ with branching fraction $76.4\%$, then the signature $pp \to {\tilde t_1} {\tilde t_1}^* \to t {\bar t} H_2 H_2 \tilde{\chi}^0_1 \tilde{\chi}^0_1$ can be sizable when luminosity is large enough. Therefore multi-b jets plus top pair searches should be useful for this channel. For the second benchmark point, the $\tilde{\chi}^0_2$ can dominantly decay to $\tilde{\chi}^0_1$ plus an off-shell Z boson with branching fraction $100\%$, then the signature $pp \to {\tilde t_1} {\tilde t_1}^* \to t {\bar t} Z^* Z^* \tilde{\chi}^0_1 \tilde{\chi}^0_1$ ($W^*$, $Z^*$ means off-shell gauge bosons) can be sizable when luminosity is large enough.

For the first benchmark point, the $\tilde{\chi}^0_3$ can dominantly decay to $Z(A_1)\chi^0_1$ with a branching fraction of $66.2\%$ ($31.3\%$), then the signature $pp \to {\tilde t_1} {\tilde t_1}^* \to t {\bar t} Z(A_1) Z (A_1) \tilde{\chi}^0_1 \tilde{\chi}^0_1$ can be sizable. For the second benchmark point, the $\tilde{\chi}^0_3$ also goes to $Z(A_1)\tilde{\chi}^0_1 $ but with a branching fraction of $83.0\%$ ($16.9\%$). Therefore, except for the $pp \to t {\bar t} \tilde{\chi}^0_1 \tilde{\chi}^0_1$ search, the search for $pp \to {\tilde t_1} {\tilde t_1}^* \to t {\bar t} Z(A_1) Z(A_1) \tilde{\chi}^0_1 \tilde{\chi}^0_1$ can be complementary to constrain these two benchmark points (here the branching fraction of $A_1 \to b {\bar b}$ can be $90\%$ for both benchmark points).

It is also remarkable that due to the fact that the lighter sbottom quarks are dominantly left-handed, its dominant decay mode is consequently ${\tilde b_1} \to t \tilde{\chi}^-_1$ with branching fractions of $82.9\%$ and $93.2\%$, respectively. For both benchmark points the chargino $\tilde{\chi}^+_1$ decays $100\%$ to $\tilde{\chi}^0_1$ and W boson. Therefore the production of $pp \to {\tilde t}_1 {\tilde t}_1^* $ can lead to a sizable final state with $b {\bar b} W^+ W^- W^+ W^- \tilde{\chi}^0_1 \tilde{\chi}^0_1$.

Obviously, the same sign lepton plus jets signature can help to constrain these two benchmark points. With more data sets accumulated at $\sqrt{s} =8$ TeV \cite{ATLsamelep}, it is expected that the direct searches of the LHC can either rule out or discover these three benchmark points.

For the third benchmark point, the decay mode ${\tilde t}_1 \to b \tilde{\chi}^+_2 $ is quite large, while $\tilde{\chi}^+_2$ dominantly goes to $Z \tilde{\chi}^+_1 $ and $W^+ \tilde{\chi}^0_1 $ with branching fractione $44.4\%$ and $41.3\%$, respectively. Therefore, the production of $pp \to {\tilde t}_1 {\tilde t}_1^* $ can lead to sizable final states like $b {\bar b} W^+ W^- \tilde{\chi}^0_1 \tilde{\chi}^0_1$. The decay mode ${\tilde b}_1 \to b \tilde{\chi}^0_2 $ is the dominated ${\tilde b}_1$ decay mode, while $\tilde{\chi}^0_2$ dominantly goes to $W^* \tilde{\chi}^-_1 $.

For the fourth benchmark points, more decay modes are open and many decay modes share large branching fractions. In the ${\tilde t}_1 \to b \tilde{\chi}^+_2 $ mode, the heavier chargino $\tilde{\chi}^+_2$ dominantly goes to $ H_2$, $H_1$, and $Z \tilde{\chi}^+_1$ with branching fractions $36.0\%$, $10.8\%$, and $14.7\%$ and $W\tilde{\chi}_1^0/\tilde{\chi}^0_2$ with branching fractions $15.0\%$ and $10.5\%$. For the sbottom, the decay chains also become quite long. It might be more challenging to exclude or discover this benchmark point.

For comparison, we tested the benchmark point 3 given in \cite{King:2012is} and found it is still alive. The main decay modes of light stops in this point are ${\tilde t}_1 \to t \tilde{\chi}^0_1 $, ${\tilde t}_1 \to t \tilde{\chi}^0_2 $ and ${\tilde t}_1 \to b \tilde{\chi}^+_2$ with branching fractions $16\%$, $29\%$, and $45\%$, while $\tilde{\chi}^0_2$ and $\tilde{\chi}^+_2$ decay to $Z^* \tilde{\chi}^0_1$ and $Z/H_2 \tilde{\chi}^+_1 $ with large branching fractions, respectively. We also tested the benchmark points given in \cite{Ellwanger:2012ke} and found that they are not excluded by the LHC direct SUSY searches. The main decay modes of light stops in these two benchmark points are ${\tilde t}_1 \to t \tilde{\chi}^0_1 $ and ${\tilde t}_1 \to b \tilde{\chi}^+_1$ with branching fractions $23\%$ and $56\%$ ($20\%$ and $68\%$). In these benchmark points, compared with the MSSM case, the branching fractions of ${\tilde b}_1 \to b \tilde{\chi}^0_1$ and ${\tilde t}_1 \to t \tilde{\chi}^0_1$ can be suppressed drastically if $\tilde{\chi}^0_1$ is singlino dominant. Such a fact leads to weaker constraints when we apply the LHC direct searches to parameter spaces with lighter stops and sbottoms.

\section{Conclusions}

The NMSSM provides a natural framework for the recently discovered
125 GeV Higgs boson. Within the NMSSM, we have analyzed the
constraints from the 125 GeV Higgs boson as well as the results
from the dark matter searches to the parameter space. We
concentrate on the LHC direct SUSY searches on the allowed
parameter points.

We have focused on a scenario where the stop/sbottom can be lighter
than 500 GeV and performed a detailed study to examine how the
SUSY direct search can constrain them by using the results based on $\sqrt{s}=$7 TeV and $2\sim 5$ fb$^{-1}$ of data. It is found that the direct SUSY searches,
especially the channels with tagged b jets, are powerful and can
put bounds to the allowed parameter space.

We would like to point out that when the inclusive signatures of both stop and sbottom pair productions are considered, the direct SUSY searches can exclude many parameter points with the left-handed stop/sbottom up to 500 GeV or so. With $\sqrt{s}=$8 TeV and 5 fb$^{-1}$ of data or more, although the kinematics could be a little different, we can expect the direct SUSY searches will push the light third generation squarks of NMSSM to narrower corners. However, for the benchmark points given in our work and Refs. \cite{Ellwanger:2012ke,King:2012is}, special strategies and kinematic variables are still needed for searching light stop/sbottom pair signatures.

\begin{acknowledgments}
The authors thank Tianjun Li and Jinmin Yang for valuable comments on the manuscript. The authors also thank Zhao-huan Yu for helpful discussions. This work is supported by the Natural Science Foundation of China under the Grants NO. 11105157, NO. 11075169, NO. 11175251 and NO. 11135009, the 973 project under Grant No. 2010CB833000, and the Chinese Academy of Science under Grant No. KJCX2-EW-W01.
\end{acknowledgments}

\setcounter{equation}{0}
\renewcommand{\theequation}{\arabic{section}.\arabic{equation}}%

\end{document}